\documentclass[twocolumn,preprintnumbers]{revtex4}
\usepackage{amsmath}
\usepackage{dcolumn}
\usepackage{bm}
\usepackage[all]{xy}
\usepackage{indentfirst}
\usepackage{multirow}
\usepackage{mathrsfs}
\usepackage{bbm}
\usepackage{euscript}
\usepackage{amssymb}
\usepackage{extarrows}
\usepackage{bbm}
\usepackage[ruled]{algorithm2e}
\usepackage{graphics}
\begin{document}
\title{Network configuration theory for all networks}

\author{Ming-Xing Luo}

\affiliation{\small{} The School of Information Science and Technology, Southwest Jiaotong University, Chengdu 610031, China
}

\newtheorem{Thm}{Theorem}
\newtheorem{Def}{Definition}
\newtheorem{Prop}{Proposition}
\newtheorem{Lem}{Lemma}
\newtheorem{Cor}{Corollary}
\newtheorem{Exa}{Example}
\newtheorem{Alg}{Algorithm}

\begin{abstract}
Entangled quantum networks provide great flexibilities and scalabilities for quantum information processing or quantum Internet. Most of results are focused on the nonlocalities of quantum networks. Our goal in this work is to explore new  characterizations of any networks with theory-independent configurations. We firstly prove the configuration inequality for any network using the fractional independent set of the associated graph. These inequalities can be built with polynomial-time complexity. The new result allows featuring correlations of any classical network depending only on its network topology. Similar inequalities hold for all entangled quantum networks with any local measurements. This shows an inherent feature of quantum networks under local unitary operations. It is then applied for verifying almost all multipartite entangled pure states with linear complexity, and witnessing quantum network topology without assumption of inputs. The configuration theory is further extended for any no-signalling networks. These results may be interesting in entanglement theory, quantum information processing, and quantum networks.
\end{abstract}
\maketitle

Bell inequality provides an important method to experimentally verify the nonlocality of entangled states \cite{EPR,Bell,CHSH}. It has been improved both in theory and experiment to reveal the inherent features of quantum theory \cite{BCP,HHH}. As one special case quantum nonlocality is explored in network scenarios \cite{BGP,BRG,Fritz,TALR}. The quantum network may consist of independent entangled systems. This implies new limits on quantum correlations beyond single-source network because each entanglement is shared by partial observers. Another interest difference is from the local joint operations on independent systems, which may result in distinctive quantum features \cite{BBC} or activate new nonlocal correlations \cite{Activ,Fritz}. These quantum resources allow distributed experimental settings for information processing in quantum networks \cite{Kimble,net1,net2}.

Although these improvements, however, the inherent or key features of quantum nonlocality in networks is very limited \cite{Gisin,TALR}. Even for classical networks, it is unknown how to characterize all the classical correlations available in networks consisting of independent variables. The independence assumptions of sources rule out the correlation polytope used for a single common source \cite{BCP}, but consist of non-convex correlation set. This imposes special difficulties in exploring the facet inequalities. It may be resolved for special networks by using Bell-type nonlinear inequalities [16-23] or quantum game \cite{Luo2019}.

\begin{figure}
\begin{center}
\resizebox{160pt}{120pt}{\includegraphics{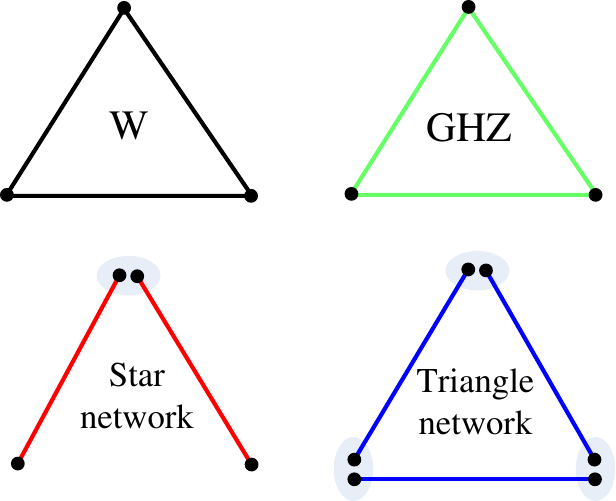}}
\end{center}
\caption{\small (Color online) Schematically tripartite networks consisting of single source, two independent sources or three independent sources.}
\label{fig1}
\end{figure}

Beyond quantum nonlocality verified with Bell inequalities, another method is to explore features of quantum correlations in networks. These limits depend only on the specific network configurations \cite{Henson}. In contrast to the standard Bell scenario, the violations of the so-called configuration inequalities, do not witness the nonlocality of quantum networks, but its network topology under local operations. One example is tripartite network with different configurations shown in Fig.\ref{fig1}. An algorithmic method is to explore information-theoretic correlations by using semidefinite programs \cite{CMG,Alex,ANDC,Tava3}. The drawback is of limits on small-size networks. Another way is from Bell-type inequalities \cite{Marc2019a}. Different from quantum nonlocality without involving any inputs \cite{Fritz,Marc2019b}, the network topology imposes fundamental limitations on achievable (classical, quantum or no-signalling) correlations \cite{CB,GPC,Marc2019a}. So far, these results hold for special quantum networks consisting of bipartite entangled resources \cite{Marc2019a,GPC,Alex}. A natural problem is how to characterize general quantum networks. This is also related to another basic problem of witnessing entanglement \cite{HHH}.

Our goal in this work is to characterize correlations derived from any networks. We firstly identify all the achievable correlations for chain networks consisting of bipartite sources by constructing new nonlinear inequalities. This can be used to verify triangle network beyond the recent result \cite{Marc2019b,Alex}. It is further extended as a general method for all networks consisting of any sources by exploring generalized Finner inequalities \cite{Finner}. Similar configuration inequalities hold for any quantum networks under any local projections beyond special networks \cite{CMG,Alex,ANDC,Tava3,Marc2019b}. This can be used to resolve the compatibility problem of quantum network topology in a device-independent manner. A specific network topology can be verified by using the observed correlations. It further motivates another remarkable application of verifying almost all multipartite entangled pure states (except for one set with zero measure) with linear complexity. Interestingly, the proposed configuration theory provides theory-independent limits of correlations in any network scenarios independent of the underlying physical model (for any possible no-signaling theory) \cite{PR,BLM,Barrett}  even for any noisy sources. It answers affirmatively a recent conjecture \cite{Marc2019a,GPC,Alex}.

\textbf{General classical networks}. Consider an $n$-partite network ${\cal N}$ consisting of independent sources $\lambda_1, \cdots, \lambda_m$. Each party $\textsf{A}_j$ shares some variables $\Lambda_j\subseteq\{\lambda_{1}, \cdots, \lambda_{m}\}$. $\Lambda$ is arbitrary and could exist prior to the measurement choices. The measurement outcome $a_j$ of $\textsf{A}_j$ depends only on its own resources $\Lambda_j$, $j=1, \cdots, n$. The generalized locally causal model \cite{Bell,Luo2018} suggests a joint probability of the measurement outcomes as
\begin{eqnarray}
P({\bf a})=\int_{\times_{j=1}^n\Omega_j} \prod_{j=1}^np(a_j|\Lambda_j) d\mu_1(\lambda_1)\cdots d\mu_m(\lambda_m),
\label{eqn-1}
\end{eqnarray}
where ${\bf a}=(a_1,\cdots{},a_n)$, $a_j\in \{0, \cdots, k_j\}$, and $(\Omega_j, \Sigma_j, \mu_j)$ denotes the measure space of $\lambda_j$, $\mu_j(\lambda_j)$ is the measure of $\lambda_j$ with the normalization condition $\int_{\Omega_j} d\mu_j(\lambda_j)=1$, and $p(a_j|\Lambda_j)$ is the probability of $a_j$ conditional on $\Lambda_j$ and satisfies $\sum_{a_j}p(a_j|\Lambda_j)=1$ for any $x_j$ and $\Lambda_i$. Different from the nonlocality [16-23], the configuration of given network can be verified using the observed correlations without assumption of inputs \cite{Henson,CMG,Alex,ANDC,Tava3,CB,GPC}. These provide fundamental limitations on correlations achievable in a given network regardless of specific sources. The only requirement is the no-signaling principle \cite{PR}.

A toy model is from tripartite networks consisting of single source \cite{GHZ,Dur}, two bipartite sources \cite{BBC} or three bipartite sources \cite{Marc2019b}, as shown in Fig.\ref{fig1}. It turns out the Finner inequality \cite{Finner,Marc2019a} given by
\begin{eqnarray}
P(a,b,c)\leq \sqrt{p(a)p(b)p(c)}
\label{fin1}
\end{eqnarray}
provides a configuration inequality for three bipartite sources. It is used for verifying the GHZ-type source \cite{Marc2019a}, but useless for other networks (Appendix A). Other methods may be useful for special networks \cite{Henson,CMG,Alex,ANDC,Tava3,CB,GPC} or sources \cite{Marc2019a}. There is few result to distinguish these networks \cite{NWR,Kraft,Luo2021,BGP,Marc2019b}. A triangle network \cite{Marc2019a} yields to the chain network \cite{BBC} regardless of one source. This intrigues a new configuration inequality for the chain network as (Appendix B)
\begin{eqnarray}
P(a,b,c)\leq p(a)^{\frac{m-k}{m}}p(b)^{\frac{k}{m}}p(c)^{\frac{m-k}{m}}
\label{fin4'}
\end{eqnarray}
with $m\geq 2$ and $k\leq m-1$. It reduces to the inequality (\ref{fin1}) for $m=2$ and $k=1$. Interestingly, the inequality (\ref{fin4'}) is efficient for verifying the W-type source of $P_{w}=p[001]+\frac{1-p}{2}([010]+[100])$ which cannot be verified by the inequality (\ref{fin1}), where $p_{xyz}[xyz]$ denotes the probability for getting the outcomes $x,y,z$. Another example is the triangle network consisting of three sources (Appendix C).

The inequality (\ref{fin1}) shows new limits of correlations beyond the inequality (\ref{fin1}). A natural problem is how to feature general networks. Our goal in what follows is to explore configuration inequalities for all networks with polynomial-time complexity. The main idea is inspired by generalized fractional independent set of graphs \cite{Schrijver}. For a given $n$-partite network ${\cal N}$, there is an associated undirected graph ${\cal G}=(V, E)$, where the vertex $i$ in $V$ denotes the party $\textsf{A}_i$, the edge $e_{ij}$ in $E$ connected with the vertexes $i$ and $j$ denotes one bipartite source shared by $\textsf{A}_i$ and $\textsf{A}_j$, and the $s$-hyper edge $e_{i_1\cdots i_s}$ in $E$ connected with the vertexes ${i_1}, \cdots, {i_s}$ denotes one $s$-partite source shared by $\textsf{A}_{i_1}, \cdots, \textsf{A}_{i_s}$.

\begin{Def}
A fractional independent set of a graph ${\cal G}$ associated with the network ${\cal N}$ is a non-negative vector of ${\bf s} =(s_1, \cdots, s_n)$ satisfying
\begin{eqnarray}
 s_j&\geq& 0,
 \nonumber\\
 \sum_{j:i\to j}s_j&\leq& 1, \forall i
 \label{clique}
\end{eqnarray}
\label{Def1}
\end{Def}

In the inequality (\ref{clique}) a nonnegative weight is setting for each node in ${\cal G}$ such that for each source $j$ the summation of the weights of nodes connected to it is at most 1. For any graph with bipartite sources, the inequality (\ref{clique}) is known as edge inequality \cite{Schrijver}. This allows an easy way to characterize all solutions by using its extreme points. It is clique inequality \cite{Schrijver} for a general graph with hyper edges, that is, multipartite sources in the corresponding network. It is NP-complete hard to find fractional clique cover numbers for any graph \cite{Schrijver}. We provide two algorithms with polynomial-time complexity for any finite networks (Appendix D).

\begin{Thm}(Informal) Let $\textbf{s}$ be a fractional independent set of graph ${\cal G}$ associated with the network ${\cal N}$. Any correlation $P$ achievable in ${\cal N}$ satisfies
\begin{eqnarray}
P({\bf a}) \leq \prod_{j=1}^nP(a_j)^{s_j}
\label{cfin1}
\end{eqnarray}
for any outcome ${\bf a}$.

\label{Thm1}
\end{Thm}

The Finner inequalities \cite{Marc2019a,Finner} provide the configuration limits for general networks consisting of bipartite sources. These kinds of restrictions are optimal for specific networks such as triangle network or cyclic networks \cite{Marc2019a,Schrijver}. Theorem 1 implies new configuration limits for all the networks. One example is an $n$-partite chain network ${\cal N}$, where each pair of $\textsf{A}_j$ and $\textsf{A}_{j+1}$ shares one bipartite source, $j=1, \cdots, n-1$. One fractional independent set is given by ${\bf s}$ with $s_i=\frac{m-k}{m}$ for odd $i$'s and $s_j=\frac{k}{m}$ for even $j$'s. The joint distribution $P({\bf a})$ satisfies the configuration inequality as
\begin{eqnarray}
P({\bf a}) \leq \prod_{\textrm{odd } j}P(a_j)^{\frac{m-k}{m}}\prod_{\textrm{even}\, j}P(a_j)^{\frac{k}{m}}
\label{chain}
\end{eqnarray}
for any $k,m$ with $1\leq k<m$ and $m\geq 2$. Another is $n$-partite star networks, where each pair of $\textsf{A}_j$ and $\textsf{A}_n$ shares one bipartite source, $j=1, \cdots, n-1$. The joint distribution $P({\bf a})$ satisfies
\begin{eqnarray}
P({\bf a}) \leq \prod_{j=1}^{n-1}P(a_j)^{\frac{m-k}{m}}P(a_n)^{\frac{k}{m}}
\label{star0}
\end{eqnarray}
where $s_j=\frac{m-k}{m}$ for $j=1, \cdots, n-1$ and $s_n=\frac{k}{m}$. Similar results hold for other networks consisting of multipartite sources. This provides a computationally efficient method for testing any network configurations beyond recent results \cite{CMG,Alex,ANDC,Tava3}. The proof is based on the generalized Finner inequality \cite{Finner} (Appendix E), which is itself interesting in applications.

\textbf{Configuration limits of all quantum networks}. Consider a general quantum network consisting of any multipartite states. From Born rule, the joint distribution after the local measurements performed on the total state $\rho$ is given by
\begin{eqnarray}
P_q({\bf a})={\rm tr}(M_{a_1}\otimes \cdots \otimes M_{a_n}\rho)
\end{eqnarray}
where $M_{a_j}$ are positive semidefinite operators satisfying $\sum_{a_j}M_{a_j}=\mathbbm{1}$ with the identity operator $\mathbbm{1}$. The quantum distribution is consistent with its from hidden variable model with single source \cite{Bell}. It also holds for quantum networks consisting of EPR states \cite{EPR,Marc2019a}. A natural problem is for exploring general quantum networks \cite{Henson,CMG,Alex,ANDC,Tava3,CB,GPC,137,138,139,140}. Our goal here is to present an efficient method for featuring all quantum networks.

\begin{Thm} (Quantum generalized Finner inequality) Let $\textbf{s}$ be a fractional independent set of graph ${\cal G}$ associated with quantum network ${\cal N}_q$. Any correlation $P_q$ achievable under local projection measurements on ${\cal N}_q$ satisfies
\begin{eqnarray}
P_q({\bf a}) \leq \prod_{j=1}^nP_{q}(a_j)^{s_j}
\label{qfin2}
\end{eqnarray}
for any outcome ${\bf a}$.

\end{Thm}

Theorem 2 means generalized Finner inequality (\ref{cfin1}) holds for quantum networks with the same topology of classical network (Appendix F). Compared with the recent result \cite{Marc2019a}, the inequality (\ref{qfin2}) holds for all quantum networks consisting of any entangled quantum states such as EPR state \cite{EPR}, multipartite entangled states \cite{GHZ,Dur} or noisy states \cite{Werner}. Remarkably, the generalized Finner inequality (\ref{cfin1}) holds for local joint measurements which can change network configurations such as entanglement swapping experiment \cite{BBC}. It means Finner inequality depends on the inherent configuration under local unitary operations. This provides a trivial feature of quantum networks different from single entanglement \cite{NWR,Kraft,Luo2021}.

\begin{Cor}
Let $\textbf{s}$ be a positive number set satisfying $\sum_{i=1}^ns_i\leq 1$. Any correlation $P_q$ achievable in all $n$-partite quantum networks satisfies
\begin{eqnarray}
P_q({\bf a}) \leq \prod_{j=1}^nP_{q}(a_j)^{s_j}
\label{qfin3}
\end{eqnarray}
for any outcome ${\bf a}$.

\label{Cor1}
\end{Cor}

As a special case of Theorem 2, the inequality (\ref{qfin3}) provides the first limit of correlations from single-source networks without inputs. Any $n$-partite quantum network ${\cal N}_q$ can be changed into a single-source network by local entangled operations. All the correlations achievable under local joint measurements on multiple-source networks are special cases of single-source networks, and then satisfy both the inequalities (\ref{qfin2}) and (\ref{qfin3}). One example is the chain network consisting of generalized EPR states \cite{EPR} with the configuration inequality (\ref{qfin3}) (Appendix G). The other is the star networks consisting of generalized EPR states \cite{EPR} with the configuration inequalities (\ref{star0}) and (\ref{qfin3}) (Appendix H). For $n$-partite cyclic networks, Theorem 2 reduces recent result \cite{Marc2019a} for an odd $n$ and new limits for an even $n$. Generally, Theorem 2 provides a general method for solving the configuration compatibility problem (Appendix I).

\textbf{Witnessing Multipartite Entanglement}. For an $n$-partite entangled pure state $|\Phi\rangle_{A_1\cdots A_n}$ on Hilbert space $\mathbb{H}_{A_1}\otimes \cdots \otimes \mathbb{H}_{A_n}$, lots of methods such as Bell inequalities \cite{Bell,CHSH,BCP} and entanglement witness \cite{HHH} are useful for verifying its entanglement. However, it is difficult to verify any multipartite entanglement. Our goal here is to address this problem using configuration inequalities in Theorem 2. Especially, for any $n$-partite entangled pure state $|\Phi\rangle$, the local joint distribution $P(a_i,a_{i+1})$ can be obtained by performing local projection measurements, $i=1, \cdots, n-1$. This provides an efficient method to verify almost all entangled pure states without their density matrices.

\begin{Cor} Almost all entangled pure states can be witnessed with linear complexity.

\label{Cor2}
\end{Cor}

Corollary 2 means any multipartite entangled pure states can be verified with linear complexity except for one set with zero measure. The main idea is from the configuration inequality (\ref{fin4'}) (Appendix J). Take a three-qubit state $|\phi\rangle_{A_1A_2A_3}$ as an example. If it is biseparable over the bipartition $A_1$ and $\{A_2, A_3\}$, it may be regarded as an equivalent chain network ${\cal N}_q$ consisting of $A_1$, $\{A_2, A_3\}$ and an auxiliary party $C$. The output $c$ of $C$ is fixed as $1$. We get the joint distribution given by $P(a_1,c=1,a_2,a_3)=P(a_1,a_2,a_3)$ which satisfies the inequality (\ref{fin4'}). This inspires an efficient method for witnessing almost all multipartite entangled pure states $|\Phi\rangle_{A_1\cdots{}A_n}$ by using distributions $P(a_i,a_{i+1})$, $i=1, \cdots, n-1$. It is linear complexity because of $n-1$ numbers of $2$-particle testing. One example is $n$-partite GHZ state \cite{GHZ} using one distribution $P(a_1,a_{2})$ from the symmetry. For an $n$-partite W state \cite{Dur} of  $|W\rangle_{A_1\cdots{}A_n}=\sum_{i=1}^n\alpha_i|1\rangle_{A_i}\otimes_{j\not=i}|0\rangle_{A_j}$, it can be verified by using the inequality (\ref{fin4'}) for $\prod_{i=1}^n\alpha_i^2\not=0$ with large $m$.

\textbf{Configuration limits of no-signalling networks}. Classical networks or quantum networks may be regarded as particular instances of general networks with no-signalling sources \cite{PR}. In this extension, the only condition is the no-signaling principle requiring local operations on local systems cannot instantaneously relay information to other systems. This shows the theory-independent restrictions on space-like separated observations on any network. Consider a no-signalling network ${\cal N}_{ns}$ with independent sources. Each source can distribute no-signalling boxes to any parties who are connected. These no-signalling boxes are then regarded as resources to generate correlated outputs. Each party can locally wire the shared boxes in a general way to determine an output \cite{PR}. This model shows a general theory for featuring most of well-known models \cite{Bell,PR}. A natural problem is to feature general constraints of no-signaling correlations. So far, various methods \cite{CB,Fritz,Tava3,Henson,GPC,137,ANDC,138,139,140} have been explored for special networks. Our consideration is to provide configuration inequalities for any no-signalling networks. Note all theories in the no-signalling framework share some axiomatic features such as the convexity of state spaces, distinguishability of local measurement and commutitity of local operations (Appendix K) \cite{BLM,Barrett}. With these axioms, we prove the present Finner inequalities hold for any no-signalling networks.

\begin{Thm} (No-signalling Finner inequality) Let $s$ be a fractional independent set of graph ${\cal G}$ associated with the no-signalling network ${\cal N}_{ns}$. Any distribution $P$ achievable in ${\cal N}_{ns}$ satisfies
\begin{eqnarray}
P_{ns}({\bf a}) \leq \prod_{j=1}^nP_{ns}(a_j)^{s_j}
\label{nsfin}
\end{eqnarray}
for any outcome ${\bf a}$.

\label{Thm3}
\end{Thm}

Theorem \ref{Thm3} shows the generalized Finner inequality holds for no-signalling networks, that is, the inequality (\ref{cfin1}) is no-signalling. This implies a method to verify distributions generated by local measurements on any no-signalling networks without inputs. Different from previous approaches \cite{CB,Fritz,Tava3,Henson,GPC,137,ANDC,138,139,140}, the configuration inequality (\ref{nsfin}) can be built for any finite networks within polynomial-time complexity. This answered a recent conjecture for the Finner inequality \cite{Marc2019a}.

\textbf{Robustness of network configurations}. The present network configurations are robust against general noises. Especially, Theorems 1-3 hold for any networks consisting of noisy sources. The main reason is from the corresponding proofs independent of specific sources (pure or noisy states in quantum scenarios for example). This provides an efficient method to verify the correlations available in noisy networks.

\begin{figure}[ht] 
\centering 
\resizebox{240pt}{120pt}{\includegraphics{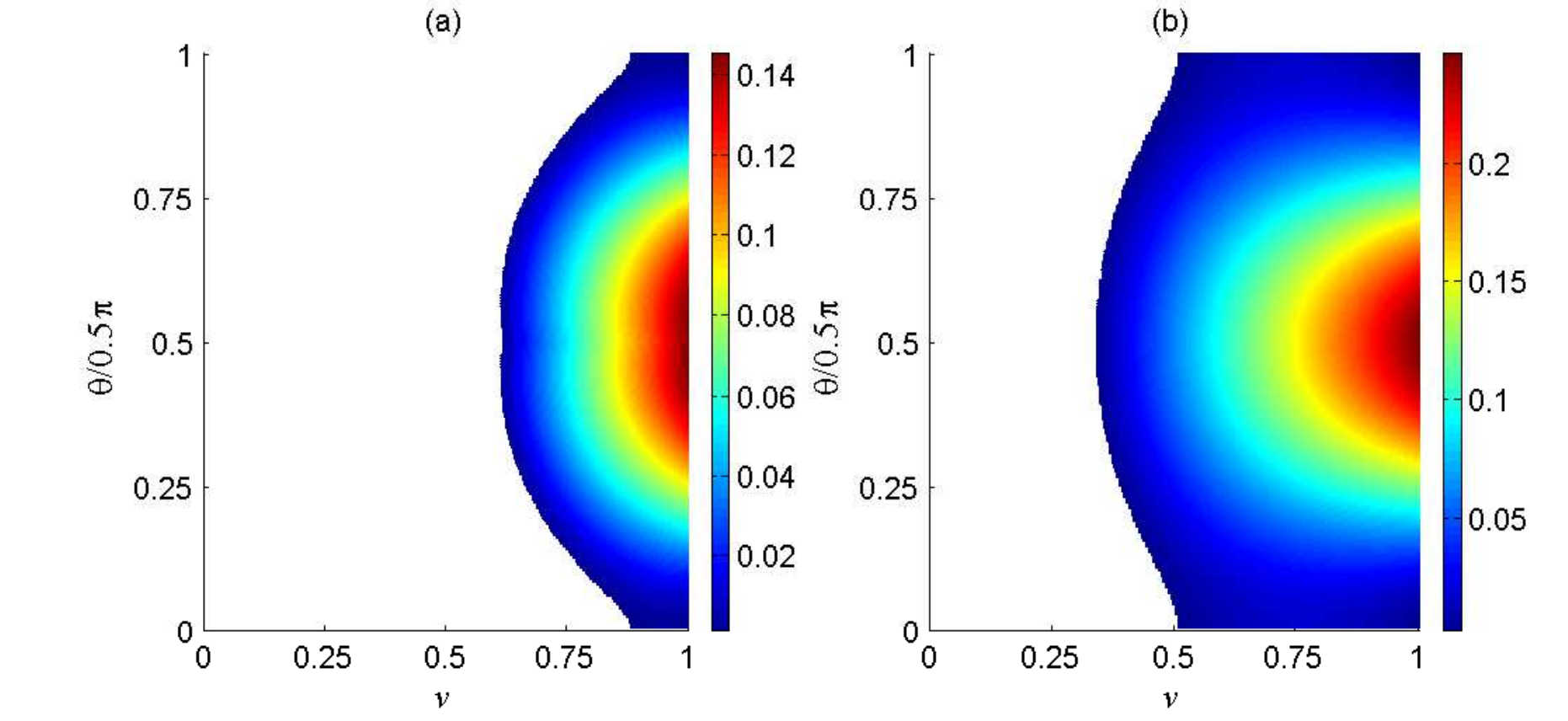}}
\caption{\small (Color online) (a) Achievable regions of $\theta$ and $v$ defined by the inequality (\ref{fin1}). (b) Achievable regions of $\theta$ and $v$ defined by the inequality (\ref{fin4}). Here, $m=1000$.}
\label{fig2}
\end{figure}

\textit{Example 1}. Consider a noisy GHZ state \cite{GHZ,Werner} as
\begin{eqnarray}
\rho_{ghz}=v|ghz\rangle\langle ghz|+\frac{1-v}{8}\mathbbm{1}
\label{nghz}
\end{eqnarray}
where $|ghz\rangle=\cos\theta|000\rangle+\sin\theta|111\rangle$ with $\theta\in (0,\frac{\pi}{2})$, and $v\in (0,1)$.  From the inequality (\ref{fin4'}) with large $m$, the visibility is given by $v^*=\frac{3-\sqrt{9-8\cos2\theta^2}}{4\cos2\theta^2}$ using the distribution $P(0,0,0)$ from local projection measurement. The achievable regions of $\theta$ and $v$ with which at least one distribution $P(a,b,c)$ violates the inequality (\ref{fin4'}) are shown in Fig.\ref{fig2}. The inequality (\ref{fin4'}) is useful for verifying noisy GHZ states for any $\theta\in (0,\frac{\pi}{2})$ and $v>0.5$ beyond the inequality (\ref{fin1}) for $v>0.88$.

\textit{Example 2}. Consider a noisy W state \cite{Dur,Werner} as
\begin{eqnarray}
\rho_{W}=v|W\rangle\langle W|+\frac{1-v}{8}\mathbbm{1}
\label{nw}
\end{eqnarray}
where $|w\rangle=\cos\theta\cos\gamma|001\rangle+\sin\theta\cos\gamma|010\rangle
+\sin\gamma|100\rangle$ and $v\in (0,1)$. The joint distribution of $P(0,0,1)$ violates the inequality (\ref{fin4'}) with large $m$ for any $v\in(0,1)$ while it is impossible for the inequality (\ref{fin1}).

\begin{figure}[ht]
\begin{center}
\resizebox{240pt}{120pt}{\includegraphics{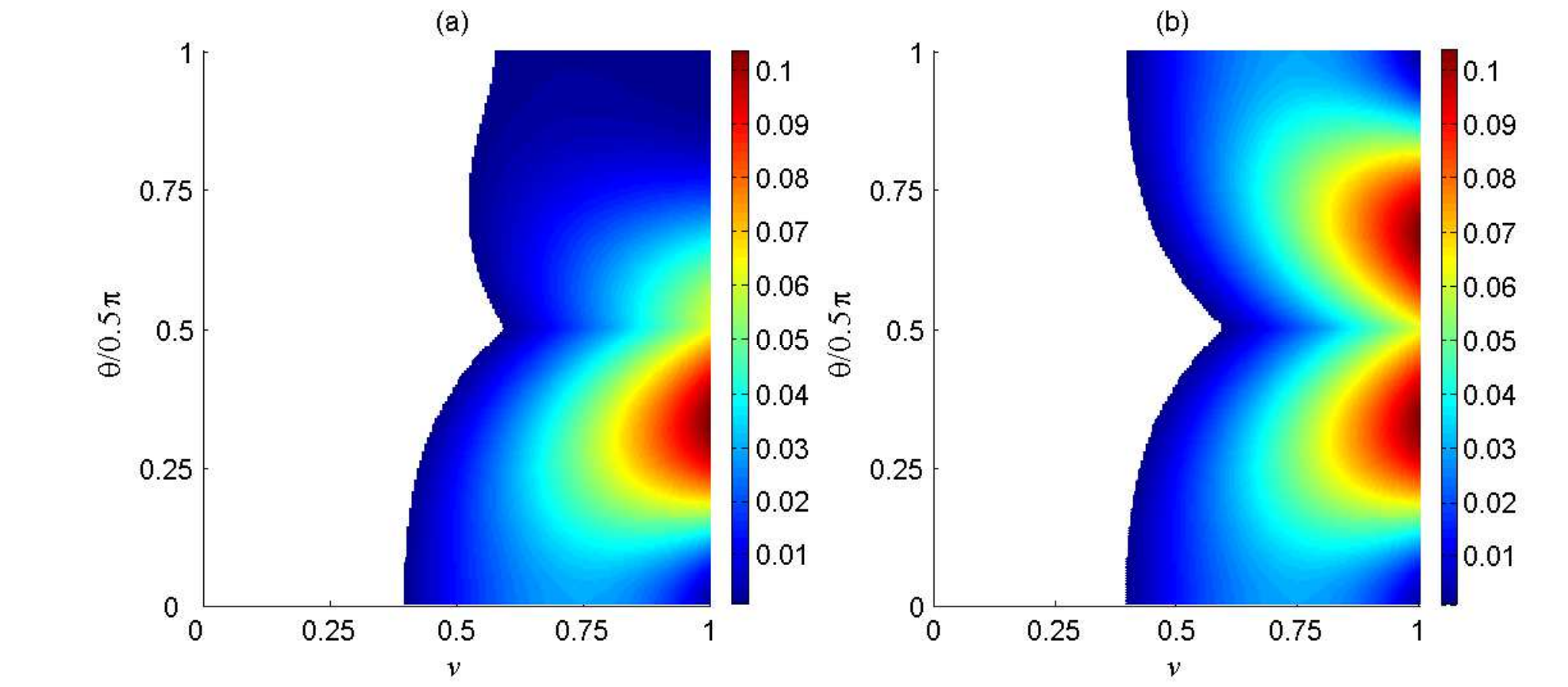}}
\end{center}
\caption{\small (Color online) Achievable regions of $\theta$ and $v$.  (a) Noisy triangle network. (b) Noisy $4$-partite star-type network. The achievable regions of $\theta$ and $v$ are obtained if at least one distribution $P(a,b,c)$ violates the inequality (\ref{fin4'}) with $m=1000$.}
\label{fig3}
\end{figure}

\textit{Example 3}. Consider a triangle network \cite{Marc2019b} consisting of three generalized noisy EPR states \cite{EPR,Werner} of
\begin{eqnarray}
\rho=v|EPR\rangle\langle{}EPR|+\frac{1-v}{4}\mathbbm{1}
\end{eqnarray}
where $|EPR\rangle=\cos\theta|00\rangle+\sin\theta|11\rangle$ with $\theta\in (0,\frac{\pi}{2})$, and $v\in (0,1)$. The inequality (\ref{fin4'}) can verify all the correlations for $\theta\in (0,\frac{\pi}{2})$ and $v\geq 0.59$, as shown in Fig.\ref{fig3}(a). Similar result holds for $n$-partite star networks, where one party $\textsf{A}_i$ is shared one noisy EPR state $\rho$ with $\textsf{A}_n$, $i=1, \cdots, n-1$. The achievable regions of $\theta$ and $v$ are shown in Fig.\ref{fig3}(b) with $n=4$.

\textit{Discussions}. The inequality (\ref{fin4'}) provides configuration limits for any no-signalling networks. An interesting feature is from large $m$. Especially, the inequality (\ref{fin4'}) is changed into a linear form as
\begin{eqnarray}
P(a,b,c)\leq p(a)p(c)
\label{fin4''}
\end{eqnarray}
when $m\to \infty$ and $p(b)\not=0$. It may be regarded as the independence of two parties in experiment. This provides the first facet inequality for any tripartite chain networks. Similar results hold from Theorem 1 (Appendix L). A natural problem is to explore other facet inequalities. This is related to the completeness of configuration inequalities. Another problem is related to verify generic entanglement such as many-body systems.

In summary, we provided an operational characterization of theory-independent correlations from networks consisting of independent sources. The main idea is inspired by the graph theory of generalized Finner inequalities. This allows featuring correlations of all classical networks depending only on its network topology. Similar results hold for all entangled quantum networks. This implies a general method for witnessing almost all entangled pure states. It is also applicable for testing entangled quantum network topology without assumption of inputs. Similar result can be extended for no-signalling networks. The present results should be interesting in entanglement theory, quantum information processing, and quantum networks.

\section*{Acknowledgements}

We thank Marc-Olivier Renou, Alejandro Pozas-Kerstjens, Tavakoli Armin, Carlos Palazuelos, Luming Duan, Donglin Deng, Shaoming Fei, and Salman Beigi. This work was supported by the National Natural Science Foundation of China (No.61772437), Fundamental Research Funds for the Central Universities (No.2682014CX095), Chuying Fellowship, and EU ICT COST CryptoAction (No.IC1306).

\appendix

\section{The Finner inequality (2)}

In this section, we show the Finner inequality (2) for the triangle network cannot be used for verifying other networks.

\subsection{W-type distributions}

Consider a generalized W state \cite{Dur} as
\begin{eqnarray}
|W\rangle=\alpha_1|001\rangle+\alpha_2|010\rangle+\alpha_3|100\rangle
\end{eqnarray}
where $\alpha_j$ satisfies $\sum_{j=1}^3\alpha_j^2=1$. By performing the projection measurement with basis $\{|0\rangle, |1\rangle\}$ on each particle, the joint distribution is given by
\begin{eqnarray}
P_{w}(a,b,c)=\alpha_1^2[001]+\alpha_2^2[010]+\alpha_3^2[100].
\label{W}
\end{eqnarray}
It follows that
\begin{eqnarray}
&&p_a(0)=1-p_3^2, p_a(1)=p_3^2,
\nonumber\\
&&p_b(0)=1-p_2^2, p_b(1)=p_2^2,
\nonumber\\
&&p_c(0)=1-p_1^2, p_c(1)=p_1^2,
\label{W1}
\end{eqnarray}
where $p_a(x), p_b(y), p_c(y)$ denote the respective marginal distribution of three parties for the outcomes $x, y, z$.

From Eqs.(\ref{W}) and (\ref{W1}) we get
\begin{eqnarray}
&&P(0,0,1)^2-p_a(0)p_b(0)p_c(1)
\nonumber
\\
&=&p_1^4-(1-p_3^2)(1-p_2^2)p_1^2
\nonumber
\\
&=&-p_3^2p_2^2p_1^2
\nonumber
\\
&\leq &0
\end{eqnarray}
for any $\alpha_i$'s. The distribution of $P(0,0,1)$ does not violate the inequality (\ref{fin1}).

Similarly, we can prove
\begin{eqnarray}
&&P(0,1,0)\leq \sqrt{p_a(0)p_b(1)p_c(0)}
\\
&&P(1,0,0)\leq \sqrt{p_a(1)p_b(0)p_c(0)}
\end{eqnarray}
for any $\alpha_i$'s. The Finner inequality (\ref{fin1}) holds for all the generalized W distributions given in Eq.(\ref{W}).

\subsection{Chain quantum networks}

Consider an entanglement swapping network consisting of two generalized EPR states $|\phi_1\rangle|\phi_2\rangle$ with $|\phi_i\rangle=\cos\theta_i|00\rangle+\sin\theta_i|11\rangle$ and $\theta_i\in (0,\frac{\pi}{2})$. Here, the parties $\textsf{A}_1$ and $\textsf{A}_2$ share $|\phi_1\rangle$ while $\textsf{A}_2$ and $\textsf{A}_3$ share $|\phi_2\rangle$. By performing the projection measurement with the basis $\{|0\rangle, |1\rangle\}$ on each qubit, the joint distribution is given by
\begin{eqnarray}
P(a,b,c)=p_0[000]+p_1[011]+p_2[120]+p_3[131]
\label{EW}
\end{eqnarray}
where the two-qubit states for $\textsf{A}_2$ are encoded into a four-dimensional states with the basis $|0\rangle:=|00\rangle, \cdots, |3\rangle=|11\rangle$, $p_0=\cos^2\theta_1\cos^2\theta_2$, $p_1=\cos^2\theta_1\sin^2\theta_2$, $p_2=\sin^2\theta_1\cos^2\theta_2$ and $p_3=\sin^2\theta_1\sin^2\theta_2$. It follows that
\begin{eqnarray}
&&p_a(0)=\cos^2\theta_1, p_a(1)=\sin^2\theta_1,
\nonumber\\
&&p_b(0)=\cos^2\theta_1\cos^2\theta_2, p_b(1)=\cos^2\theta_1\sin^2\theta_2,
\nonumber\\
&&p_b(2)=\sin^2\theta_1\cos^2\theta_2, p_b(3)=\sin^2\theta_1\sin^2\theta_2,
\nonumber\\
&&p_c(0)=\cos^2\theta_2, p_c(1)=\sin^2\theta_2.
\label{EW1}
\end{eqnarray}

For the distributions in Eqs.(\ref{EW}) and (\ref{EW1}), from the forward evaluation we get
\begin{eqnarray}
 P(a,b,c)=\sqrt{p(a)p(b)p(c)}
 \label{finner1ac}
\end{eqnarray}
for any $a,b,c\in \{0,1\}$. The Finner inequality (\ref{fin1}) cannot verify any chain networks even without noises.

\subsection{Biseparable quantum networks}

Consider a one-side entanglement network consisting of one-qubit state $|\phi_1\rangle=\cos\theta_1|0\rangle+\sin\theta_1|1\rangle$ and one generalized EPR state $|\phi_2\rangle$ with $|\phi_2\rangle=\cos\theta_2|00\rangle+\sin\theta_2|11\rangle$, $\theta_1,\theta_2\in (0,\frac{\pi}{2})$. Here, the party $\textsf{A}_1$ has $|\phi_1\rangle$ while the parties $\textsf{A}_2$ and $\textsf{A}_3$ share $|\phi_2\rangle$. By performing the projection measurement on each particle, the joint distribution is given by
\begin{eqnarray}
P(a,b,c)=p_0[000]+p_1[011]+p_2[100]+p_3[111]
\label{bs}
\end{eqnarray}
where $p_0=\cos^2\theta_1\cos^2\theta_2$, $p_1=\cos^2\theta_1\sin^2\theta_2$, $p_2=\sin^2\theta_1\cos^2\theta_2$, and $p_3=\sin^2\theta_1\sin^2\theta_2$.  It follows that
\begin{eqnarray}
&&p_a(0)=\cos^2\theta_1, p_a(1)=\sin^2\theta_1,
\nonumber\\
&&p_b(0)=\cos^2\theta_2, p_b(1)=\sin^2\theta_2,
\nonumber\\
&&p_c(0)=\cos^2\theta_2, p_c(1)=\sin^2\theta_2.
\label{bs1}
\end{eqnarray}
For the probability distributions in Eqs.(\ref{bs}) and (\ref{bs1}), we can get
\begin{eqnarray}
 P(a,b,c)\leq \sqrt{p(a)p(b)p(c)}
 \label{finner1aa}
\end{eqnarray}
for any $a,b,c\in \{0,1\}$. The Finner inequality (\ref{fin1}) cannot verify any biseparable quantum networks even without noises.

\subsection{Fully separable quantum networks}

Consider one fully separable network consisting of three single-qubit states $\otimes_{i=1}^3|\phi_i\rangle$ with $|\phi_i\rangle=\cos\theta_i|0\rangle+\sin\theta_i|1\rangle$, $\theta_i\in (0,\frac{\pi}{2})$. Here, the party $\textsf{A}_i$ has $|\phi_i\rangle$, $i=1,2,3$. By performing the projection measurement on each qubit, the joint distribution is given by
\begin{eqnarray}
P(a,b,c)=\sum_{i,j,k=0}^1p_{ijk}[ijk]
\label{fs}
\end{eqnarray}
where $p_{ijk}$ is defined by $p_{ijk}=\cos^{2-2i}\theta_i\sin^{2i}\theta_i\cos^{2-2j}\theta_j\sin^{2j}\theta_j
\cos^{2-2k}\theta_k\sin^{2k}\theta_k$. It follows that
\begin{eqnarray}
&&p_a(0)=\cos^2\theta_1, p_a(1)=\sin^2\theta_1,
\nonumber\\
&&p_b(0)=\cos^2\theta_2, p_b(1)=\sin^2\theta_2,
\nonumber\\
&&p_c(0)=\cos^2\theta_3, p_c(1)=\sin^2\theta_3.
\label{fs1}
\end{eqnarray}
For the probability distributions in Eqs.(\ref{fs}) and (\ref{fs1}), we can get
\begin{eqnarray}
 P(a,b,c)\leq \sqrt{p(a)p(b)p(c)}
 \label{finner1ab}
\end{eqnarray}
for any $a,b,c\in \{0,1\}$. The Finner inequality (\ref{fin1}) cannot verify any fully separable networks.

\subsection{General cases}

Consider a triangle network consisting of three independent sources $\lambda_1, \lambda_2, \lambda_3$, where each pair of two parties share one source. The joint probability distribution of $P(a,b,c)$ is defined by
\begin{eqnarray}
P(a,b,c)&=&\int_{\Omega}p(a|\lambda_1,\lambda_2)p(b|\lambda_2,\lambda_3)
p(c|\lambda_1,\lambda_3)\mu_1(\lambda_1)
\nonumber\\
&&\times \mu_2(\lambda_2)\mu_3(\lambda_3)d\lambda_1
d\lambda_2d\lambda_3
\label{Ba1}
\end{eqnarray}
where $p(a|\lambda_1,\lambda_2)$ denotes the indicator function of the first party with outcome $a$ conditional on the connected sources $\lambda_1$ and $\lambda_2$, $p(b|\lambda_2,\lambda_3)$ and $p(c|\lambda_1,\lambda_3)$ are defined similarly, $\mu_i(\lambda_i)$ denotes the probability distribution of source $\lambda_i$, $i=1, 2, 3$.

Consider a chain network consisting of three parties $A_1, A_2, A_3$, where the parties $A_1$ and $A_3$ are respectively connected the source $\lambda_1$ and $\lambda_2$ while $A_2$ is connected with two sources. Denote $a, b, c$ as the respective output of three parties. The joint probability distribution of $P(a,b,c)$ is defined by
\begin{eqnarray}
P(a,b,c)&=&\int_{\Omega}p(a|\lambda_1)p(b|\lambda_1,\lambda_2)
p(c|\lambda_2)
\nonumber\\
&&\times\mu_1(\lambda_1)\mu_2(\lambda_2)d\lambda_1
d\lambda_2
\label{B111}
\end{eqnarray}
where $p(a|\lambda_1)$ denotes the indicator function of the first party with outcome $a$ conditional on the connected source $\lambda_1$, $p(b|\lambda_1,\lambda_2)$ and $p(c|\lambda_2)$ are defined similarly, $\mu_i(\lambda_i)$ denotes the probability distribution of source $\lambda_i$, $i=1, 2$. By regarding the source $\lambda_1$ in Eq.(\ref{Ba1}) as constant, the joint distribution in Eq.(\ref{Ba1}) yields to its in Eq.(\ref{B111}). All the correlations obtained from chain networks can be obtained from triangle networks. This can be extended to general networks.

\textbf{Proposition S1}. \textit{Consider two $n$-partite networks ${\cal N}_1$ and ${\cal N}_2$ consisting of some independent sources. Suppose ${\cal N}_1$ is a sub-network of ${\cal N}_2$ (or ${\cal N}_1\subset {\cal N}_2$). All the correlations  $\{P_1(a_1\cdots{}a_n)|{\cal N}_1\}$ available in ${\cal N}_1$ can be generated from ${\cal N}_2$, that is
\begin{eqnarray}
\{P_1(a_1\cdots{}a_n)|{\cal N}_1\}\subset \{P_2(a_1\cdots{}a_n)|{\cal N}_2\}.
\label{B112}
\end{eqnarray}}

From Proposition S1, all the correlations of $\{P_1(a_1\cdots{}a_n)|{\cal N}_1\}$ satisfy any configuration inequality which holds for $\{P_2(a_1\cdots{}a_n)|{\cal N}_2\}$. Hence, the Finner inequality of triangle network cannot be used to verify the correlations derived from chain networks.

\section{Proof of Inequality (3)}

Consider an $n$-partite long-distance chain network consisting of $n$ parties $\textsf{A}_1, \cdots, \textsf{A}_n$. Each pair of two parties $\textsf{A}_j$ and $\textsf{A}_{j+1}$ shares one source $\lambda_j$, $j=1, \cdots, n-1$. Our goal in this section is to prove the generalized version of the inequality (\ref{fin4'}) as
\begin{eqnarray}
P({\bf a})&\leq & \min\{ \prod_{\textrm{even}\, i}p(a_i)^{\frac{m-k}{m}}\prod_{\textrm{odd}\, j}p(a_j)^{\frac{k}{m}},
\nonumber
\\
&&\prod_{\textrm{odd}\, i}p(a_i)^{\frac{m-k}{m}}\prod_{\textrm{even}\, j}p(a_j)^{\frac{k}{m}}
\}
\label{fin4}
\end{eqnarray}
for any $m$ and $k$ with $m\geq 2$ and $k\leq m-1$, where ${\bf a}=(a_1,\cdots{},a_n)$, and $p(a_i)$ denotes the marginal distribution of the outcome $a_i$ for the party $\textsf{A}_i$, $i=1, \cdots, n$.

\textbf{Proof of the inequality (\ref{fin4})}. We firstly prove the following inequality
\begin{eqnarray}
P({\bf a})&\leq & \prod_{\textrm{even}\, i}p(a_i)^{\frac{m-k}{m}}\prod_{\textrm{odd}\, j}p(a_j)^{\frac{k}{m}}
\label{B1}
\end{eqnarray}
for any $m\geq 2$. In fact, let $(\Omega_j, q_{j}(\lambda_j), \mu(\lambda_j))$ be the measurable space of the source $\lambda_j$, where $q_{j}(\lambda_j)$ denotes the probability distribution of $\lambda_j$, and $\mu(\lambda_j)$ denotes the measure of $\lambda_j$, $j=1, \cdots, n-1$. From Eq.(1) we have
\begin{eqnarray}
P({\bf a})
&=&\int_{\Omega_1\times\cdots \times \Omega_n }\!\!\!\!\!\!\!\!\!\!\!\!p(a_1|\lambda_1)\prod_{j=2}^{n-1}p(a_j|\lambda_{j-1},\lambda_j)p(a_n|\lambda_{n-1})
\nonumber\\
&&\times
q_1(\lambda_1)\cdots{}q_{n-1}(\lambda_{n-1})d\mu(\lambda_1)
\cdots{}d\mu(\lambda_{n-1})
\nonumber
\\
&=&
\int_{\Omega_1\times\cdots \times \Omega_n }\prod_{\textrm{even}\, i} p(a_i|\lambda_{i-1},\lambda_i)q_{i-1}(\lambda_{i-1})^{\frac{m-k}{m}}
\nonumber\\
&&\times q_i(\lambda_{i})^{\frac{m-k}{m}}\prod_{\textrm{odd}\, i} p(a_j|\lambda_{j-1},\lambda_j)q_{j-1}(\lambda_{j-1})^{\frac{k}{m}}
\nonumber\\
&&\times{}q_j(\lambda_{j})^{\frac{k}{m}}
d\mu(\lambda_1)\cdots{}d\mu(\lambda_{n-1})
\label{B3}
\\
&\leq &
\prod_{\textrm{even}\, i}(\int_{\Omega_{i-1}\times\Omega_{i}} p(a_i|\lambda_{i-1},\lambda_i)^{\frac{m-k}{m}} q_{i-1}(\lambda_{i-1})
  \nonumber\\
&&  \times{}q_i(\lambda_i)d\mu(\lambda_1)\cdots{}d\mu(\lambda_{n-1}))^{\frac{m-k}{m}}
\nonumber\\
&&\times
      \prod_{\textrm{odd}\, j}(\int_{\Omega_{j-1}\times\Omega_{j}} p(a_j|\lambda_{j-1},\lambda_j)^{\frac{k}{m}} q_{j-1}(\lambda_{j-1})
\nonumber\\
&&\times{}q_i(\lambda_i)d\mu(\lambda_1)\cdots{}d\mu(\lambda_{n-1}))^{\frac{k}{m}}
\label{B4}
\\
&=&
\prod_{\textrm{even}\, i}(\int_{\Omega_{i-1}\times\Omega_{i}} p(a_i|\lambda_{i-1},\lambda_i)q_{i-1}(\lambda_{i-1})
  \nonumber\\
&&  \times{}q_i(\lambda_i)d\mu(\lambda_1)\cdots{}d\mu(\lambda_{n-1}))^{\frac{m-k}{m}}
\nonumber\\
&&\times
      \prod_{\textrm{odd}\, j}(\int_{\Omega_{j-1}\times\Omega_{j}} p(a_j|\lambda_{j-1},\lambda_j)q_{j-1}(\lambda_{j-1})
\nonumber\\
&&\times{}q_i(\lambda_i)d\mu(\lambda_1)\cdots{}d\mu(\lambda_{n-1}))^{\frac{k}{m}}
\label{B5}
\\
&=&
\prod_{\textrm{even}\, i}p(a_i)^{\frac{m-1}{m}}\prod_{\textrm{odd}\, j}p(a_j)^{\frac{k}{m}}
\label{B6}
\end{eqnarray}
In Eq.(\ref{B3}), $\lambda_0$ and $\lambda_n$ are two constants, and $m\geq 2$. The inequality (\ref{B4}) follows from the H\"{o}lder inequality:
\begin{eqnarray}
\int{}fg{}dx\leq (\int{}f{}^{p}dx)^{\frac{1}{p}}(\int{}g{}^{q}dx)^{\frac{1}{q}}
\end{eqnarray}
with positive constants $p$ and $q$ satisfying $\frac{1}{p}+\frac{1}{q}=1$. Eq.(\ref{B5}) follows from the equalities: $p(a_j|\lambda_{j-1},\lambda_j)^{s_j}=p(a_j|\lambda_{j-1},\lambda_j)\in \{0,1\}$ with positive constant $s_j$, where  $p(a_j|\lambda_{j-1},\lambda_j)$ is indicator function, that is, $p(a_j|\lambda_{j-1},\lambda_j)\in \{0,1\}$, $j=1, \cdots, n$. Eq.(\ref{B6}) follows from the equalities:
\begin{eqnarray}
p(a_j)&=&\int{}_{\Omega_{j-1}\times \Omega_j}p(a_j|\lambda_{j-1},\lambda_j)q_{j-1}
(\lambda_{j-1})
\nonumber\\
&&\times{}q_j(\lambda_j)d\mu(\lambda_{j-1})d\mu(\lambda_{j})
\end{eqnarray}
for $j=1, \cdots, n$.

Similarly, we can prove
\begin{eqnarray}
P({\bf a})&\leq & \prod_{\textrm{odd}\, i}p(a_i)^{\frac{m-k}{m}}\prod_{\textrm{even}\, j}p(a_j)^{\frac{k}{m}}
\label{B7}
\end{eqnarray}
for any $m\geq 2$ and $k\leq m-1$. This completes the proof.

\section{Verifying triangle network}

In this section, we show the inequality (3) can be used to verify the correlations available in triangle networks. Consider a triangle network ${\cal N}$ consisting of three independent sources with the probability distributions
\begin{eqnarray}
P(x_i,y_i)=p_i[00]+q_i[11], i=1, 2, 3.
\label{Ca1}
\end{eqnarray}
Here, the party $\textsf{A}_1$ shares two different sources $x_1$ and $y_3$, the party $\textsf{A}_2$ shares two different sources $y_1$ and $x_2$, and the party $\textsf{A}_3$ shares two different sources $y_2$ and $x_3$. Let $a=x_1y_3$, $b=y_1x_2$ and $c=y_2x_3$. If each party performs independently sampling on each source, the joint distribution is given
\begin{eqnarray}
P(a,b,c)&=&p_1p_2p_3[000]+p_1q_2p_3[012]+q_1p_2p_3[120]
\nonumber\\
&&+q_1q_2p_3[132]+p_1p_2q_3[201]+p_1q_2q_3[213]
\nonumber\\
&&+q_1p_2q_3[321]+q_1q_2q_3[333]
\label{C2}
\end{eqnarray}
where two bit series $xy$ are encoded into the four states with the basis $\{0, 1, 2, 3\}$. It follows that
\begin{eqnarray}
&& p_a(0)=p_1p_3, p_a(1)=q_1p_3,
\nonumber\\
&& p_a(2)=p_1q_3, p_a(3)=q_1q_3,
\nonumber\\
&& p_b(0)=p_1p_2, p_b(1)=p_1q_2,
\nonumber\\
&& p_b(2)=q_1p_2, p_b(3)=q_1q_2,
\nonumber\\
&& p_c(0)=p_2p_3, p_c(1)=p_2q_3,
\nonumber\\
 && p_c(2)=q_2p_3, p_c(3)=q_2q_3.
\label{C3}
\end{eqnarray}
From Eqs.(\ref{C2}) and (\ref{C3}) we obtain
\begin{eqnarray}
v_{000}&:=&P(0,0,0)-p_a(0)^{\frac{m-1}{m}}p_b(0)^{\frac{1}{m}}p_c(0)^{\frac{m-1}{m}}
\nonumber\\
&=& p_1p_2p_3-p_1p_2p_3^{\frac{2m-2}{m}}
\nonumber\\
&>&0
\label{C4}
\end{eqnarray}
for $m>2$ and $k=1$ and $p_1,p_2,p_3\in (0, 1)$. The probability distribution of $P(0,0,0)$ violates the inequality (\ref{fin4'}) for any $p_1,p_2,p_3\in (0,1)$ and large $m$. However, it does not violate the inequality (\ref{fin1}).

\section{Fractional independent sets}

Given an $n$-partite network ${\cal N}$ consisting of parties $\textsf{A}_1, \cdots, \textsf{A}_n$ who shares any independent sources $\lambda_1, \cdots, \lambda_m$ (classical variables or quantum states for example). There is an undirected graph ${\cal G}=(V, E)$ associated with ${\cal N}$, where the vertex $v_i$ in $V$ denotes one party  $\textsf{A}_i$, the edge $e_{ij}$ in $E$ connected with the vertexes $v_i$ and $v_j$ denotes one bipartite source shared by the parties $\textsf{A}_i$ and $\textsf{A}_j$, and the $s$-hyper edge $e_{i_1\cdots i_s}$ in $E$ connected with the vertexes $v_{i_1}, \cdots, v_{i_s}$ denotes one $s$-partite source shared by the parties $\textsf{A}_{i_1}, \cdots, \textsf{A}_{i_s}$. It is general difficult to find optimal fractional independent sets for the graph ${\cal G}$ associated with the given network ${\cal N}$. Here, we provide two efficient algorithms for solving this problem regardless of the optimal solutions. The first one is an application of bipartite graphs. The second is more adapt to network configurations. Both algorithms can provide fractional independent sets for each network even if they may be not optimal.

\begin{minipage}{8.2cm}
\begin{algorithm}[H]
	\centering
	\caption{Fractional independent set}
\begin{itemize}
\item[]
\item[Input:] An undirected graph ${\cal G}=(V, E)$.
\item[Output:] A fractional independent set $\textbf{s}=(s_1, \cdots, s_n)$
\item{} Set $x_j=\frac{1}{m_j}$ for each $m_j$-hyper edge $e_j\in E$;
\item{} Set $s_i=\min\{s_j|j\to i\}$;
\item{} Set $\textbf{s}=(s_1, \cdots, s_n)$.
\end{itemize}
\label{alg1}
\end{algorithm}
\end{minipage}

\begin{figure}
\begin{center}
\includegraphics{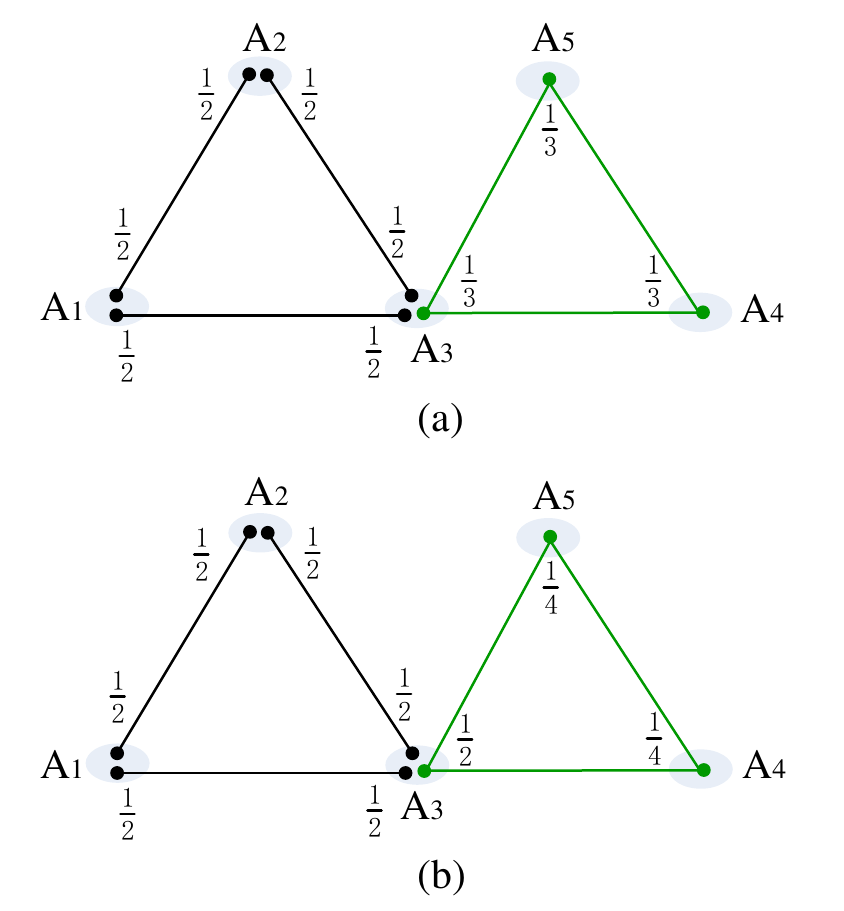}
\end{center}
\caption{\small (Color online) 5-partite network. $\textsf{A}_1, \textsf{A}_2$ and $\textsf{A}_3$ share a triangle network consisting of three bipartite sources. $\textsf{A}_3, \textsf{A}_4$ and $\textsf{A}_5$ share a tripartite source corresponding to the $3$-hyper edge in the graph.}
\label{figs1}
\end{figure}

For given an $n$-partite network consisting of $m$ sources, the total time complexity is $O(mnN)$, where $N$ denotes the number of edges. It follows Algorithm 1 has polynomial time-complexity in terms of sources and edges.

\textbf{Example S1}. Take the network shown in Fig.\ref{figs1} as example. There are five parties $\textsf{A}_1,\cdots, \textsf{A}_5$. $\textsf{A}_1, \textsf{A}_2$ and $\textsf{A}_3$ share a triangle network consisting of three bipartite sources. $\textsf{A}_3, \textsf{A}_4$ and $\textsf{A}_5$ share a tripartite source associated with a hyper edge in graph. Here, we firstly set $\frac{1}{2}$ for each edge while $\frac{1}{3}$ for the hyper edge. And then, we update the covering number of $\textsf{A}_3$ as $s_3=\min\{\frac{1}{2},\frac{1}{3}\}=\frac{1}{3}$, $s_i=\frac{1}{2}$ for $\textsf{A}_i$ with $i=1, 2$, and $s_j=\frac{1}{3}$ for $\textsf{A}_j$ with $j=4,5$. Finally, the fractional independent set $\textbf{s}=(s_1, \cdots, s_6)$ is given by
\begin{eqnarray}
\textbf{s}=(\frac{1}{2}, \frac{1}{2}, \frac{1}{3}, \frac{1}{3}, \frac{1}{3})
\label{D1}
\end{eqnarray}
as shown in Fig.\ref{figs1}(a).

\begin{minipage}{8.2cm}
\begin{algorithm}[H]
	\centering
	\caption{Fractional independent set}
\begin{itemize}
\item[]
\item[Input] An undirected graph ${\cal G}=(V, E)$.
\item[Output] A fractional independent set $\textbf{s}=(s_1, \cdots, s_n)$
\item{} Decompose ${\cal G}$ into a combination of different subgraphs ${\cal G}_j$, that is, ${\cal G}=\cup_{j=2}^t{\cal G}_j$, where ${\cal G}_2$ denotes the bipartite graph, and ${\cal G}_j$ denotes the $j$-hyper edge associated with $j$-partite source;
\item{} Set a fractional independent set $\textbf{s}_j$ for the subgraph ${\cal G}_j$;
\item{} Update $s_i$ for the $i$-th node as $s_i=\min\{s_k\in \textbf{s}_j|k\to i\}$;
\item{} Set $\textbf{s}=(s_1, \cdots, s_n)$.
\end{itemize}
\label{alg2}
\end{algorithm}
\end{minipage}

The main difference between Algorithms 1 and 2 is the covering set for each subgraph. In Algorithm 2, one can choose different fractional independent sets depending on the network configurations.

\textbf{Example S2}. For Algorithm 2, set $\frac{1}{2}$ for each edge of $\textsf{A}_1, \textsf{A}_2$ and $\textsf{A}_3$ in Fig.\ref{figs1} while $\{\frac{1}{2}, \frac{1}{4}, \frac{1}{4}\}$ for three nodes of the hyper edge. And then, we update the covering number as $s_i=\frac{1}{2}$ for $i=1, 2, 3$, and $s_j=\frac{1}{4}$ for $j=4,5$. Finally, a fractional independent set $\textbf{s}=(s_1, \cdots, s_6)$ is given by
\begin{eqnarray}
\textbf{s}=(\frac{1}{2}, \frac{1}{2}, \frac{1}{2}, \frac{1}{4}, \frac{1}{4})
\label{D2}
\end{eqnarray}
which is different from the fractional independent set given in Eq.(\ref{D1}).

For given an $n$-partite network consisting of $m$ sources, the total time complexity of Algorithm 2 is $O(mLN)$, where $N$ denotes the number of edges and $L$ denotes the maximal time complexity for setting fractional independent set $\textbf{s}_j$. Note $L$ is upper bounded by some constant depending on all the sub-graphs. This is because ${\cal G}_j\subset {\cal G}$. It follows that Algorithm 2 has polynomial time-complexity in terms of sources and edges.

A more involved method depends on the connectedness of nodes. A general principle is to set large fractional index for those nodes with large connected edges. This may be completed by using the independent set \cite{Luo2018}. The total time is polynomial in terms of sources and edges.

\begin{figure}
\begin{center}
\resizebox{200pt}{220pt}{\includegraphics{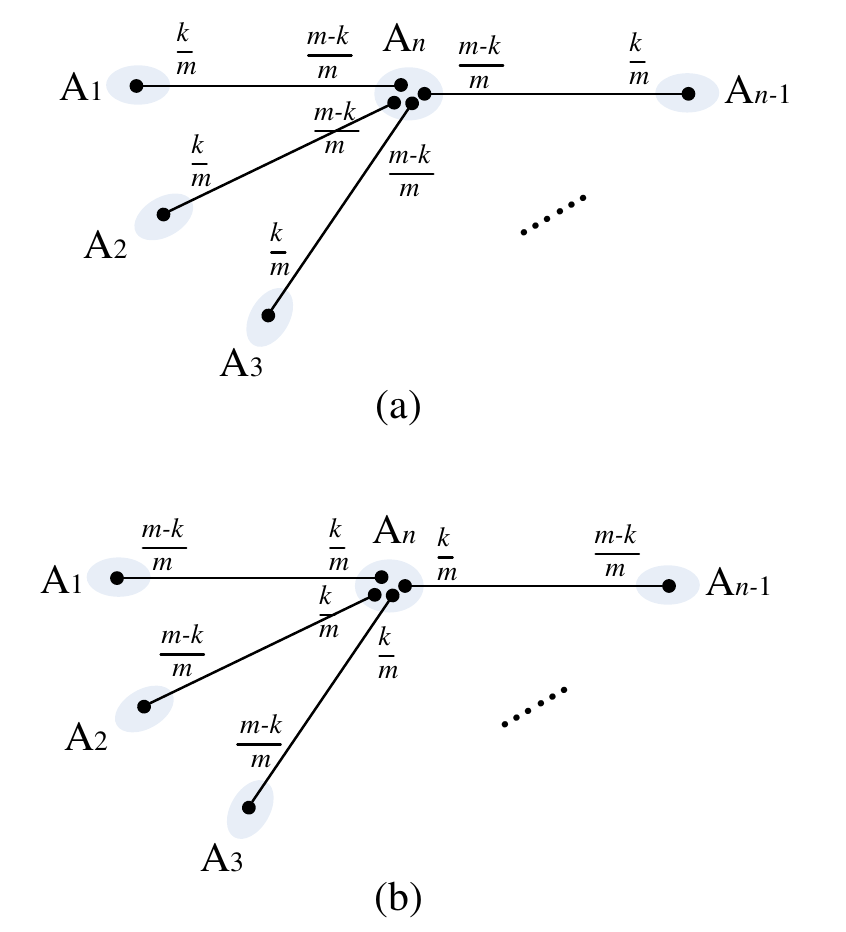}}
\end{center}
\caption{\small (Color online) An $n$-partite star network. The party $\textsf{A}_n$ shares one independent bipartite source with $\textsf{A}_j$, $j=1, \cdots, n-1$. Here, $m\geq 2$.}
\label{figs2}
\end{figure}

\textbf{Example S3 (Star Network)}. Consider an $n$-partite star network, as shown in Fig.\ref{figs2}. Here, one party $\textsf{A}_n$ has $n-1$ connected edges while there is one connected edge for each one $\textsf{A}_j$ of other parties, $j=1, \cdots, n-1$. Hence, we can get $s_{n}=\frac{m-k}{m}$ for $\textsf{A}_n$ and $s_{j}=\frac{k}{m}$ for $\textsf{A}_j$, $j=1, \cdots, n-1$. The fractional independent set $\textbf{s}=(s_{1}, \cdots, s_{n-1}, s_n)$ is given by
\begin{eqnarray}
\textbf{s}=(\frac{k}{m}, \cdots, \frac{k}{m}, \frac{m-k}{m})
\label{D3}
\end{eqnarray}
with $m\geq 2$ and $k\leq m-1$, as shown in Fig.\ref{figs2}(a). Another method is to get the fractional independent set $\textbf{s}$ as
\begin{eqnarray}
\textbf{s}=(\frac{m-k}{m}, \cdots, \frac{m-k}{m}, \frac{k}{m})
\label{D4}
\end{eqnarray}
as shown in Fig.\ref{figs2}(b). Both fractional independent sets satisfy Definition 1.

\begin{figure}
\begin{center}
\resizebox{200pt}{110pt}{\includegraphics{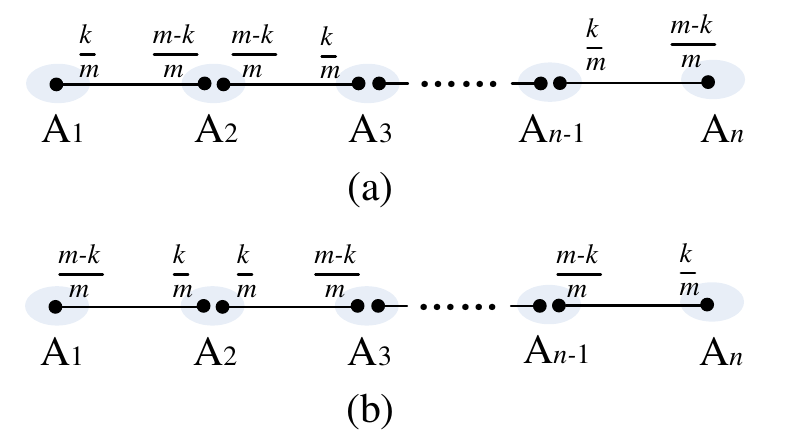}}
\end{center}
\caption{\small (Color online) An $n$-partite chain network. Each pair of two parties $\textsf{A}_j$ and $\textsf{A}_{j+1}$ shares one bipartite source, $j=1, \cdots, n-1$. Here, $m\geq 2$ and $n$ is an even integer.}
\label{figs3}
\end{figure}

\textbf{Example S4 (Chain network)}. Consider an $n$-partite chain network. Here, each pair of two parties $\textsf{A}_j$ and $\textsf{A}_{j+1}$ shares one source, $j=1, \cdots, n-1$. The set of $\{\textsf{A}_{2j}|j=1,\cdots, \lceil\frac{n}{2}\rceil\}$ is an independent set \cite{Luo2018}, that is, any pair of them have not shared sources. Hence, we can get $s_{2j}=\frac{m-k}{m}$ for $\textsf{A}_{2j}$, and $s_{{2j+1}}=\frac{k}{m}$ for $\textsf{A}_{2j+1}$, $j=1,\cdots, \lceil\frac{n}{2}\rceil$, where $\lceil{x}\rceil$ denotes the minimal integer no less than $x$, $m\geq 2$ and $k\leq m-1$. The fractional independent set of $\textbf{s}=(s_{1}, s_{2}, s_{3}, \cdots, s_{{n}})$ is given by
\begin{eqnarray}
\textbf{s}=(\frac{k}{m}, \frac{m-k}{m}, \frac{k}{m},  \cdots, \frac{m-k}{m})
\label{D5}
\end{eqnarray}
for an even integer $n$, as shown in Fig.\ref{figs3}(a), and
\begin{eqnarray}
\textbf{s}=(\frac{m-k}{m}, \frac{k}{m}, \frac{m-k}{m},  \cdots, \frac{k}{m}, \frac{m-k}{m})
\label{D6}
\end{eqnarray}
for an odd integer $n$, as shown in Fig.\ref{figs3}(b). Another method is to set $s_{2j}=\frac{k}{m}$ for $\textsf{A}_{2j}$, and $s_{{2j+1}}=\frac{m-k}{m}$ for $\textsf{A}_{2j+1}$, $j=1,\cdots, \lceil\frac{n}{2}\rceil$. The fractional independent set of $\textbf{s}$ is given by
\begin{eqnarray}
\textbf{s}=(\frac{m-k}{m}, \frac{k}{m}, \frac{m-k}{m},  \cdots, \frac{k}{m})
\label{D7}
\end{eqnarray}
for an even integer $n$, as shown in Fig.\ref{figs3}(b), and
\begin{eqnarray}
\textbf{s}=(\frac{m-k}{m}, \frac{k}{m}, \frac{m-k}{m},  \cdots, \frac{k}{m}, \frac{k}{m})
\label{D8}
\end{eqnarray}
for an integer odd $n$, where $2k\leq m$. Both fractional independent sets satisfy Definition 1.

\begin{figure}
\begin{center}
\resizebox{160pt}{210pt}{\includegraphics{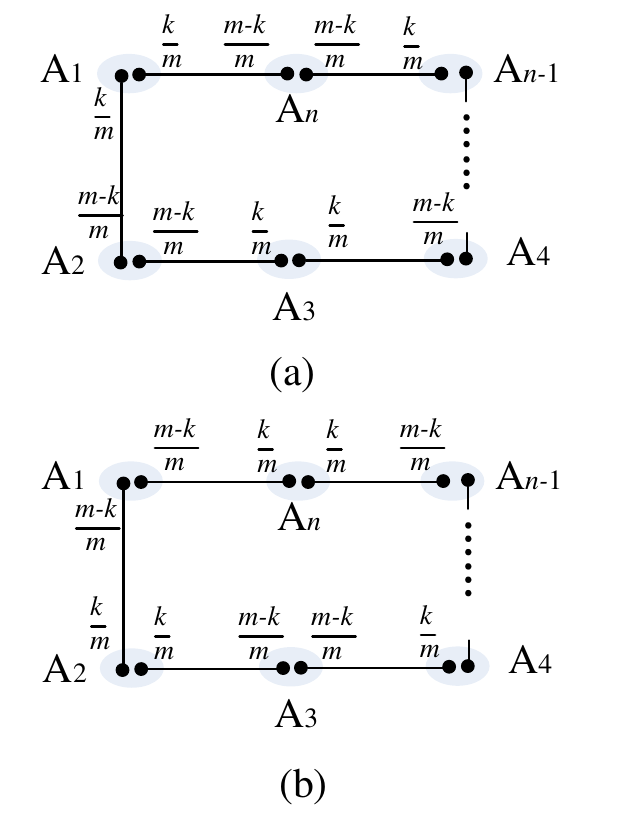}}
\end{center}
\caption{\small (Color online) An $n$-partite cyclic network. Each pair of two parties  $\textsf{A}_j$ and $\textsf{A}_{j+1}$ (or $\textsf{A}_1$ and $\textsf{A}_{n}$ ) shares one bipartite source, $j=1, \cdots, n-1$. Here, $m\geq 2$ and $n$ is even integer.}
\label{figs4}
\end{figure}

\textbf{Example S5 (Cyclic network)}. Consider an $n$-partite cyclic network, shown as Fig.\ref{figs4}. Here, each pair of two parties $\textsf{A}_j$ and $\textsf{A}_{j+1}$ (or $\textsf{A}_1$ and $\textsf{A}_{n}$ ) shares one bipartite source, $j=1, \cdots, n-1$. Similar to $n$ chain network shown in Fig.\ref{figs3}, we get the fractional independent sets given in Eqs.(\ref{D5})-(\ref{D8}).

\begin{figure}
\begin{center}
\resizebox{200pt}{360pt}{\includegraphics{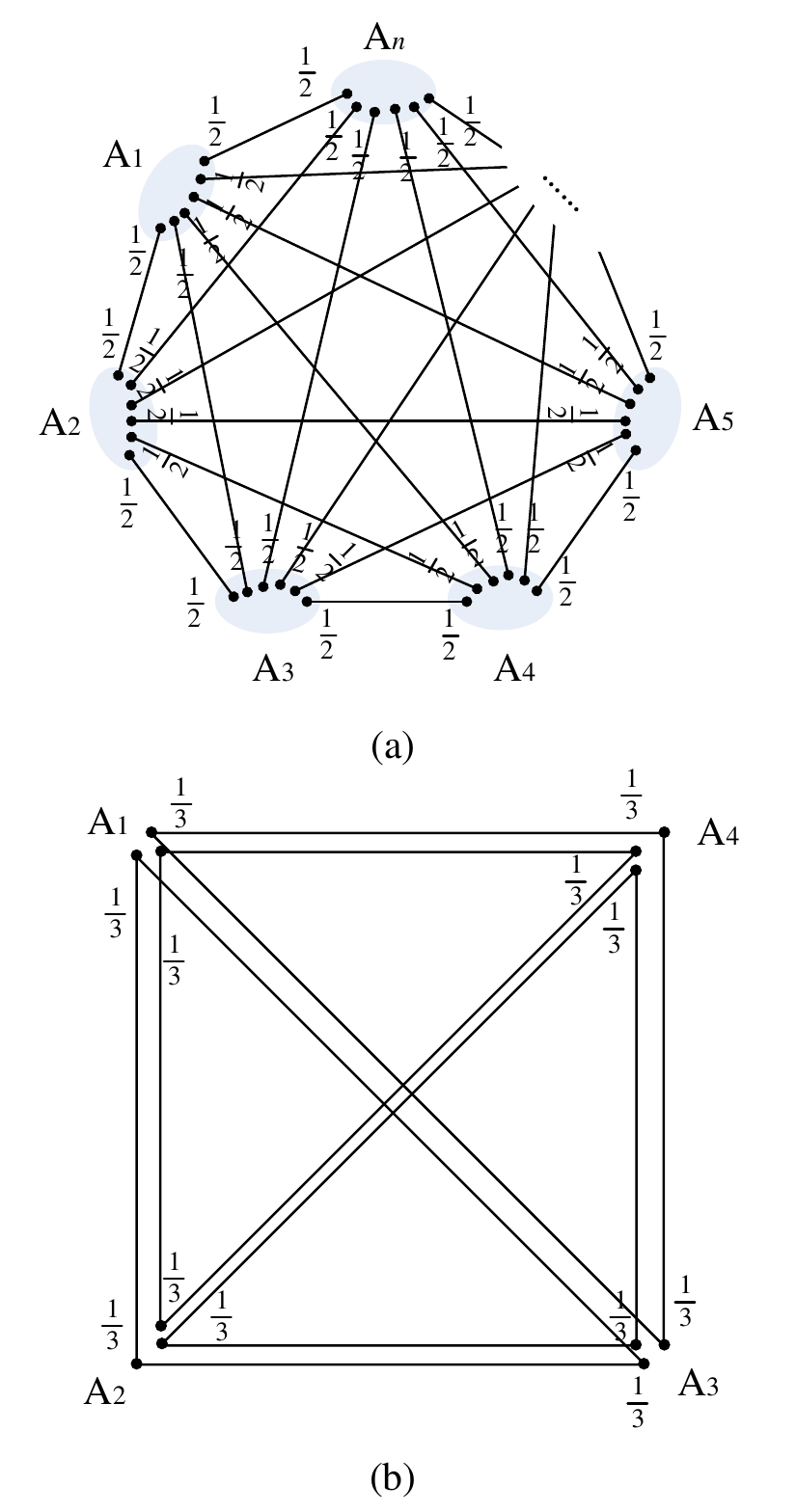}}
\end{center}
\caption{\small (Color online) (a) An $n$-partite complete network. Each pair of two parties $\textsf{A}_i$ and $\textsf{A}_{j}$ shares one bipartite source, $1\leq i\not=j\leq n$. (b) A $4$-partite complete network. Each three parties $\textsf{A}_i, \textsf{A}_{j},\textsf{A}_{k}$ shares one tripartite source, $1\leq i\not=j\not=k\leq 4$. Here, $m\geq 2$.}
\label{figs5}
\end{figure}

\textbf{Example S6 (Complete network)}. Consider an $n$-partite complete network (associated with a complete hyper graph), shown in Fig.\ref{figs5}, where each $m$ parties $\textsf{A}_{i_1}, \cdots, \textsf{A}_{i_m}$ shares one source, $1\leq i_1, \cdots, i_m\leq  n$. We set $s_{i}=\frac{1}{m}$ for each party $\textsf{A}_{i}$. The fractional independent set of $\textbf{s}=(s_{1}, s_{2}, \cdots, s_{{n}})$ is given by
\begin{eqnarray}
\textbf{s}=(\frac{1}{m}, \cdots, \frac{1}{m})
\label{D9}
\end{eqnarray}
which satisfies Definition 1. For the special case of $m=2$, it reduces to the Finner inequality \cite{Finner}, as shown in Fig.\ref{figs5}(a). Another example of $m=3$ is shown in Fig.\ref{figs5}(b).

\section{Proof of Theorem 1}

Suppose $(\Omega_{i}, {\cal A}_i, \mu_i), i\in I_n=\{1, \cdots, n\}$ are measure spaces. Let $(\Omega_{S}, {\cal A}_S, \mu_S)$ be the product space of $(\Omega_{i}, {\cal A}_i, \mu_i)$ with $i\in S$, that is, $\Omega_{S}=\times_{i\in S}\Omega_i, {\cal A}_S=\otimes_{i\in S}{\cal A}_i$ and $\mu_S=\otimes_{i\in S}\mu_i$ for any set $S\subset I_n$. Let $L^{p}(\Omega_{S}, {\cal A}_{S}, \mu_{S})$ be the space consisting of equivalence classes of measurable functions $f: \Omega_{S}\to \mathbb{R}$  such that $\int_{\Omega_{S}}|f|^pd\mu<\infty$. Theorem 1 is formalized as

\textbf{Theorem 1'}. \textit{Let $m\in \mathbb{N}$, $M=\{1, \cdots, m\}$, $\emptyset\not=S_j\subseteq{}I_n$, $M_j=\{j\in M: i\in S_j\}$ for $i\in I_n$ such that $\sum_{j\in M_i}s_{ij}\leq 1$ for all $i\in I_n$, and $0<s_{ij}<1$. Suppose $f_j\in L^{\frac{1}{q_j}}(\Omega_{S_j}, {\cal A}_{S_j}, \mu_{S_j})$ with $q_j=\min_j\{s_t,t\in S_j\}$ and $|f_j|\leq 1$, $j\in M$. We have $\prod_{j}|f_j|\in L^{1}(\Omega_{I_n}, {\cal A}_{S_n}, \mu_{S_n})$ and
\begin{eqnarray}
\int\prod_{j\in M}|f_j|d\mu_{I_n}\leq \prod_{j\in M}(\int|f_j|^{\frac{1}{q_j}})^{q_j}
\label{E1}
\end{eqnarray}
Moreover, if $f_j\in \{0, \pm 1\}$, that is, $f_j$ being characteristic function, we get
\begin{eqnarray}
\int\prod_{j\in M}|f_j|d\mu_{I_n}\leq \prod_{j\in M}(\int|f_j|^{\frac{1}{q_j}})^{q_j}
\label{E2}
\end{eqnarray}
where $q_j=\min_j\{s_t,t\in S_j\}$}.

\textbf{Proof}. The proof is inspired by the Fubini Theorem \cite{Finner} and completed by induction on $n$. For $n=1$, from the generalized H\"{o}lder inequality \cite{Finner} it follows that
\begin{eqnarray}
\int\prod_{j\in M}|f_j|d\mu_{I_n}\leq \prod_{j\in M}(\int|f_j|^{\frac{1}{p_j}})^{p_j}
\label{E4}
\end{eqnarray}
with $\sum_jp_j=1$ and $p_j\in (0,1)$. Combined with the norm inequality of
\begin{eqnarray}
(\int |f_j|^{\frac{1}{s_j}}d\mu_{S_j})^{s_j} \leq (\int |f_j|^{\frac{1}{q_j}}d\mu_{S_j})^{q_j}
\label{E5}
\end{eqnarray}
with $q_j\leq s_j$, it follows the inequality (\ref{E1}).

Now, assume the inequality (\ref{E1}) holds for $n-1$ with $n\geq 2$. Let $i\in I_n$, $J_i=I_n\backslash\{i\}$, $\emptyset\not=S_i\subseteq{}I_n$, and $M_i=\{i\in M: j\in S_i\}$.

From the Fubini Theorem \cite{Finner} we get
\begin{eqnarray}
\int_{\Omega_{I_n}}\prod_{j\in M}|f_j|d\mu_{I_n}
&=&
\int_{\Omega_{J_i}}\int_{\Omega_i}\prod_{j\in M}|f_j|d\mu_{i}d\mu_{J_i}
\nonumber
\\
&=&\!\!\!\int_{\Omega_{J_i}}\prod_{j\in M\backslash{}M_i}\!\!\!|f_j|(\int_{\Omega_i}\prod_{j\in M_i}|f_j|d\mu_{i})d\mu_{J_i}
\nonumber
\\
&\leq &
\int_{\Omega_{J_i}}\prod_{j\in M\backslash{}M_i}|f_j|d\mu_{J_i}
\nonumber
\\
&&\times(\int_{\Omega_i}\prod_{j\in M_i}|f_j|^{\frac{1}{s_{ij}}}d\mu_{i})^{s_{ij}}
\label{E6}
\end{eqnarray}
which follows from the H\"{o}lder inequality. By decomposing $j\in M_i$ as $j\in M_i:S_j=\{i\}$ and $t\in M_i:S_t\not=\{i\}$, it follows that
\begin{eqnarray}
 \int_{\Omega_{I_n}}\prod_{j\in M}|f_j|d\mu_{I_n}
&\leq &\prod_{j\in M_i: \atop{S_j=\{i\}}}(\int_{\Omega_i}|f_j|^{\frac{1}{s_{ij}}}d\mu_{i})^{s_{ij}}
\nonumber
\\
&&\times\int_{\Omega_{J_i}}\prod_{j\in M\backslash{}M_i}|f_j|\prod_{t\in M_i: \atop{S_t\not=\{i\}}}
\nonumber
\\
&&\times(\int_{\Omega_i}|f_k|^{\frac{1}{s_{it}}}d\mu_{i})^{s_{it}}d\mu_{J_i}
\nonumber\\
&\leq &\prod_{j\in M_i: \atop{S_j=\{i\}}}(\int_{\Omega_i}|f_j|^{\frac{1}{q_{j}}}d\mu_{i})^{q_{j}}
\nonumber
\\
&&\times\int_{\Omega_{J_i}}\prod_{j\in M\backslash{}M_i}|f_j|
\nonumber
\\
&&\times\prod_{t\in M_i: \atop{S_t\not=\{i\}}}(\int_{\Omega_i}|f_k|^{\frac{1}{q_{t}}}d\mu_{i})^{q_{t}}d\mu_{J_i}
\nonumber
\\
\label{E7}
\end{eqnarray}
from the norm inequality (\ref{E5}). From the assumption of $n-1$, it follows from the inequality (\ref{E7})
\begin{eqnarray}
\int_{\Omega_{I_n}}\prod_{j\in M}|f_j|d\mu_{I_n}
&\leq &\prod_{j\in M_i:\atop{S_j=\{i\}}}(\int_{\Omega_i}|f_j|^{\frac{1}{q_j}}d\mu_{i})^{q_j}
\nonumber
\\
&&\times\prod_{j\in M\backslash{}M_i}(\int_{\Omega_{S_j}}|f_j|^{\frac{1}{q_j}}d\mu_{S_j})^{q_j}
\nonumber
\\
&&\times\!\!\!\prod_{j\in M_i: \atop{S_j\not=\{i\}}}(\int_{\Omega_{S_j\backslash\{i\}}}\!\!\!\!\!\!|f_j|^{\frac{1}{q_j}}d\mu_{i})^{q_j}d\mu_{S_j\backslash\{i\}}
\nonumber\\
&=& \prod_{j\in M}(\int_{\Omega_{S_j}}|f_j|^{\frac{1}{q_j}}d\mu_{S_j})^{q_j}
\label{E8}
\end{eqnarray}
This has completed the proof of the inequality (\ref{E1}). The inequality (\ref{E2}) is a special case of the inequality (\ref{E1}). It completes the proof.

Different from the Finner inequality \cite{Finner}, the necessary condition may be not true because of the inequality $\sum_{j\in M_i}s_{ij}\leq 1$.

\section{Proof of Theorem 2}

The generalized Finner inequality holds for all quantum networks. It can be formalized as follows:

\textbf{Theorem 2'(Quantum generalized Finner inequality)}. \textit{Let $s=\{s_1, \cdots, s_n\}$ be a fractional independent set of graph ${\cal G}$ associated with quantum network ${\cal N}_q$. Let $f_j$ be any real positive local post-processing (function) of the classical output (under the local projections with computation basis) of the party $\textsf{A}_j$. Then, any correlation $P$ achievable in ${\cal N}_q$ satisfies
\begin{eqnarray}
\mathbb{E}_q[\prod_{j=1}^nf_j]\leq \prod_j\|f_j\|_{\frac{1}{s_j}}
\label{Gth1}
\end{eqnarray}
where $\|f\|=\mathbb{E}[f^{1/s}]^s$ and the expectation $\mathbb{E}(\cdot{})$ is with respect to $P$. In particular, if $f_j$ is the indicator function of the output of $\mathsf{A}_j$ associated with the outcome $a_j$ (under any projections with general orthogonal basis), we have
\begin{eqnarray}
P_q(a_1, \cdots, a_n) \leq \prod_{j=1}^nP_{q}(a_j)^{s_j}
\label{Gth2}
\end{eqnarray}}

A well-known result for single source is given by

\textbf{Lemma 1}. \textit{Any correlation $P$ achievable in $n$-partite states can be generated by $n$-partite variable.}

\textbf{Proof}. Different from previous proof for general network consisting of bipartite sources \cite{Marc2019a}, we do not take use of special features such as the convexity or extra point of fractional independent sets. Our goal is to present a general method with less restriction. This is essential for extensions in no-signalling model. In fact, for each $m$-partite state $\rho_j$, we define a set of $\overline{s}_j:=\{s_{i\to j}\}$ for all the parties $\textsf{A}_i$'s who share the state $\rho_j$, where $s_{i\to j}\in [0,1]$ and satisfies $\sum_{i}s_{i\to j}\leq 1$. This set is further regarded as a fractional independent set of one hyper-edge in the associated graph of ${\cal N}_q$. Consider any set of ${\bf s}=\cup_j\overline{s}_j=(s_1, \cdots, s_n)$ being a total fractional independent set of the associated graph ${\cal G}$, where $s_j=\min\{s_{i\to j}\}$.

\textbf{Case 1}. Proof of the inequality (\ref{Gth1})

Similar to previous discussions \cite{Marc2019a}, for $s_j=0$, we get
\begin{eqnarray}
\|f_j\|_{\frac{1}{s_j}}=\|f_j\|_{\infty}
=\max_{a_j:P_{A_j}(a_j)\not=0}f(a_j)
\label{G1}
\end{eqnarray}
and
\begin{eqnarray}
\mathbb{E}_q[\prod_{i=1}^nf_i]\leq \mathbb{E}_q[\prod_{i\not=j}f_i]\|f_{j}\|_{\infty}
\label{G2}
\end{eqnarray}
If the generalized Finner inequality (\ref{Gth1}) holds by ignoring the $n$-th party (for the joint distribution $P(a_1,\cdots, a_n)$), it also holds including this party by putting $s_j=0$. Hence, we can restrict all the weights to be $\textbf{s}=(s_1, \cdots, s_n)$ such that $s_j\in (0,1]$ for all $j$'s.

Moreover, for $s_j=1$, the $j$-th party cannot share any source with others. Here, we get
\begin{eqnarray}
\mathbb{E}_q[\prod_{i=1}^nf_i]\leq \mathbb{E}_q[\prod_{i\not=j}f_j]
\mathbb{E}_q[f_j]
\label{G3}
\end{eqnarray}
where $f_j$ satisfies $\|f_j\|_{\frac{1}{s_j}}=\|f_j\|=\mathbb{E}_q[f_j]$. As a result, all the parties with weights being equal to 1 can be ignored.

To sum up, assume $\textbf{s}=(s_1, \cdots, s_n)$ such that $s_j\in (0,1)$ for all $j$'s. In this case, it is sufficient to prove
\begin{eqnarray}
 \mathbb{E}_q[\prod_{j=1}^nf_j]\leq \prod_j\|f_j\|_{\frac{1}{s_j}}
\label{G4}
\end{eqnarray}

Note $f_j$ is an arbitrary positive function on the output of the $j$-th party $\textsf{A}_j$. Composing the projection measurement operators of $\textsf{A}_j$ with this classical post-processing, assume $f_j$ is the outcome of some quantum observable $X_j$ that the $j$-th party applies on quantum systems in her hand.

Suppose $X_j$ is decomposed into the superposition of complete orthogonal projection operators. In this case, we assume
\begin{eqnarray}
 X_j=\sum_{a_j}f_j(a_j)M_{a_j}
\label{G5}
\end{eqnarray}
where $\{M_{a_j}=|a_j\rangle\langle a_j|\}$ are orthogonal projection operators satisfying $M_{a_j=x_1}M_{a_j=x_2}=\delta_{x_1x_2}M_{a_j=x_1}$ with the delta function $\delta_{x_1x_2}$, that is, $\delta_{x_1x_2}=1$ for $x_1=x_2$ and $\delta_{x_1x_2}=0$ for $x_1\not=x_2$.

From the Born rule, we get
\begin{eqnarray}
\mathbb{E}_q[\prod_{j=1}^nf_j]={\rm tr}[\otimes_{j}X_j\rho]
\label{G7}
\end{eqnarray}
where $\rho=\otimes_i\rho_i$ and $\rho_i$ are pure states. Otherwise, consider the purification $|\Phi\rangle_{A_1\cdots A_n;B}$ of $\rho$ with auxiliary particle $B$. It is easy to check
\begin{eqnarray}
\mathbb{E}_q[\prod_{j=1}^nf_j]={\rm tr}[\mathbbm{1}_B\otimes(\otimes_{j=1}^nX_j)|\Phi\rangle\langle\Phi|]
\label{G8}
\end{eqnarray}
where $\mathbbm{1}_B$ denotes the identity operator on the particle $B$.

From Eqs.(\ref{G7}) and (\ref{G8}), we get
\begin{eqnarray}
\mathbb{E}_q[\prod_{j=1}^nf_j]
&=&{\rm tr}[
\otimes_{j=1}^n\sum_{a_j}f_j(a_j)|a_j\rangle\langle a_j|\rho
]
\nonumber
\\
&=&
{\rm tr}[\sum_{{\bf a}}f_1(a_1)\cdots{}f_n(a_n)
\otimes_{j=1}^n|a_j\rangle\langle{}a_j|\rho]
\nonumber
\\
&=&\sum_{{\bf a}}\gamma_{{\bf a}} f_1(a_1)\cdots{}f_n(a_n)
\label{G9}
\\
&\leq &\mathbb{E}_c[f_1\cdots{}f_n]
\label{G10}
\\
&\leq &\prod_{j=1}^n\|f_j\|_{\frac{1}{s_j}}
\label{G11}
\end{eqnarray}
In Eq.(\ref{G9}), we have $\gamma_{{\bf a}}={\rm tr}[\otimes_{j=1}^n|a_j\rangle\langle{}a_j|\rho]$. From Eq.(\ref{G5}), it follows $\sum_{i_1\cdots i_n}\gamma_{i_1\cdots{}i_n}=1$. It means $\{\gamma_{i_1\cdots{}i_n}\}$ can be regarded as a classical probability distribution $\{P_c({\bf a})\}_{{\bf a}=i_1\cdots{}i_n}$ with $P_c({\bf a})=\gamma_{{\bf a}}^2$ for any $a_1, \cdots, a_n$. Note the local projection measurement $\{\otimes_{j=1}^n|a_j\rangle\langle{}a_j|\}$ does not change the network configuration of ${\cal N}_q$. From Lemma 1, any correlation achievable in single state can be generated by one classical variable. Moreover, $f_j(a_j)\geq 0$. These facts imply the correlation of $\gamma_{{\bf a}}$ is achievable in classical network ${\cal N}$ with the same network configuration of ${\cal N}_q$. This implies the inequality (\ref{G10}) from Theorem 1'. $\mathbb{E}_c[f_1\cdots{}f_n]$ in the inequality (\ref{G10}) denotes the classical expect of the positive function $f_1\cdots{}f_n$ associated with the classical probability distribution $\{P_c({\bf a})\}$. The equality follows if all the orthogonal bases of $\{|a_j\rangle\langle a_j|, \forall j\}$ are complete. This completes the proof of the inequality (\ref{Gth1}).

\begin{figure*}[ht]
\begin{center}
\resizebox{460pt}{240pt}{\includegraphics{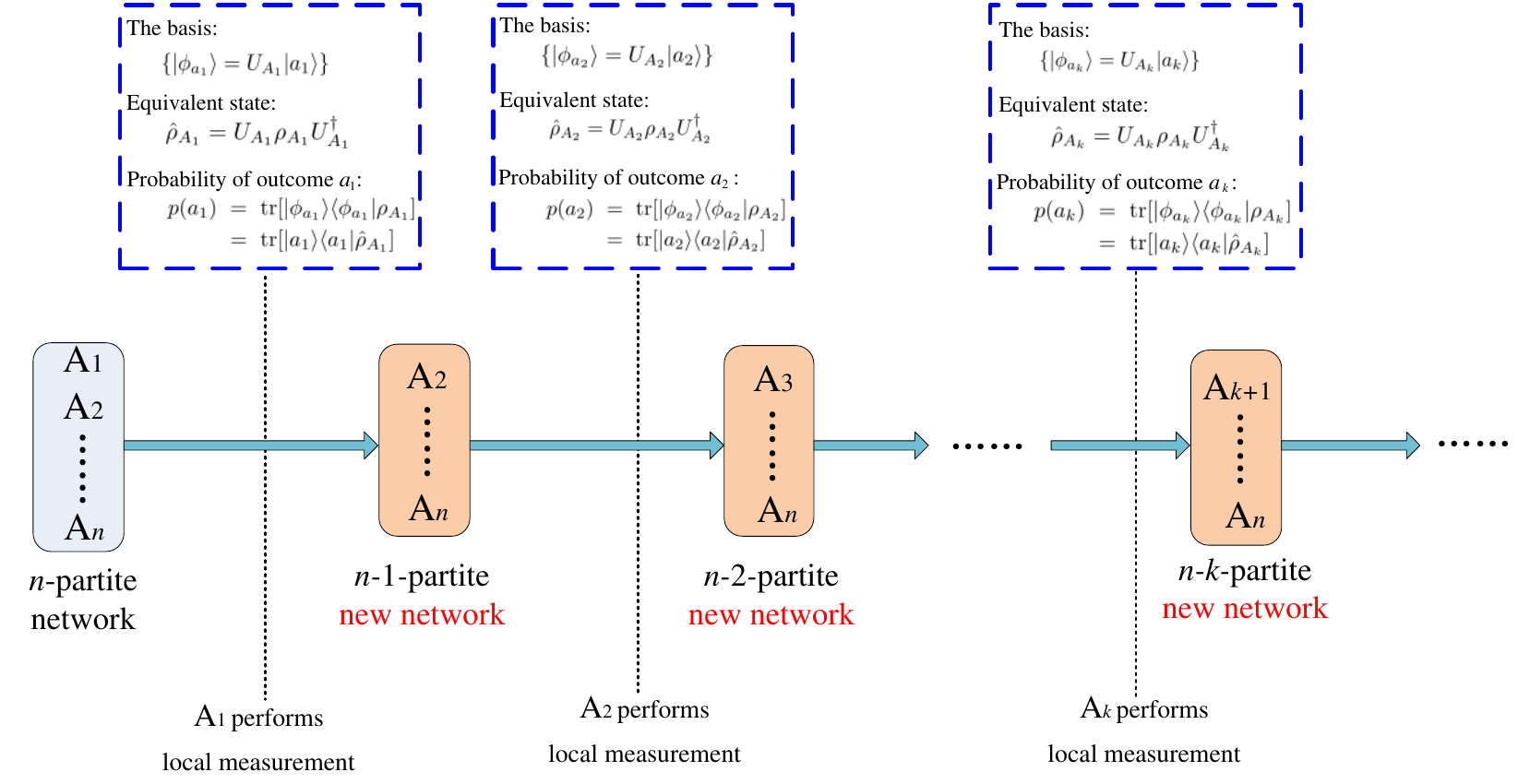}}
\end{center}
\caption{\small (Color online) Schematic procedure for generating the distributions in Eqs.(\ref{nw4})-(\ref{nw7}).}
\label{figs6a}
\end{figure*}

\textbf{Case 2}. Proof of the inequality (\ref{Gth2}) with local projection measurements under the computation basis.

In this case, we have
\begin{eqnarray}
&&P(a_1,\cdots, a_n)={\rm tr}[\otimes_{j=1}^n|a_j\rangle\langle a_j|\rho],
\\
&&P(a_j)={\rm tr}[|a_j\rangle\langle a_j|\otimes\mathbbm{1}_{\overline{j}}\rho], j=1, \cdots, n
\end{eqnarray}
when $X_j=|a_j\rangle\langle a_j|$, where $\mathbbm{1}_{\overline{j}}$ denotes the identity operator performed on all the systems not owned by the party $\textsf{A}_j$, $j=1, \cdots, n$. From the inequality (\ref{G11}), it follows that
\begin{eqnarray}
P_q(a_1, \cdots, a_n) \leq \prod_{j=1}^nP_{q}(a_j)^{s_j}
\label{Gtha2}
\end{eqnarray}
This has proved the inequality (\ref{Gth2}) for local projection measurements.

\textbf{Case 3}. Proof of the inequality (\ref{Gth2}) with local projection measurements.

In what follows, we prove the inequality (\ref{Gth2}) holds for any local joint projection measurements $\{|\phi_{a_j}\rangle=U_{A_i}|a_j\rangle\}$ with the computation basis $\{|a_j\rangle\}$. For given an $n$-partite quantum network ${\cal N}_q$ shared by parties $\textsf{A}_{1}, \cdots, \textsf{A}_n$, suppose the total state is given by $\rho_{A_{1}\cdots A_n}$, where the party $\textsf{A}_{j}$ own the system $A_j$, $j=1, \cdots, n$. From Born rule, the joint probability is given by
\begin{eqnarray}
P(a_1,\cdots, a_n)={\rm tr}[\otimes_{j=1}^n|\phi_{a_j}\rangle\langle \phi_{a_j}|\rho]
\label{nw1}
\end{eqnarray}
From Bayes' Rule, it follows that
\begin{eqnarray}
P(a_1,\cdots, a_n)=p(a_1)p(a_2|a_1)\cdots{}p(a_n|a_1\cdots{}a_{n-1})
\label{nw2}
\end{eqnarray}
Hence, for given a distribution $\{P(a_1,\cdots, a_n)\}$, we get $n$ distributions given by
\begin{eqnarray}
P_{a_j}=\{p(a_j|a_{1}\cdots{}a_{j-1})\}, j=1, \cdots, n
\label{nw3}
\end{eqnarray}
where $a_0=1$. The main idea is to generate the distributions $P_{a_j}$'s by performing local projection measurements on ${\cal N}_q$ assisted by some additional states for each party. So, the network configuration is unchanged. The proof is completed by induction on the Bayes' Rule in Eq.(\ref{nw2}), shown as Fig.\ref{figs6a}.
\begin{itemize}
\item{} Firstly, we generate the distribution of $\{p(a_1)\}$. From the assumption of the local measurement, we get
\begin{eqnarray}
p(a_1)&=&{\rm tr}[|\phi_{a_1}\rangle\langle \phi_{a_1}|\rho_{A_1}]
\nonumber
\\
&=& {\rm tr}[|a_1\rangle\langle a_1|\hat{\rho}_{A_1}]
\label{nw4}
\end{eqnarray}
where $\hat{\rho}_{A_1}=U_{A_1}\rho_{A_1}U_{A_1}^\dag$ and $\rho_{A_1}$ is the reduced density matrix of the system $A_1$. After this local measurement, the $n$-partite network ${\cal N}_q$ is changed into $n-1$-partite new network ${\cal N}_q^{(n-1)}$, where each party of $\textsf{A}_2, \cdots, \textsf{A}_n$ may perform local unitary operation only on its own particles which are entangled with the particles owned by the party $\textsf{A}_1$  conditional on the outcome $a_1$. These operations do not change the configuration of the reduced $n-1$-partite network after the local measurement of $\textsf{A}_1$ being performed.

\item{}Secondly, the party $\textsf{A}_2$ performs the local measurement under the basis $\{|\phi_{a_2}\rangle\}$, where its reduced density matrix is denoted by $\rho_{A_2}$. Similar to Eq.(\ref{nw4}), we get
\begin{eqnarray}
p(a_2|a_1)&=&{\rm tr}[|\phi_{a_2}\rangle\langle \phi_{a_2}|\rho_{A_2}]
\nonumber
\\
&=& {\rm tr}[|a_2\rangle\langle a_2|\hat{\rho}_{A_2}]
\label{nw5}
\end{eqnarray}
where $\hat{\rho}_{A_2}=U_{A_2}\rho_{A_2}U_{A_2}^\dag$. After this local measurement, the $n-1$-partite network ${\cal N}_q^{(n-1)}$ is changed into an $n-2$-partite new network ${\cal N}_q^{(n-2)}$, where each party of $\textsf{A}_3, \cdots, \textsf{A}_n$ may perform local unitary operation only on its own particles which are entangled with the particles owned by the party $\textsf{A}_2$ conditional on the outcome $a_2$. These operations do not change the configuration of the reduced $n-2$-partite network after the local measurement of $\textsf{A}_2$ being performed.

\item{} This procedure can be iteratively performed. Take the party $\textsf{A}_k$ for example. After the party $\textsf{A}_{k-1}$ performs the local measurement under the basis $\{|\phi_{a_{k-1}}\rangle\}$, the party $\textsf{A}_k$ performs the local measurement under the basis $\{|\phi_{a_2}\rangle\}$, where its reduced density matrix is denoted by $\rho_{A_k}$. Similar to Eq.(\ref{nw4}), we get
\begin{eqnarray}
p(a_k|a_1\cdots{}a_{k-1})&=&{\rm tr}[|\phi_{a_k}\rangle\langle \phi_{a_k}|\rho_{A_k}]
\nonumber
\\
&=& {\rm tr}[|a_k\rangle\langle a_k|\hat{\rho}_{A_k}]
\label{nw6}
\end{eqnarray}
where $\hat{\rho}_{A_k}=U_{A_k}\rho_{A_k}U_{A_k}^\dag$. After this local measurement, the $n-1$-partite network ${\cal N}_q^{(n-k+1)}$ is changed into an $n-k$-partite new network ${\cal N}_q^{(n-k)}$, where each party of $\textsf{A}_{k+1}, \cdots, \textsf{A}_n$ may perform local unitary operation on its own particles which are entangled with the particles owned by $\textsf{A}_k$ conditional on the outcome $a_k$. These operations do not change the configuration of the reduced $n-k$-partite network after the local measurement of $\textsf{A}_k$ being performed.

\item{} The procedure is completed after the party $\textsf{A}_{n}$ performs the local measurement under the basis $\{|\phi_{a_{n}}\rangle\}$ in order to get
\begin{eqnarray}
p(a_n|a_1\cdots{}a_{n-1})&=&{\rm tr}[|\phi_{a_n}\rangle\langle \phi_{a_n}|\rho_{A_n}]
\nonumber
\\
&=& {\rm tr}[|a_n\rangle\langle a_n|\hat{\rho}_{A_n}]
\label{nw7}
\end{eqnarray}
\end{itemize}
where $\hat{\rho}_{A_n}=U_{A_n}\rho_{A_n}U_{A_n}^\dag$.

\begin{figure*}[ht]
\begin{center}
\resizebox{440pt}{200pt}{\includegraphics{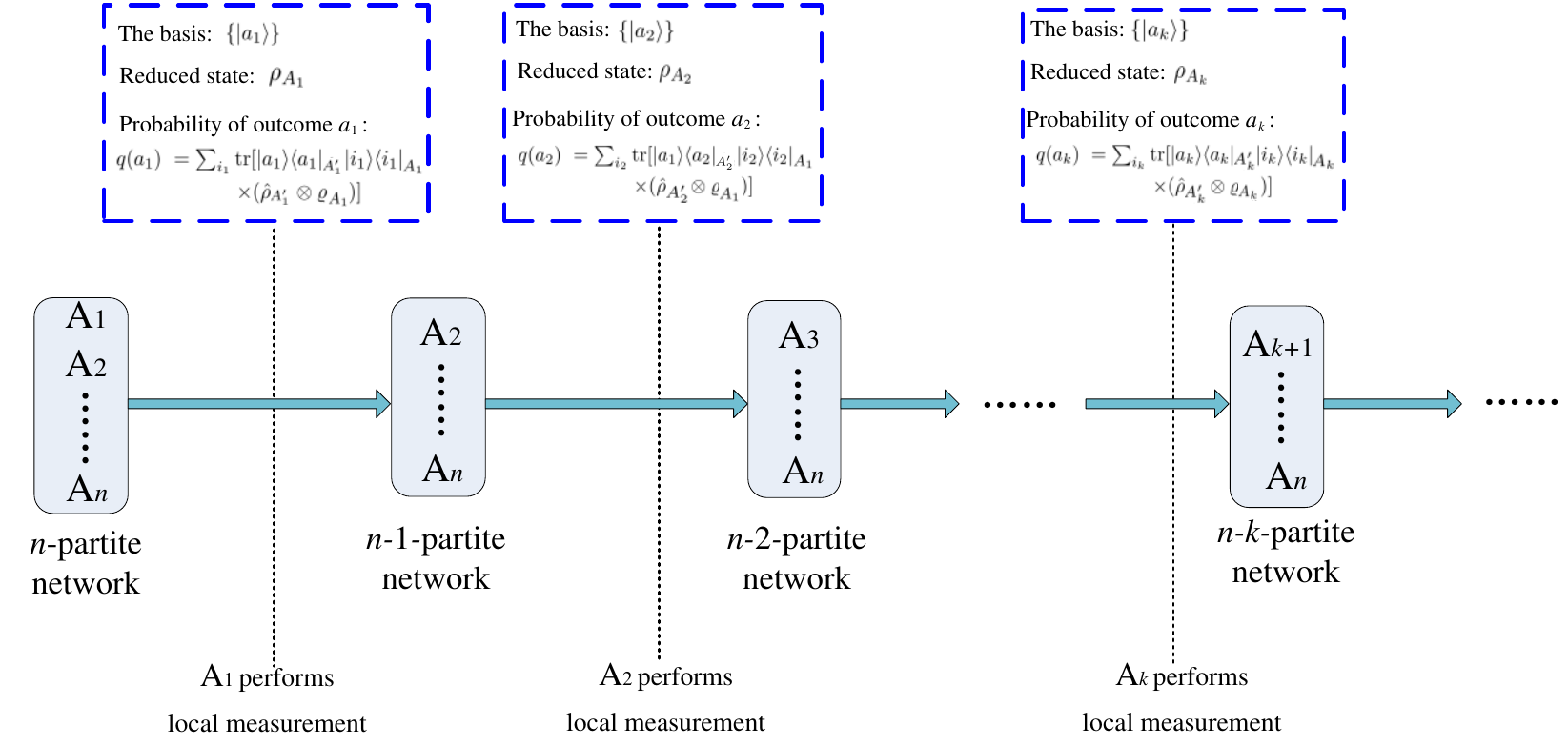}}
\end{center}
\caption{\small (Color online) Schematic procedure for generating the distributions in Eq.(\ref{nw8}).}
\label{figs6b}
\end{figure*}

So far, we have generated the joint distribution of $\{P(a_1,\cdots, a_n)\}$ by rebuilding $n$ distributions $p(a_1), \cdots, p(a_n|a_1\cdots{}a_{n-1})$ given in Eqs.(\ref{nw4})-(\ref{nw7}).

Now, suppose each party $\textsf{A}_k$ has an additional system $A_k'$ in the state $\hat{\rho}$ defined in Eq.(\ref{nw6}). A new $n$-party network ${\cal N}_q'$ is defined by considering all the systems of $A_1A_1', \cdots, A_nA_{n}'$, where each party $\textsf{A}_k$ owns two systems of $A_k$ and $A_k'$. The configuration of ${\cal N}_q'$ is same as its of the original network ${\cal N}_q$ without additional systems. In what follows, we construct new distributions of $q(a_1), \cdots, q(a_n|a_1\cdots{}a_{n-1})$ by performing local projection measurement on ${\cal N}_q'$ and post-processing of measurement outcomes such that
\begin{eqnarray}
q(a_k|a_1\cdots{}a_{k-1})=p(a_k|a_1\cdots{}a_{k-1}), j=1, \cdots, n
\label{nw8}
\end{eqnarray}
The rebuilding procedure is shown in Fig.\ref{figs6b}. The details are given as follows.
\begin{itemize}
\item{} Firstly, we generate the distribution of $\{q(a_1)\}$. The party $\textsf{A}_1$ performs the local projection measurement under the computation basis $\{|a_1\rangle_{A_1'}|i_1\rangle_{A_1} \}$ on its own systems $A_1'A_1$, where $A_1'$ is an axillary system in the state  $\hat{\rho}_{A_1'}$ defined in Eq.(\ref{nw4}). We get
\begin{eqnarray}
q(a_1)&=&\sum_{i_1}{\rm tr}[|a_1\rangle\langle a_1|_{A_1'}\otimes |i_1\rangle\langle i_1|_{A_1} (\hat{\rho}_{A_1'}\otimes\rho_{A_1})]
\nonumber
\\
&=& {\rm tr}[|a_1\rangle\langle a_1|\hat{\rho}_{A_1}]
\nonumber
\\
&=&p(a_1)
\label{nw9}
\end{eqnarray}
where $\rho_{A_1}$ is the reduced density matrix of the system $A_1$. Here, the summation can be regarded as a linearly post-processing of measurement outcomes. After this local measurement, the $n$-partite network ${\cal N}_q$ is changed into an $n-1$-partite network ${\cal N}_{n-1}$, where each party of $\textsf{A}_2, \cdots, \textsf{A}_n$ may perform local unitary operation only on its own particles that are entangled with the particles owned by the party $\textsf{A}_1$ conditional on the outcome $a_1$. These operations do not change the configuration of the reduced $n-1$-partite network after the local measurement of the party $\textsf{A}_1$ being performed.

\item{} Secondly, the party $\textsf{A}_2$ performs the local measurement under the basis $\{|a_2\rangle_{A_2'}|i_1\rangle_{A_2}\}$ on own systems $A_2'$ and $A_2$, where $A_2'$ is an axillary system in the state $\hat{\rho}_{A_2'}$ defined in Eq.(\ref{nw5}). Similar to Eq.(\ref{nw9}), we get
\begin{eqnarray}
q(a_2|a_1)&=&\sum_{i_2}{\rm tr}[|a_2\rangle\langle a_2|_{A_2'}\otimes |i_2\rangle\langle i_2|_{A_2} (\hat{\rho}_{A_2'}\otimes\rho_{A_2})]
\nonumber
\\
&=& {\rm tr}[|a_2\rangle\langle a_2|\hat{\rho}_{A_2}]
\nonumber
\\
&=&p(a_2|a_1)
\label{nw10}
\end{eqnarray}
where $\rho_{A_2}$ is the reduced density matrix of $A_2$. After this local measurement, the $n-1$-partite network ${\cal N}_{n-1}$ is changed into $n-2$-partite new network ${\cal N}_{n-2}$, where each party of $\textsf{A}_3, \cdots,\textsf{A}_n$ may perform local unitary operation on its own particles that are entangled with the particles owned by the party $\textsf{A}_2$ conditional on the outcome $a_2$. These operations do not change the configuration of the reduced $n-2$-partite network after the local measurement of $\textsf{A}_2$ being performed.

\item{}This procedure is iteratively performed. After the party $\textsf{A}_{k-1}$ performs the local measurement, the party $\textsf{A}_k$ performs the local measurement under the basis $\{|a_{k}\rangle_{A_{k}'}|i_{k}\rangle_{A_{k}}\}$ on the systems $A_k'$ and $A_k$, where $A_k'$ is an auxiliary system in the state $\hat{\rho}_{A_k'}$ defined in Eq.(\ref{nw6}). So, we get
\begin{eqnarray}
&&q(a_k|a_1\cdots{}a_{k-1})
\nonumber
\\
&=&
\sum_{i_k}{\rm tr}[|a_k\rangle\langle a_k|_{A_k'}\otimes |i_k\rangle\langle i_k|_{A_k} (\hat{\rho}_{A_k'}\otimes\rho_{A_k})]
\nonumber
\\
&=& {\rm tr}[|a_k\rangle\langle a_k|\hat{\rho}_{A_k}]
\nonumber
\\
&=&p(a_k|a_1\cdots{}a_{k-1})
\label{nw11}
\end{eqnarray}
where $\rho_{A_k}$ is the reduced density matrix of $A_k$. After this local measurement, the $n-k+1$-partite network ${\cal N}_{n-k+1}$ is changed into an $n-k$-partite new network ${\cal N}_{n-k}$, where each party of $\textsf{A}_{k+1}, \cdots,\textsf{A}_n$ may perform local unitary operation only on its own particles that are entangled with the particles owned by $\textsf{A}_k$ conditional on the outcome $a_k$. These operations do not change the configuration of the reduced $n-k$-partite network after the local measurement of $\textsf{A}_k$ being performed. The procedure is completed after the party $\textsf{A}_{n}$ performs the local measurement under the basis $\{|a_{n}\rangle_{A_{n}'}|i_{n}\rangle_{A_{n}}\}$ on the systems $A_n'$ and $A_n$.
\end{itemize}

From Eqs.(\ref{nw9})-(\ref{nw11}), we can generate all the distributions $\{p(a_k|a_1\cdots{}a_{k-1})\}$ for $k=1, \cdots, n$ by performing local projection measurements under the computation basis on the extended network ${\cal N}'_q$. From Eq.(\ref{nw2}), it means the joint distribution $\{p(a_1,\cdots{}a_{n})\}$ is rebuilt by performing local projection measurements under the computation basis on the extended network ${\cal N}'_q$ with the same configuration as ${\cal N}_q$. Hence, from Case 2 stated above, the distribution of $\{p(a_1,\cdots{}a_{n})\}$ should satisfy the inequality (\ref{Gtha2}). This has completed the proof. $\Box$

\begin{figure*}[ht]
\begin{center}
\resizebox{400pt}{230pt}{\includegraphics{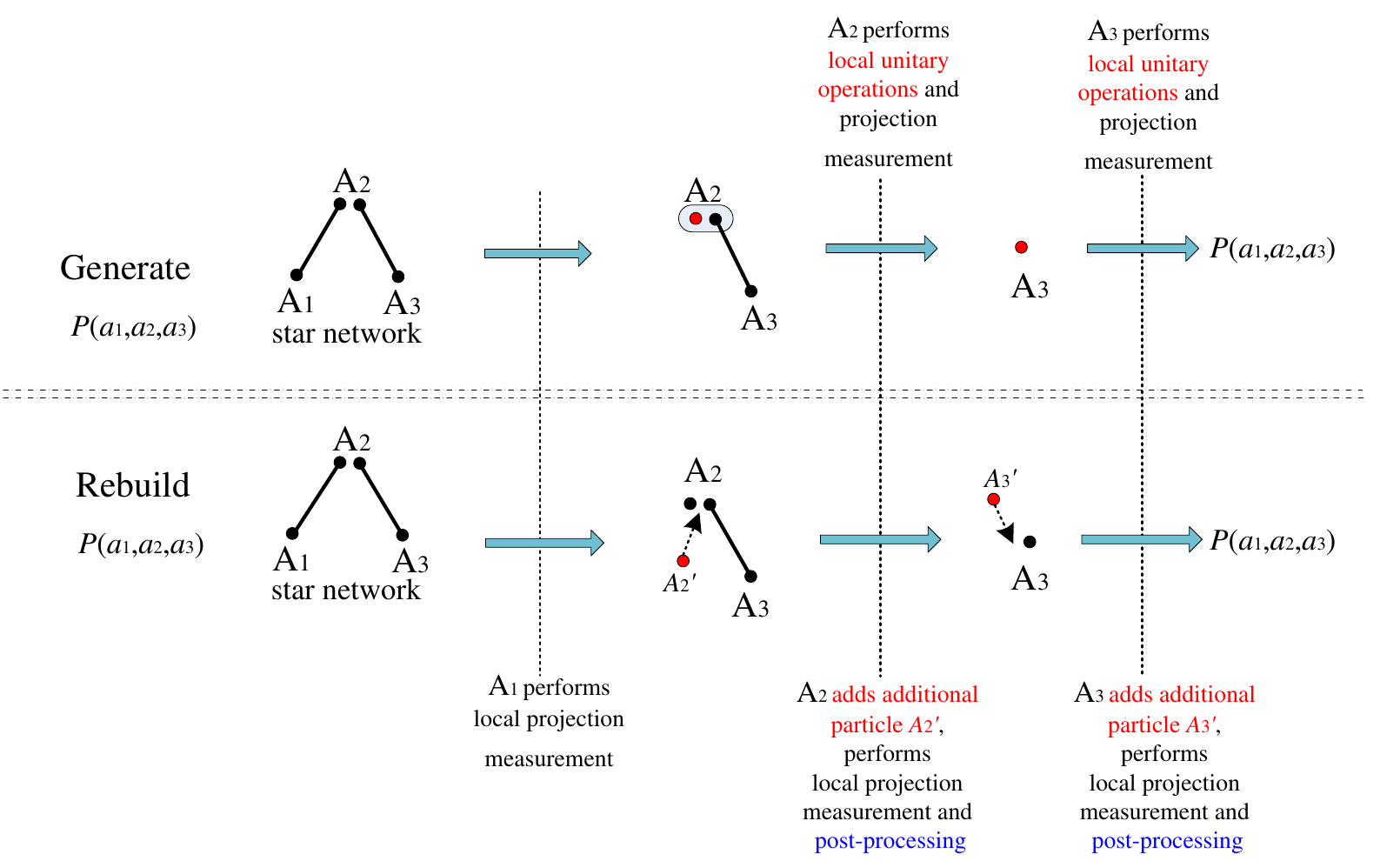}}
\end{center}
\caption{\small (Color online) Tripartite star quantum network.  The upper sub-figure is used to generate the joint distribution $P(a_1,a_2,a_3)$ by performing local joint projection measurements. The lower sub-figure is used to rebuild $P(a_1,a_2,a_3)$ by performing local projection measurements with the same measurement order in the upper sub-figure.}
\label{figs6c}
\end{figure*}

{\bf Examples S7}. Consider a tripartite star quantum network shown in Fig.\ref{figs6c}, where each pair of two adjacent parties $\textsf{A}_i$ and $\textsf{A}_{i+1}$ shares one bipartite entanglement, $i=1, 2$. The upper sub-figure is used to generate the joint distribution $P(a_1,a_2,a_3)$ from Eq.(\ref{nw2}) using the measurement order from $\textsf{A}_1$, $\textsf{A}_{2}$ to $\textsf{A}_3$. $p(a_1)$ and $p(a_3|a_1a_2)$ are completed by local projection measurements. $p(a_2|a_1)$ is completed by the party $\textsf{A}_{2}$. One method is to perform local joint projection measurement under the basis $\{U|k,s\rangle,\forall k, s\}$ (such as Bell basis) with any unitary matrix $U$. The other is firstly to perform the unitary operation $U$, and then performs local projection measurements. The lower sub-figure is used to rebuild $P(a_1,a_2,a_3)$ from Eq.(\ref{nw8}) with the same measurement order in the upper sub-figure. $q(a_1)=p(a_1)$ is completed by local projection measurements, where the local operations of $\textsf{A}_{1}$ do not change the network configuration. $p(a_2|a_1)$ is completed by the party $\textsf{A}_{2}$. From Eq.(\ref{nw5}), $\textsf{A}_{2}$ firstly adds one additional particle $A_2'$ in the state $\hat{\rho}_{A_2'}$ corresponding to the reduced state of $\textsf{A}_{2}$ in the upper sub-figure. And then, he performs local projection measurement and post-processing from Eq.(\ref{nw10}). Similarly, $\textsf{A}_{3}$ can get $q(a_3|a_1a_2)=p(a_3|a_1a_2)$ without changing of network configuration from Eq.(\ref{nw11}). It means the quantum Finner inequalities hold for correlations from star networks under any local joint measurements.

Another method is analytically to verify the inequality (\ref{fin4}) for any given local measurements such as Bell basis. Especially, consider a tripartite network consisting of two generalized EPR states \cite{EPR} given by $|\Phi\rangle_{12}\otimes |\Phi\rangle_{34}$ with $|\Phi\rangle=\cos\theta|00\rangle+\sin\theta|11\rangle$, $\theta\in (0, \frac{\pi}{2})$. The parties $\textsf{A}_1$ and $\textsf{A}_3$ perform projection measurement under the computation basis $\{|0\rangle, |1\rangle\}$. The party $\textsf{A}_2$ performs joint measurement under the Bell basis $\{|0\rangle:=\frac{1}{\sqrt{2}}(|00\rangle+ |11\rangle), |1\rangle:=\frac{1}{\sqrt{2}}(|01\rangle+|10\rangle), |2\rangle:=\frac{1}{\sqrt{2}}(|01\rangle-|10\rangle),|3\rangle:=\frac{1}{\sqrt{2}}(|00\rangle- |11\rangle)\}$. The joint probability $P(a_1,a_2,a_3)$ of the measurement outcome $a_1a_2a_3$ is given by
\begin{eqnarray}
P(0,0,0)&=&P(0,3,0)=\frac{1}{2}\cos^4\theta
\nonumber\\
P(0,1,1)&=&P(0,2,1)=P(1,1,0)=P(1,2,0)
\nonumber\\
&=&\frac{1}{2}\cos^2\theta\sin^2\theta
\nonumber\\
P(1,0,1)&=&P(1,3,1)=\frac{1}{2}\sin^4\theta
\label{ex1}
\end{eqnarray}

From Eq.(\ref{ex1}), it follows that
\begin{eqnarray}
&&P(0,0,0)-(p_{a_1}(0)p_{a_3}(0))^{\frac{m-1}{m}}p_{a_2}(0)^{\frac{1}{m}}
\nonumber\\
&=&\frac{1}{2}\cos^4\theta-\frac{1}{2^{\frac{1}{m}}}\cos^{\frac{4m-4}{m}}\theta(\cos^4\theta+\sin^4\theta)^{\frac{1}{m}}
\nonumber\\
&\leq &\frac{1}{2}\cos^4\theta-\frac{1}{2^{\frac{2}{m}}}\cos^{\frac{4m-4}{m}}\theta
\nonumber\\
&\leq &0
\end{eqnarray}
for any $m\geq 2$ and $\theta\in (0, \frac{\pi}{2})$, where we have used the inequality:
$\cos^4\theta+\sin^4\theta\geq \frac{1}{2}(\cos^2\theta+\sin^2\theta)^2=\frac{1}{2}$. Similar results hold for $P(0,3,0), P(1,0,1)$ and $P(1,3,1)$.

Moreover, we have
\begin{eqnarray}
&&P(0,1,1)-(p_{a_1}(0)p_{a_3}(1))^{\frac{m-1}{m}}p_{a_2}(1)^{\frac{1}{m}}
\nonumber\\
&=&\frac{1}{2}\cos^2\theta\sin^2\theta-\frac{1}{2^{\frac{1}{m}}}\cos^2\theta\sin^2\theta
\nonumber\\
&\leq &0
\end{eqnarray}
for any $m\geq 2$ and $\theta\in (0, \frac{\pi}{2})$. Similar results hold for $P(0,2,1), P(1,1,0)$ and $P(1,2,0)$. So, we have proved the distribution $P(a_1,a_2,a_3)$ satisfies the Finner inequality: $P(a_1,a_2,a_3)\leq (p({a_1})p(a_3))^{\frac{m-1}{m}}p(a_2)^{\frac{1}{m}}$ for any $m\geq 2$. Similar proof holds for the inequality of $P(a_1,a_2,a_3)\leq (p({a_1})p(a_3))^{\frac{1}{m}}p(a_2)^{\frac{m-1}{m}}$. It follows that the distribution derived from local joint projection measurement on star network satisfies the Finner inequality (\ref{fin4}). For general measurements one can verify the inequality (\ref{fin1}) by numeric evaluations.

\begin{figure*}[ht]
\begin{center}
\resizebox{360pt}{230pt}{\includegraphics{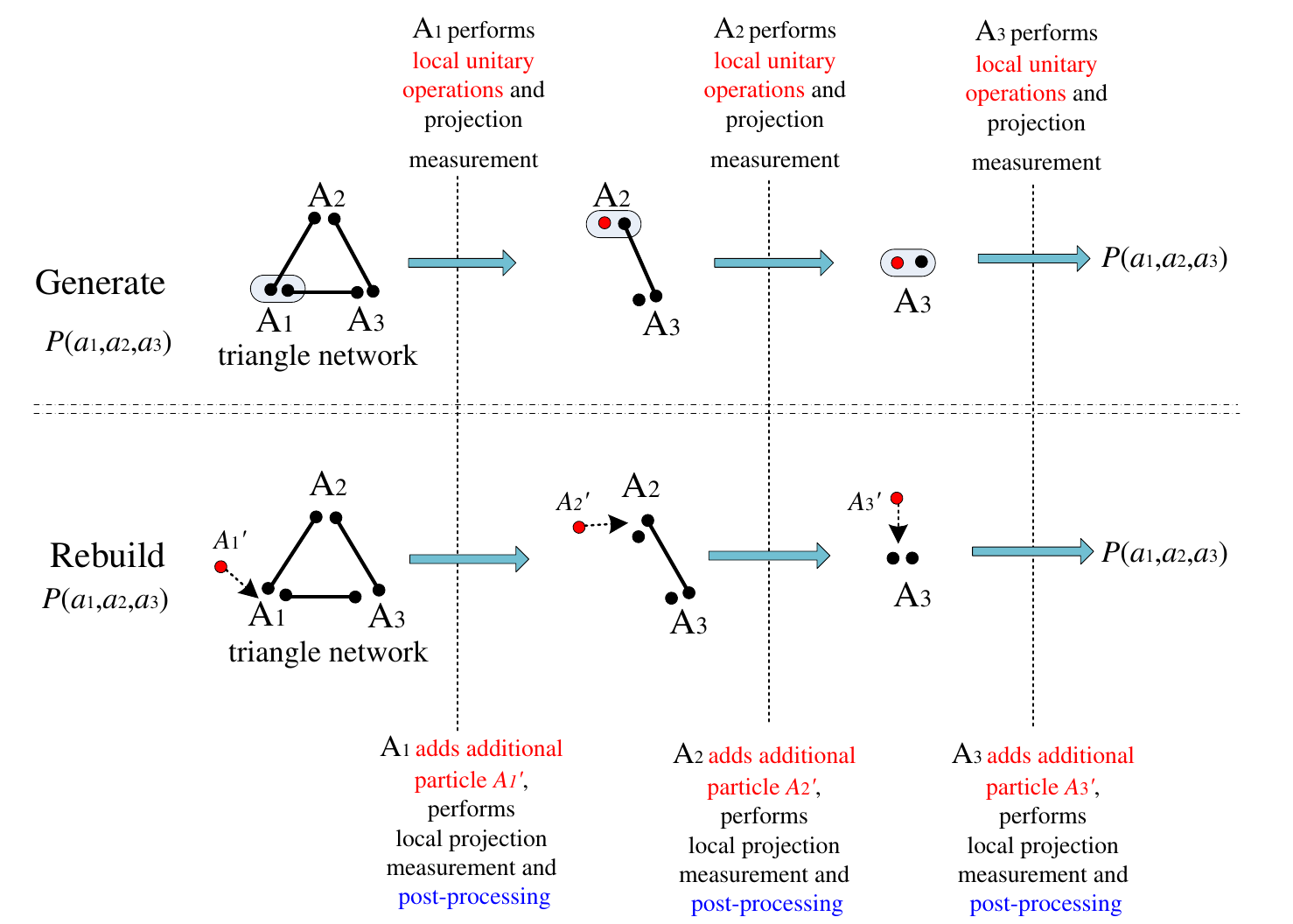}}
\end{center}
\caption{\small (Color online) Triangle quantum network. The upper sub-figure is used to generate the joint distribution $P(a_1,a_2,a_3)$ by performing local joint projection measurements. The lower sub-figure is used to rebuild $P(a_1,a_2,a_3)$ by performing local projection measurements with the same measurement order in the upper sub-figure.}
\label{figs6d}
\end{figure*}

{\bf Examples S8}. Consider a tripartite cyclic quantum network shown in Fig.\ref{figs6d}, where each pair of two parties $\textsf{A}_i$ and $\textsf{A}_{j}$ shares one bipartite entanglement, $1\leq i\not=j\leq 3$. The upper sub-figure is used to generate the joint distribution $P(a_1,a_2,a_3)$ by performing local joint projection measurements. From Eq.(\ref{nw2}), the measurement is completed by using the measurement order from $\textsf{A}_1$, $\textsf{A}_{2}$ to $\textsf{A}_3$. One method is to perform local joint projection measurement under the basis $\{U_i|k,s\rangle,\forall k, s\}$ (such as Bell basis) with any unitary matrix $U_i$. The other is firstly to perform the unitary operation $U_i$, and then performs local projection measurements. The lower sub-figure is used to rebuild $P(a_1,a_2,a_3)$ with the same measurement order in the upper sub-figure. $p(a_1)$ is completed by the party $\textsf{A}_{1}$. From Eq.(\ref{nw4}), $\textsf{A}_{1}$ firstly adds one additional particle $A_1'$ in the state $\hat{\rho}_{A_1'}$ corresponding to the reduced state of $\textsf{A}_{1}$ in the upper sub-figure. And then, he performs local projection measurement and post-processing from Eq.(\ref{nw9}). $q(a_2|a_1)=p(a_2|a_1)$ is completed by the party $\textsf{A}_{2}$. From Eq.(\ref{nw5}), $\textsf{A}_{2}$ firstly adds one additional particle $A_2'$ in the state $\hat{\rho}_{A_2'}$ corresponding to the reduced state of $\textsf{A}_{2}$ in the upper sub-figure. And then, he performs local projection measurement and post-processing from Eq.(\ref{nw10}). Similarly, $\textsf{A}_{3}$ can get $q(a_3|a_1a_2)=p(a_3|a_1a_2)$  without changing the network configuration. The quantum generalized Finner inequality holds for correlations under any local joint measurements on triangle networks.

Another method is analytically to verify the inequality (\ref{fin1}) for any given local measurements. Especially, consider a triangle network consisting of three generalized EPR states \cite{EPR} given by $|\Phi\rangle_{12}\otimes |\Phi\rangle_{34}|\Phi\rangle_{56}$, where $|\Phi\rangle$ is defined in Example S7. Each party $\textsf{A}_i$ performs joint measurement \cite{Marc2019a} under the basis $\{|0\rangle:=|00\rangle, |1\rangle:=\cos\gamma|01\rangle+\sin\gamma|10\rangle, |2\rangle:=\sin\gamma|01\rangle-\cos\gamma|10\rangle, |3\rangle:=|11\rangle\}$ with $\gamma\in (0, \frac{\pi}{2})$. The joint probability $P(a_1,a_2,a_3)$ for the measurement outcome $a_1a_2a_3$ is given by
\begin{eqnarray}
P(0,0,0)& =&\cos^6 \theta
\nonumber\\
P(1,1,0) &=& P(0,2,1) =\sin^4\gamma\cos^4\theta\sin^2\theta
\nonumber\\
P(1,2,0) & =& P(2,1,0) = P(0,2,2) = P(2,0,1)
\nonumber\\
&=& P(0,1,1) = P(1,0,2)
\nonumber\\
&=& \sin^2\gamma \cos^2\gamma \cos^4\theta \sin^2\theta
\nonumber\\
P(2,2,0) & =&P(0,1,2) =P(1,0,1) =\cos^4\gamma \cos^4\theta\sin^2\theta
\nonumber\\
P(1,3,1) & =& P(2,0,2)= P(3,1,2)=P(2,2,3)
\nonumber\\
&=&\sin^4\gamma \cos^2\theta\sin^4\theta
\nonumber\\
P(2,3,1) & =& P(2,1,3)= P(3,1,1) = P(1,3,2)
\nonumber\\
&=& P(3,2,2)=P(1,2,3)
\nonumber\\
&=&\sin^2\gamma\cos^2\gamma\cos^2\theta\sin^4\theta
\nonumber\\
P(3,2,1) & =& P(2,3,2)=P(1,1,3)=\cos^4\gamma\cos^2\theta\sin^4 \theta
\nonumber\\
P(3,3,3) & =& \sin^6\theta
\label{ex2}
\end{eqnarray}
From Eq.(\ref{ex2}), it follows that
$p_{a_i}(0)=\cos^4\theta$, $p_{a_i}(1)=p_{a_i}(2)=\sin^2\theta\cos^2\theta$, and $p_{a_i}(3)=\sin^4\theta$ (independent of $\gamma$) for $i=1, 2, 3$. So, we get
\begin{eqnarray}
P(0,0,0)-\sqrt{p_{a_1}(0)p_{a_2}(0)p_{a_3}(0)}=0
\end{eqnarray}
Similarly, we can prove the inequality (\ref{fin1}) for the distribution $P(a_1,a_2,a_3)$. So, the distribution derived from local joint projection measurements on triangle network satisfies the Finner inequality (\ref{fin1}). For general measurements, we can verify the inequality (\ref{fin1}) by numeric evaluations (code available for requirement).

{\bf Note 1}. From Theorem 2', the quantum generalized Finner inequality holds for local joint measurement. In entanglement swapping experiment, such as tripartite star network shown in Fig.\ref{figs6c}, the middle party can perform joint measurement under Bell basis. This is equivalent to local projection measurement after a two-particle entangling operation. The quantum generalized Finner inequality holds for new networks even if this local joint operation can change the network configuration. It means the quantum generalized Finner inequality provides correlation invariance under local unitary operations.

{\bf Note 2}. Theorem 2' does not contradict to the recent result \cite{Marc2019a} in which authors provide quantum correlations beyond classical correlations in a triangle network consisting of three generalized EPR states \cite{EPR}. The possible reason is the quantum Finner inequalities do not imply all the facets of correlations from the networks. Hence, an interesting problem is to explore network nonlocality.

\section{$n$-partite chain quantum networks}

Consider an $n$-partite chain network ${\cal N}_q$, as shown in Fig.\ref{figs3}, consisting of $n$ generalized EPR states \cite{EPR} of
\begin{eqnarray}
|\phi_i\rangle=\cos\theta_i|00\rangle+\sin\theta_i|11\rangle
\label{F11ab}
\end{eqnarray}
where the parties $\textsf{A}_i$ and $\textsf{A}_{i+1}$ share $|\phi_i\rangle$, $i=1, \cdots, n-1$. After the projection measurement being performed by each party $\textsf{A}_1$ and $\textsf{A}_n$,  and Bell basis $\{|0\rangle:=\frac{1}{\sqrt{2}}(|00\rangle+|11\rangle), |1\rangle:=\frac{1}{\sqrt{2}}(|01\rangle+|10\rangle), |2\rangle:=\frac{1}{\sqrt{2}}(|01\rangle-|10\rangle), |3\rangle:=\frac{1}{\sqrt{2}}(|00\rangle-|11\rangle)\}$ for $\textsf{A}_j$ with $j=2, \cdots, n-1$, the joint distribution of the outcome ${\bf 0}$ is given by
\begin{eqnarray}
P_{ch}({\bf 0})=\frac{\prod_{i=1}^n\cos^2\theta_i}{2^{n-2}}[{\bf 0}]
\label{g11b}
\end{eqnarray}
It is easy to evaluate
\begin{eqnarray}
&&p_{a_1}(0)=\cos^2\theta_1,
\nonumber\\
&&p_{a_n}(0)=\cos^2\theta_n,
\nonumber\\
&&p_{a_i}(0)=\frac{1}{2}\cos^2\theta_i\cos^2\theta_{i+1}, i=2, \cdots, n-1
\label{g11c}
\end{eqnarray}
From Eqs.(\ref{g11b}) and (\ref{g11c}), it follows that
\begin{eqnarray}
v_{{\bf 0}}&:=&P_{ch}({\bf 0})-\prod_{i=1}^np_{a_i}(0)^{s_i}
\nonumber\\
&=&\frac{1}{2^{n-2}}\prod_{i=1}^n\cos^2\theta_i-
\frac{1}{2^{\sum_{i=2}^{n-1}s_i}}
\nonumber\\
&&\times\cos^{2s_1}\theta_1\cos^{2s_n}\theta_n\prod_{i=2}^{n-1}\cos^{2s_i+2s_{i+1}}\theta_i
\nonumber\\
&\leq &0
\label{g11d}
\end{eqnarray}
for any $s_i$ and $\theta_1, \cdots, \theta_n\in (0, \frac{\pi}{2})$, where we have used the inequalities: $\sum_{i=2}^{n-1}s_i\leq 1$, $s_1, s_n\leq 1$, and $s_i+s_{i+1}\leq 1$, $i=2, \cdots, n-1$. Similar results hold for other local measurements. This implies the inequality (\ref{qfin3}) is the configuration inequality for chain networks.

\section{$n$-partite star quantum networks}

Consider an $n$-partite star network ${\cal N}_q$, as shown in Fig.\ref{figs2}, consisting of $n$ generalized EPR states \cite{EPR} given in Eq.(\ref{F11ab}), where the parties $\textsf{A}_i$ and $\textsf{A}_n$ share $|\phi_i\rangle$, $i=1, \cdots, n-1$. Under the projection measurement for each party $\textsf{A}_i$ with $i=1, \cdots, n-1$, and joint measurement with the $n-1$-partite Bell basis $\{|i\rangle:=\frac{1}{\sqrt{2}}(|i\rangle+|2^{n-1}-1-i\rangle), |2^{n-2}+i\rangle:=\frac{1}{\sqrt{2}}(|i\rangle+|2^{n-1}-1-i\rangle), i=0, \cdots, 2^{n-2}\}$, the joint distribution of the outcome ${\bf 0}$ is given by
\begin{eqnarray}
P_{st}({\bf 0})=\frac{1}{2^{n-1}}\prod_{i=1}^n\cos^2\theta_i[{\bf 0}]
\label{h11b}
\end{eqnarray}
It is easy to evaluate
\begin{eqnarray}
&&p_{a_i}(0)=\cos^2\theta_i,  i=1, \cdots, n-1
\nonumber\\
&&p_{a_n}(0)=\frac{1}{2^{n-1}}\prod_{i=1}^{n-1}\cos^2\theta_i
\label{h11c}
\end{eqnarray}
From Eqs.(\ref{h11b}) and (\ref{h11c}), it follows that
\begin{eqnarray}
v_{{\bf 0}}&:=&P_{st}({\bf 0})-\prod_{i=1}^np_{a_i}(0)^{s_i}
\nonumber\\
&=&\frac{1}{2^{n-1}}\prod_{i=1}^n\cos^2\theta_i-\frac{1}{2^{ns_n-s_n}}\prod_{i=1}^{n-1}
\cos^{2s_i}\theta_i\cos^{2s_n}\theta_n
\nonumber\\
&\leq &0
\label{h11d}
\end{eqnarray}
for any $s_i$ and $\theta_1, \cdots, \theta_n\in (0, \frac{\pi}{2})$, where we have used the inequalities: $ns_n-s_n\leq n-1$, $s_i\leq 1$, $i=1, \cdots, n$. Similar results hold for other measurements. It follows that the inequality (\ref{qfin3}) is the configuration inequality for star networks.

\section{Verifying the configuration of quantum networks}

In this section, we firstly present a general algorithm to verify the configuration of quantum networks in the first subsection. And then, we prove the correlations available in any networks consisting of bipartite entangled pure states and GHZ states violate specific configuration inequality in the second subsection. This provides a general method to construct different networks with incompatible configuration inequalities. Moreover, we provide some special networks including tripartite single-source networks, chain networks and 4-party network in the last subsection.

\subsection{Algorithm for verifying the configuration of quantum networks}

Given an $n$-partite quantum network ${\cal N}_q$ consisting of any multipartite entangled states, the configuration compatibility problem is related to test the configuration of ${\cal N}_q$ by using observed correlations.

\textbf{The configuration compatibility problem}. \textit{Given a correlation $P({\bf a})$, the configuration compatibility problem is to determine whether $P({\bf a})$ is compatible with the a given quantum network ${\cal N}_q$, that is, whether $P({\bf a})$ can be generated by specific local measurements on ${\cal N}_q$.}

Theorem 2 is useful for solving the configuration compatibility problem by testing the inequality (9) with the given correlation $P({\bf a})$. This may be further applied to verify quantum networks. Unfortunately, it is impossible to verify the specific configuration of general quantum networks with polynomial-time complexity because there are exponential numbers of different configurations for given a network. Our goal here is to provide a procedure to complete this task regardless of time complexity. It may be useful for small-size networks. Another application is for configuration compatibility problem.

\begin{minipage}{8.2cm}
\begin{algorithm}[H]
	\centering
	\caption{Verifying network configuration}
\begin{itemize}
\item[Input:] An $n$-partite network ${\cal N}_q$.
\item[Output:] Configuration of network.
\item{} Consider another $n$-partite network ${\cal N}_q'$.
\item{} Find a fractional independent set $\textbf{s}=(s_1, \cdots, s_n)$ for ${\cal N}_q'$.
\item{} Obtain the configuration inequality from Theorem 2 for ${\cal N}_q'$.
\item{} Get the joint distribution $P({\bf a})$ for ${\cal N}_q$.
\item{} Verify whether $P({\bf a})$ violates the present configuration inequality or not.
\end{itemize}

\label{alg3}
\end{algorithm}
\end{minipage}

This algorithm provides a general method to test the configuration of given networks. It depends on the proposed configuration inequality in Theorem 2. Take the network ${\cal N}_q'$ in Fig.\ref{figs6}(b) as an example. Different from Case 1, all the parties $\textsf{A}_{k+1}, \cdots, \textsf{A}_{n}$ in Fig.\ref{figs6}(b) can be regarded as one party $\textsf{B}$. In this case, from Eq.(\ref{D3}) and Theorem 2 we get the following configuration inequality
\begin{eqnarray}
P({\bf a}) \leq \prod_{i=1}^kp(a_i)^{\frac{1}{m}}p(a_{k+1}, \cdots, a_n)^{\frac{m-1}{m}}
\label{FF0a}
\end{eqnarray}

\begin{figure}
\begin{center}
\resizebox{220pt}{120pt}{\includegraphics{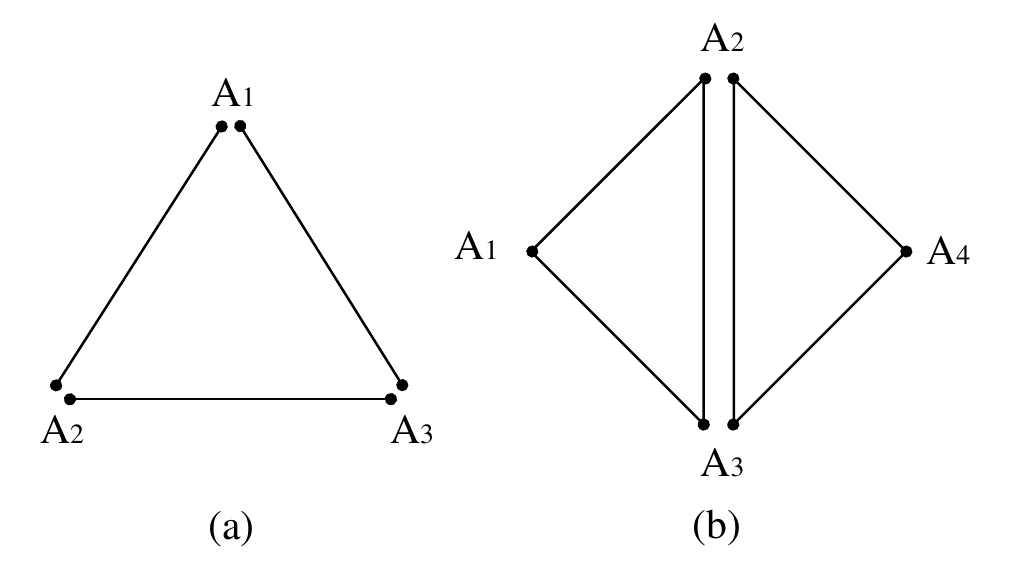}}
\end{center}
\caption{\small (Color online) (a) Triangle network consisting of three generalized EPR states. (b) $4$-partite network consisting of two generalized GHZ states.}
\label{figs7}
\end{figure}

\subsection{Any networks consisting of bipartite entangled pure states and GHZ states}

In this subsection, we prove for an $n$-partite network ${\cal N}_q$ consisting of any bipartite entangled pure states \cite{EPR} or generalized GHZ states \cite{GHZ}, its joint distribution can violate specific configuration inequality associated with other networks.

{\bf Proposition S2}. \textit{There exists an $n$-partite network ${\cal N}_q'$ such that the joint distribution $P({\bf 0})$ achievable in ${\cal N}_q$ violates the configuration inequality of ${\cal N}_q'$}.

{\bf Proof of Corollary 3}. Consider an $n$-partite quantum network ${\cal N}_q$ consisting of any bipartite entangled pure states \cite{EPR} or generalized GHZ states \cite{GHZ} as: $|\Phi_1\rangle, \cdots, |\Phi_m\rangle$. Assume each party $\textsf{A}_j$ owns some particles contained in entangled systems $|\Phi_{j_{1}}\rangle, \cdots, |\Phi_{j_{s_{j}}}\rangle$. Note ${\cal N}_q$ may be disconnected, that is, consisting of some disconnected networks. All the particles owned by one party can be regarded as one high-dimensional state. For example, two qubits with basis $\{|00\rangle, \cdots, |11\rangle\}$ can be regarded as $4$-dimensional system with basis $\{|0\rangle, \cdots, |3\rangle\}$. Here, each entanglement $|\Phi_j\rangle$ can be locally transformed into
\begin{eqnarray}
|\Phi_j\rangle=\sqrt{q_j}|0\rangle^{\otimes \ell_j}+\sum_{i\not=0}\alpha_{ij}|i\rangle^{\otimes \ell_j}
\label{FF1}
\end{eqnarray}
For simplicity, assume $|\Phi_j\rangle$ is shared by $\ell_j$ parties. Otherwise, all the particles in $|\Phi_j\rangle$ owned by one party are regarded as one high-dimensional system. Each party performs projective measurement under the computation basis $\{|0\rangle, \cdots, |d-1\rangle\}$ for a $d$-dimensional system. It is easy to evaluate the joint distribution of $P({\bf 0})$ given by
\begin{eqnarray}
P({\bf 0})=\prod_{j=1}^mq_j
\label{FF2}
\end{eqnarray}

In what follows, we need to find specific network ${\cal N}_q'$ and its fractional independent set ${\bf s}$. The main idea is inspired by the $k$-independent set \cite{Luo2018}. Especially, for each network ${\cal N}_q$, we find a $k$-independent set $\{\textsf{A}_{1}, \cdots, \textsf{A}_k\}$ (for example) with polynomial time complexity \cite{Luo2018}, where each party $\textsf{A}_{i}$ ($1\leq i\leq k$) does not share any entanglement with other party $\textsf{A}_{j}$ ($1\leq j\not=i\leq k$). Different from the result \cite{Luo2018}, $k$ may be any integer, that is, $k\geq 1$.

From Eq.(\ref{FF1}), for each party $\textsf{A}_j$ the probability for the outcome $a_j=0$ is given by
\begin{eqnarray}
p_{a_j}(0)=\prod_{t=1}^{s_j}q_{j_t}
\label{FF3}
\end{eqnarray}
where $q_{j_t}$ is derived from the local measurement on $|0\rangle^{\otimes \ell_{j_t}}$ of $|\Phi_{j_t}\rangle$.

\begin{figure}
\begin{center}
\resizebox{240pt}{120pt}{\includegraphics{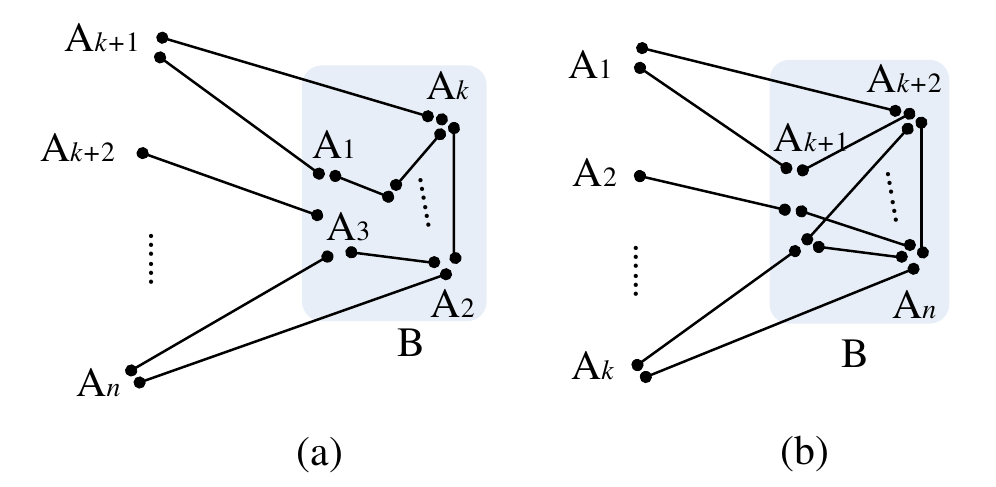}}
\end{center}
\caption{\small (Color online) A star network. The parties $\textsf{A}_{k+1}, \cdots, \textsf{A}_{n}$ are independent, and all the parties $\textsf{A}_{1}, \cdots, \textsf{A}_{k}$ are regarded as one party $\textsf{B}$. (a)  $k\leq \frac{n}{2}$. (b) $k>\frac{n}{2}$. The parties $\textsf{A}_{1}, \cdots, \textsf{A}_{k}$ are independent, and all the parties $\textsf{A}_{k+1}, \cdots, \textsf{A}_{n}$ are regarded as one party $\textsf{B}$.}
\label{figs6}
\end{figure}

Define a new network ${\cal N}_q'$ consisting of all the bipartite entangled states, as shown in Fig.\ref{figs6}. Especially, for $k\leq \frac{n}{2}$, that is, $k\leq n-k$, ${\cal N}_q'$ is a star network shown in Fig.\ref{figs6}(a), where all the parties $\textsf{A}_{k+1}, \cdots, \textsf{A}_{n}$ are independent. For $k>\frac{n}{2}$, that is, $k>n-k$, ${\cal N}_q'$ is another star network shown in Fig.\ref{figs6}(b), where all the parties $\textsf{A}_{1}, \cdots, \textsf{A}_{k}$ are independent.

From Eq.(\ref{D3}), define ${\bf s}$ as follows
\begin{eqnarray}
{\bf s}&=&({\footnotesize {\underbrace{\frac{1}{m}, \cdots, \frac{1}{m}}\atop k}, { \underbrace{\frac{m-1}{m}, \cdots, \frac{m-1}{m}} \atop{n-k}}}), k\leq \frac{n}{2}
\label{FF4}
\\
{\bf s}&=&({\footnotesize {\underbrace{\frac{m-1}{m}, \cdots, \frac{m-1}{m}}\atop k}, {\underbrace{\frac{1}{m}, \cdots, \frac{1}{m}} \atop n-k}}), k>\frac{n}{2}
\label{FF5}
\end{eqnarray}
Consider the network in Fig.\ref{figs6}(a). For each bipartite source shared by two parties $\textsf{A}_i$ and $\textsf{A}_j$ in $\{\textsf{A}_{1}, \cdots, \textsf{A}_{k}\}$, we have $s_i+s_j\leq 1$. Moreover, for each bipartite source shared by two parties $\textsf{A}_i,\textsf{A}_j$ with $\textsf{A}_i\in \{\textsf{A}_{1}, \cdots, \textsf{A}_{k}\}$ and $\textsf{A}_j\in \{\textsf{A}_{k+1}, \cdots, \textsf{A}_{n}\}$, we have $s_i+s_j=1$. It means ${\bf s}$ in Eq.(\ref{FF4}) is the fractional independent set of the network in Fig.\ref{figs6}(a). Similarly, we can prove ${\bf s}$ in Eq.(\ref{FF5}) is the fractional independent set of the network in Fig.\ref{figs6}(b). In applications, ${\cal N}_q'$ may be consisted of multipartite entangled states.

From Theorem 2, the correlation $P({\bf a})$ derived from local measurements on the network ${\cal N}_q'$ satisfies the configuration inequality
\begin{eqnarray}
P({\bf a})\leq \prod_{i=1}^kp(a_i)^{\frac{1}{m}}\prod_{j=k+1}^np(a_j)^{\frac{m-1}{m}}, \mbox{ if } k\leq \frac{n}{2},
\label{FF6}
\\
P({\bf a})\leq \prod_{i=1}^kp(a_i)^{\frac{m-1}{m}}\prod_{j=k+1}^np(a_j)^{\frac{1}{m}}, \mbox{ if } k> \frac{n}{2}.
\label{FF7}
\end{eqnarray}

Define four sets as follows:
\begin{itemize}
\item{} ${S}_1$ consists of all the bipartite entangled states shared by one party in the independent set $\{\textsf{A}_1, \cdots, \textsf{A}_k\}$ and one party not in the independent set.
\item{}${S}_2$ consists of all the bipartite entangled states shared by two parties not in the independent set.
\item{} $S_3$ denotes all the multipartite entangled states shared by one party in the the independent set $\{\textsf{A}_1, \cdots, \textsf{A}_k\}$ and other parties  not in the independent set.
\item{} $S_4$ denotes all the multipartite entangled states shared by all parties not in the independent set.
\end{itemize}
It is easy to prove $S_1, S_2, S_3$ and $S_4$ consist of a four-partite partition of all entangled states $|\Phi_1\rangle, \cdots, |\Phi_m\rangle$.

In what follows, we prove the joint distribution $P({\bf 0})$ in Eq.(\ref{FF2}) violates the inequality (\ref{FF6}) or (\ref{FF7}). For $k\leq \frac{n}{2}$, from Eqs.(\ref{FF2}) and (\ref{FF3}) we have
\begin{eqnarray}
v_{{\bf 0}}&:=&P({\bf 0})-\prod_{i=1}^kp_{a_i}(0)^{\frac{1}{m}}\prod_{j=k+1}^np_{a_j}(0)^{\frac{m-1}{m}}
\nonumber
\\
&=&\prod_{j=1}^mq_j
-\prod_{i=1}^k(\prod_{t=1}^{s_i}q_{i_t})^{\frac{1}{m}}\prod_{j=k+1}^n(\prod_{t=1}^{s_j}q_{j_t})^{\frac{m-1}{m}}
\nonumber
\\
&=&\prod_{j=1}^mq_j-\prod_{i\in S_1}q_i\prod_{i\in S_2}q_i^{\frac{2m-2}{m}}
\nonumber\\
&&\times \prod_{j\in{}S_3}q_j^{1+\frac{\ell_j-2}{m}}\prod_{j\in{}S_4}q_j^{\frac{\ell_j(m-1)}{m}}
\label{FF8}
\\
&=&\prod_{j=1}^mq_j(1-\prod_{i\in S_2}q_i^{\frac{m-2}{m}}
\nonumber\\
&&\times \prod_{j\in{}S_3}q_j^{\frac{\ell_j-2}{m}}\prod_{j\in{}S_4}q_j^{\frac{\ell_jm-m-\ell_j}{m}})
\nonumber\\
&>&0
\label{FF9}
\end{eqnarray}
In Eq.(\ref{FF8}), for any entangled state $|\Phi_i\rangle\in S_1$ shared by two parties, we have $q_i=q_{i}^{\frac{1}{m}}q_{i}^{\frac{m-1}{m}}$, where one weight is $\frac{1}{m}$ and other is $1-\frac{1}{m}$. For any entangled state $|\Phi_i\rangle\in S_2$ shared by two parties, we have $q_i^{\frac{2m-2}{m}}=q_{i}^{\frac{m-1}{m}}q_{i}^{\frac{m-1}{m}}$, where both weights are $1-\frac{1}{m}$. For any entangled state $|\Phi_j\rangle\in S_3$ shared by $\ell_j$ parties, we get $q_j^{1+\frac{\ell_j-2}{m}}=q_{j}^{\frac{m-1}{m}}q_{j}^{\frac{1}{m}\times (\ell_j-1)}$, where one weight is $1-\frac{1}{m}$ and all the others are $\frac{1}{m}$. For any entangled state $|\Phi_j\rangle\in S_4$ shared by $\ell_j$ parties, it follows $q_j^{\frac{\ell_j(m-1)}{m}}=q_{j}^{\frac{m-1}{m}\times \ell_j}$, where all the weights are $1-\frac{1}{m}$. The inequality (\ref{FF9}) follows from the inequalities: $\prod_{i\in S_2}q_i^{\frac{m-2}{m}}\leq 1$, $\prod_{j\in{}S_3}q_j^{\frac{\ell_j-2}{m}}<1$, $\prod_{j\in{}S_4}q_j^{\frac{\ell_jm-m-\ell_j}{m}}<1$ for large $m\geq 3$. From the inequality (\ref{FF9}), it follows the joint distribution $P({\bf 0})$ in Eq.(\ref{FF2}) violates the inequality (\ref{FF6}).

Similarly, we can prove the joint distribution $P({\bf 0})$ in Eq.(\ref{FF2}) violates the inequality (\ref{FF7}) for $ k\leq \frac{n}{2}$. This completes the proof. $\Box$

\subsection{Some examples}

\textbf{Example S9 (Tripartite single-source network)}. Consider a tripartite network shown in Fig.1. For the given correlation of $P(a,b,c)$ from local measurements on single-source network such as generalized GHZ state \cite{GHZ} or W state \cite{Dur}. The goal in this subsection is to verify $P(a,b,c)$ violates some configuration inequalities compatible with the tripartite networks consisting of two sources or three sources, that is, chain networks or triangle networks consisting of three generalized EPR states \cite{EPR}.

Note for a single-source network, from Corollary 1 the correlation should satisfy
\begin{eqnarray}
P({\bf a}) \leq \prod_{i=1}^3p(a_i)^{s_i}
\label{FF10a}
\end{eqnarray}
with $\sum_{i=1}^3s_i\leq 1$.

Firstly, consider a chain network ${\cal N}_q$ consisting of two generalized EPR states (\ref{F11a}), where the parties $\textsf{A}_1$ and $\textsf{A}_2$ share $|\phi_1\rangle$ while the parties $\textsf{A}_2$ and $\textsf{A}_3$ share $|\phi_1\rangle$. From Appendix G, the inequality (\ref{FF10a}) is compatible with the chain network.

In what follows, consider a triangle network ${\cal N}_q$, as shown in Fig.\ref{figs7}(a), consisting of three generalized EPR states \cite{EPR} of
\begin{eqnarray}
|\phi_i\rangle=\cos\theta_i|00\rangle+\sin\theta_i|11\rangle
\label{F11a}
\end{eqnarray}
where each party shares two qubits from two EPR states, and $\theta_i\in (0, \frac{\pi}{2})$. After the projection measurement under the computation basis
$\{|0\rangle:=|00\rangle, |1\rangle:=|01\rangle, |2\rangle:=|10\rangle, |3\rangle:=|11\rangle\}$, the joint distribution is given by Eq.(\ref{C2}) with $p_i=\cos^2\theta_i$ and $q_i=\sin^2\theta_i$, $i=1, 2, 3$. From Eqs.(\ref{C2}) and (\ref{C3}), it follows that
\begin{eqnarray}
v_{000}&:=&P_{tri}(0,0,0)-p_a(0)^{s_1}p_b(0)^{s_2}p_c(0)^{s_3}
\nonumber\\
&=& p_1p_2p_3-p_1^{s_1+s_2}p_2^{s_2+s_3}p_3^{s_1+s_3}
\nonumber\\
&<&0
\label{F14a}
\end{eqnarray}
for any $s_i$ with $s_i\not=0$, and $\theta_1, \theta_2, \theta_3\in (0, \frac{\pi}{2})$, where $s_i+s_j\leq 1$. The distribution $P_{tri}(0,0,0)$ satisfies the inequality (\ref{FF10a}).

Similar result holds for general two-qubit joint measurement. Take the Bell basis
$\{|0\rangle:=\frac{1}{\sqrt{2}}(|00\rangle+|11\rangle), |1\rangle:=\frac{1}{\sqrt{2}}(|01\rangle+|10\rangle), |2\rangle:=\frac{1}{\sqrt{2}}(|01\rangle-|10\rangle), |3\rangle:=\frac{1}{\sqrt{2}}(|00\rangle-|11\rangle)\}$ as an example. The joint distribution is given
\begin{eqnarray}
P_{tri}({\bf 0})=\frac{1}{8}(\prod_{i=1}^3\cos\theta_i+\prod_{i=1}^3\sin\theta_i)^2[000]
\label{Fab}
\end{eqnarray}
It is easy to evaluate
\begin{eqnarray}
&&p_{a_1}(0)=\frac{1}{2}\cos^2\theta_1\cos^2\theta_3,
\\
&&p_{a_2}(0)=\frac{1}{2}\cos^2\theta_1\cos^2\theta_2,
\\
&&p_{a_3}(0)=\frac{1}{2}\cos^2\theta_2\cos^2\theta_3.
\label{Fbc}
\end{eqnarray}
From Eqs.(\ref{Fab}) and (\ref{Fbc}), it follows that
\begin{eqnarray}
v_{000}&:=&P_{tri}(0,0,0)-\prod_{i=1}^3p_{a_i}(0)^{s_i}
\nonumber\\
&=&\frac{1}{8}(\prod_{i=1}^3\cos\theta_i+\prod_{i=1}^3\sin\theta_i)^2
\nonumber\\
&&-\frac{1}{4}\cos^{2s_1+2s_2}\theta_1\cos^{2s_2+2s_3}\theta_2\cos^{2s_1+2s_3}\theta_3
\nonumber\\
&<&\frac{1}{8}(\prod_{i=1}^3\cos\theta_i+\prod_{i=1}^3\sin\theta_i)^2
\nonumber\\
&&-\frac{1}{4}\cos^{2}\theta_1\cos^{2}\theta_2\cos^{2}\theta_3
\nonumber\\
&<&\frac{1}{8}\prod_{i=1}^3\sin^2\theta_i+\frac{1}{4}\prod_{i=1}^3\sin\theta_i\cos\theta_i
-\frac{1}{8}\prod_{i=1}^3\cos^2\theta_i
\nonumber\\
&<&0
\label{h11s}
\end{eqnarray}
for any $s_i$ with $s_i\not=0$, and $\theta_1, \theta_2, \theta_3\in (\frac{\pi}{4}, \frac{\pi}{2})$. The distribution $P_{tri}(0,0,0)$ satisfies the inequality (\ref{FF10a}). Similar results hold for other distributions.

To sum up, the inequality (\ref{FF10a}) should be satisfied by all networks.

\textbf{Example S10 (Chain network)}. Consider a tripartite network shown in Fig.1. For the given correlation $P(a,b,c)$ from local measurements on the chain network consisting of two sources. The goal in this subsection is to verify $P(a,b,c)$ violates some configuration inequalities compatible with the tripartite network consisting of one source or three sources, that is, generalized GHZ states \cite{GHZ} or W states \cite{Dur}, or triangle network consisting of three generalized EPR states \cite{EPR}. Actually, this can be completed by using the inequality (3).

It is easy to verify correlations from local measurements on the generalized GHZ states $|ghz\rangle=\cos\theta|000\rangle+\sin\theta|111\rangle$ with $\theta\in (0,\frac{\pi}{2})$ violates the inequality (3). Consider a generalized W state \cite{Dur} given by
\begin{eqnarray}
|w\rangle&=&\cos\theta\cos\gamma|001\rangle+\sin\theta\cos\gamma|010\rangle
\nonumber\\
&&+\sin\gamma|100\rangle
\label{Fd1}
\end{eqnarray}
where $\theta,\gamma\in (0,\frac{\pi}{2})$. By performing the projection measurement under the basis $\{|0\rangle, |1\rangle\}$ for each party, the joint distribution is given by
\begin{eqnarray}
P_w(a,b,c)&=&\cos^2\theta\cos^2\gamma[001]+\sin^2\theta\cos^2\gamma[010]
\nonumber\\
&&
+\sin^2\gamma[100]
\label{Fd2}
\end{eqnarray}
It follows that
\begin{eqnarray}
p_a(0)&=&\cos^2\gamma, p_a(1)=\sin^2\gamma
\nonumber\\
p_b(0)&=&1-\sin^2\theta\cos^2\gamma
\nonumber\\
p_b(1)&=&\sin^2\theta\cos^2\gamma
\nonumber\\
p_c(0)&=&1-\cos^2\theta\cos^2\gamma
\nonumber\\
p_a(1)&=&\cos^2\theta\cos^2\gamma
\label{Fd3}
\end{eqnarray}

From Eqs.(\ref{Fd2}) and (\ref{Fd3}), we get
\begin{eqnarray}
v_{001}&:=&P_w(0,0,1)-p_a(0)^{\frac{m-1}{m}}p_b(0)^{\frac{1}{m}}p_c(1)^{\frac{m-1}{m}}
\nonumber
\\
&=&\cos^2\theta\cos^2\gamma-\cos^{\frac{2m-2}{m}}\gamma(1-\sin^2\theta\cos^2\gamma)^{\frac{1}{m}}
\nonumber
\\
&&\times\cos^{\frac{2m-2}{m}}\theta\cos^{\frac{2m-2}{m}}\gamma
\nonumber\\
&>&\cos^2\theta\cos^2\gamma-\cos^{\frac{2m-2}{m}}\theta\cos^{\frac{4m-4}{m}}\gamma
\\
\label{Fd4}
&>&0
\label{Fd5}
\end{eqnarray}
if $\theta$ and $\gamma$ satisfy the following inequality
\begin{eqnarray}
|\cos\theta|<|\cos\gamma|^{m-2}
\label{Fd6}
\end{eqnarray}
where the inequality (\ref{Fd4}) follows the inequality of $1-\sin^2\theta\cos^2\gamma<1$ for $\theta,\gamma\in (0,\frac{\pi}{2})$. The distribution $P_w(0,0,1)$ in Eq.(\ref{Fd2}) violates the configuration inequality (\ref{FF6}) if the inequality (\ref{Fd6}) holds.

Moreover, from Eqs.(\ref{Fd2}) and (\ref{Fd3}) we get
\begin{eqnarray}
v_{010}&:=&P_w(0,1,0)-p_a(0)^{\frac{m-1}{m}}p_b(1)^{\frac{1}{m}}p_c(0)^{\frac{m-1}{m}}
\nonumber
\\
&=&\sin^2\theta\cos^2\gamma-\cos^{2}\gamma\sin^{\frac{2}{m}}\theta
\nonumber
\\
&&\times(1-\cos^2\theta\cos^2\gamma)^{\frac{m-1}{m}}
\nonumber\\
&>&0
\label{Fd7}
\end{eqnarray}
if $\theta$ and $\gamma$ satisfy the following inequality
\begin{eqnarray}
\sin^2\theta-\sin^2\theta\cos^2\theta\cos^2\gamma<1
\label{Fd8}
\end{eqnarray}
So, the distribution $P_w(0,1,0)$ in Eq.(\ref{Fd2}) violates configuration inequality (3) if the inequality (\ref{Fd8}) holds.

Similarly, we get
\begin{eqnarray}
v_{100}&:=&P_w(1,0,0)-p_a(1)^{\frac{m-1}{m}}p_b(0)^{\frac{1}{m}}p_c(0)^{\frac{m-1}{m}}
\nonumber
\\
&=&\sin^2\gamma-\sin^{\frac{2m-2}{m}}\gamma(1-\sin^2\theta\cos^2\gamma)^{\frac{1}{m}}
\nonumber
\\
&&\times (1-\cos^2\theta\cos^2\gamma)^{\frac{m-1}{m}}
\\
&>&0
\label{Fd9}
\end{eqnarray}
if $\theta$ and $\gamma$ satisfy the following inequality
\begin{eqnarray}
\sin^{2}\gamma-\sin^{2}\gamma\cos^2\gamma\cos^2\theta<1
\label{Fd10}
\end{eqnarray}
The distribution $P_w(1,0,0)$ in Eq.(\ref{Fd2}) violates the configuration inequality (3) if the inequality (\ref{Fd10}) holds.

\begin{figure}
\begin{center}
\resizebox{240pt}{120pt}{\includegraphics{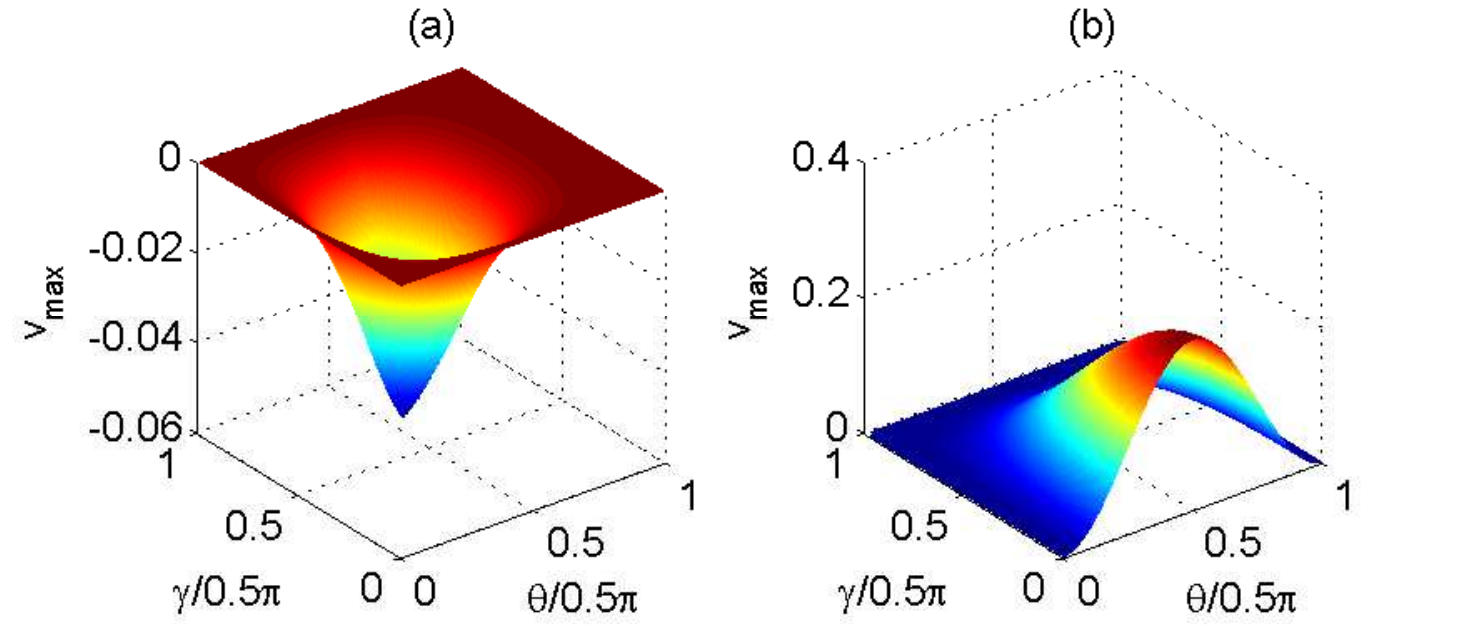}}
\end{center}
\caption{\small (Color online) Violations of W states (a) According to the inequality (\ref{fin1}). (b) According to the inequality (\ref{fin4}). Here, $m=1000$ and $v_{max}$ denotes the maximal violation over all the distributions $p(a,b,c)$.}
\label{figs9}
\end{figure}

From the inequalities (\ref{Fd6}), (\ref{Fd8}) and (\ref{Fd10}), the distribution of $P_w$ in Eq.(\ref{Fd2}) violates the inequality (3) for $\theta,\gamma\in (0,\frac{\pi}{2})$. The maximal violation of
\begin{eqnarray}
v_{\max}=\max_{a,b,c}\{P_q(a,b,c)- p(a)^{\frac{m-1}{m}}p(b)^{\frac{1}{m}}p(c)^{\frac{m-1}{m}}\}
\label{Fd10a}
\end{eqnarray}
is shown in Fig.\ref{figs9}(b). The inequality (3) can be used for verifying all the generalized W states. However, from Fig.\ref{figs9}(a), the inequality (\ref{fin1}) is useless for verifying the correlation of $P_w$ in Eq.(\ref{Fd2}).

\begin{figure}
\begin{center}
\resizebox{220pt}{120pt}{\includegraphics{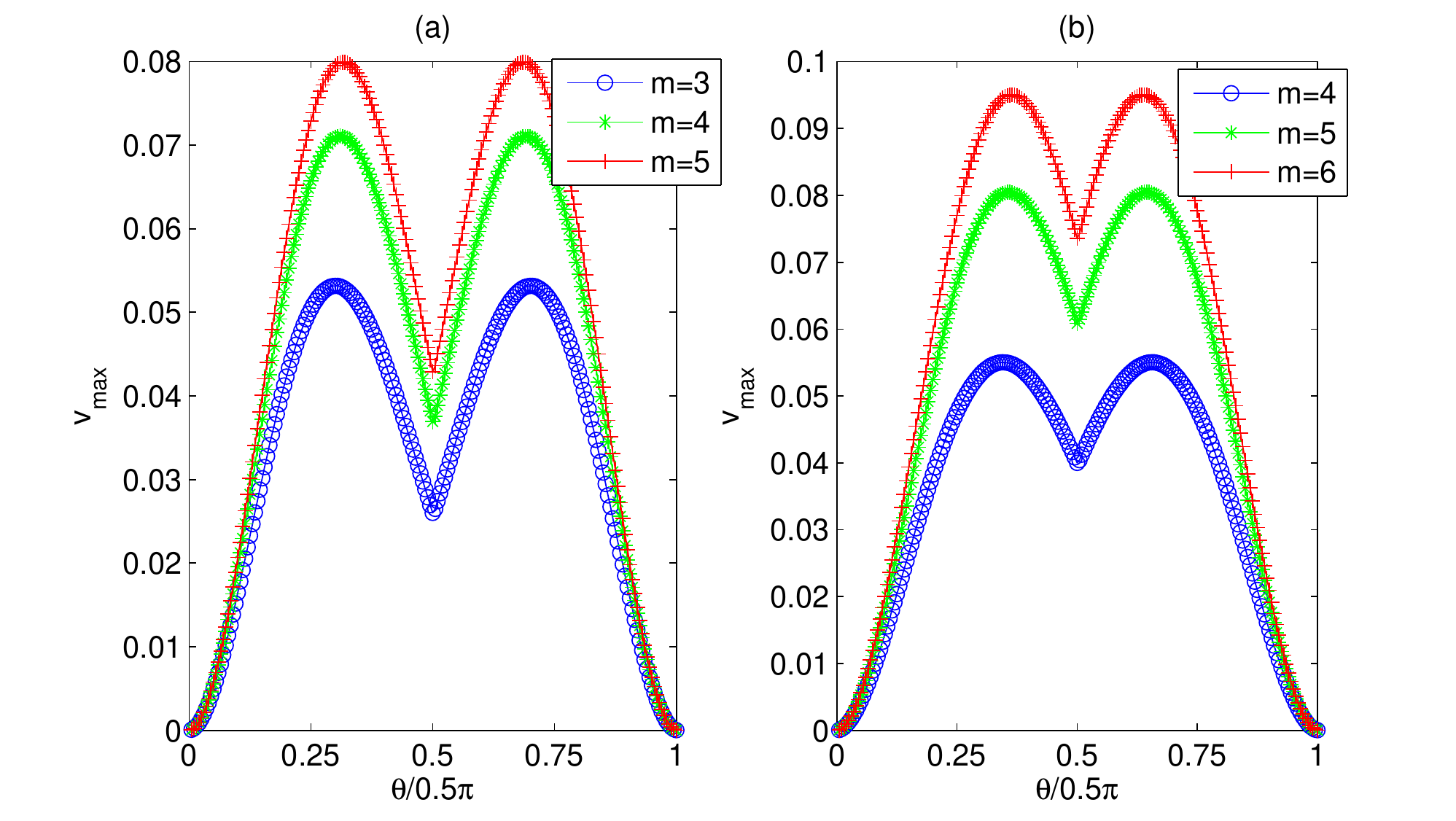}}
\end{center}
\caption{\small (Color online) Violations of networks (a) Triangle network consisting of three generalized EPR states. Here, $\theta_1=\theta_2=\theta_3$.  (b) 4-partite network consisting of two generalized GHZ states. Here, $\theta_1=\theta_2$. $v_{\max}$ is defined in Eq.(\ref{Fd10a}).}
\label{figs10a}
\end{figure}

In what follows, consider a triangle network ${\cal N}_q$, as shown in Fig.\ref{figs7}(a), consisting of three generalized EPR states \cite{EPR} in Eq.(\ref{F11a}), where each party shares two qubits from two EPR states. After the projection measurement under the basis $\{|0\rangle:=|00\rangle, |1\rangle:=|00\rangle, |2\rangle:=|10\rangle, |3\rangle:=|11\rangle\}$, the correlation is given by Eq.(\ref{C2}) with $p_i=\cos^2\theta_i$ and $q_i=\sin^2\theta_i$, $i=1, 2, 3$. From Eqs.(\ref{C2}) and (\ref{C3}), it follows that
\begin{eqnarray}
v_{000}&:=&P_{tri}(0,0,0)-p_a(0)^{\frac{m-1}{m}}p_b(0)^{\frac{1}{m}}p_c(0)^{\frac{m-1}{m}}
\nonumber\\
&=& p_1p_2p_3-p_1p_2p_3^{\frac{2m-2}{m}}
\nonumber\\
&>&0
\label{Fd14}
\end{eqnarray}
for $m>2$ and $\theta_1, \theta_2, \theta_3\in (0, \frac{\pi}{2})$. The distribution of $P_{tri}(0,0,0)$ violates the inequality (3). Similar results hold for other correlations. The maximal violation of $v_{\max}$ in Eq.(\ref{Fd10a}) is shown in Fig.\ref{figs10a}(a). The inequality (3) is useful for verifying all the correlations derived from ${\cal N}_3$ in Fig.\ref{figs7}(a).

To sum up, the inequality (3) is useful for verifying the chain network consisting of two sources.

\textbf{Example S11 (4-partite network)}. Consider a 4-partite network ${\cal N}_4$, there are lots of different configurations which may consist of single sources, two sources, three sources or four sources. Here, we consider one network topology, as shown in Fig.\ref{figs7}(b), consisting of two generalized GHZ states \cite{GHZ}: $|ghz_1\rangle_{123}|ghz_2\rangle_{456}$, where $|ghz_i\rangle$ is given by
\begin{eqnarray}
|ghz_i\rangle=\cos\theta_i|000\rangle+\sin\theta_i|111\rangle
\label{Fd15}
\end{eqnarray}
for $\theta_i\in (0,\frac{\pi}{2})$, $i=1, 2$. Here, two parties $\textsf{A}_1$ and $\textsf{A}_4$ hold respectively one qubit 1 and 6, while $\textsf{A}_2$ and $\textsf{A}_3$ hold respectively two qubits $(2,4)$ and $(3,5)$. From Theorem 2, a new configuration inequality is given by
\begin{eqnarray}
p(a_1,\cdots,a_4)\leq (p(a_1)p(a_2))^{\frac{m-2}{m}}(p(a_3)p(a_4))^{\frac{1}{m}}
\label{Fd16}
\end{eqnarray}
for $m\geq 3$. This inequality is useful for verifying the distributions derived from ${\cal N}_4$.

Especially, $\textsf{A}_1$ and $\textsf{A}_4$ perform local projection measurements under the basis $\{|0\rangle, |1\rangle\}$. $\textsf{A}_2$ and $\textsf{A}_3$ perform local projection measurements under the basis $\{|0\rangle:=|00\rangle, |1\rangle:=|01\rangle, |2\rangle:=|10\rangle, |3\rangle:=|11\rangle\}$. The joint distribution is given by
\begin{eqnarray}
P(a_1,\cdots,a_4)&=&\cos^2\theta_1\cos^2\theta_2[0000]
\nonumber\\
&&+\cos^2\theta_1\sin^2\theta_2[0111]
\nonumber\\
&&+\sin^2\theta_1\cos^2\theta_2[1220]
\nonumber\\
&&+\sin^2\theta_1\sin^2\theta_2[1331]
\label{Fd17}
\end{eqnarray}
It follows that
\begin{eqnarray}
&&p_{a_1}(0)=\cos^2\theta_1, p_{a_1}(1)=\sin^2\theta_1
\nonumber\\
&&p_{a_2}(0)=p_{a_3}(0)=\cos^2\theta_1\cos^2\theta_2,
\nonumber\\
&&p_{a_2}(1)=p_{a_3}(1)=\cos^2\theta_1\sin^2\theta_2
\nonumber\\
&&p_{a_2}(2)=p_{a_3}(2)=\sin^2\theta_1\cos^2\theta_2,
\nonumber\\
&&p_{a_2}(3)=p_{a_3}(3)=\sin^2\theta_1\sin^2\theta_2,
\nonumber\\
&&p_{a_4}(0)=\cos^2\theta_2, p_{a_4}(1)=\sin^2\theta_2
\label{Fd18}
\end{eqnarray}

From Eqs.(\ref{Fd17}) and (\ref{Fd18}), we get
\begin{eqnarray}
v_{0000}&:=&P(0,0,0,0)-(p_{a_1}(0)p_{a_2}(0))^{\frac{m-2}{m}}(p_{a_3}(0)p_{a_4}(0))^{\frac{1}{m}}
\nonumber\\
&=&\cos^2\theta_1\cos^2\theta_2-\cos^{\frac{6m-12}{m}}\theta_1\cos^{\frac{6m-12}{m}}\theta_2
\nonumber\\
&>&0
\label{Fd19}
\end{eqnarray}
for $m\geq 3$ and $\theta_i\in (0,\frac{\pi}{2})$, $i=1, 2$. The distribution $P(0,0,0,0)$ violates the inequality (\ref{Fd16}). The maximal violation of
\begin{eqnarray}
v_{\max}&=&\max_{a_1,\cdots, a_4}\{P(a_1,a_2,a_3,a_4)- (p(a_1)p(a_2))^{\frac{m-2}{m}}
\nonumber\\
&&\times (p(a_3)p(a_4))^{\frac{1}{m}}\}
\label{Fd20}
\end{eqnarray}
is shown in Fig.\ref{figs10a}(b). The inequality (\ref{Fd16}) is useful for verifying all the correlations derived from ${\cal N}_4$ in Fig.\ref{figs7}(b).

\section{Witnessing any multipartite entangled pure states}

In this section, we firstly prove Corollary 2 in Subsec.J.1 which implies an efficient method for witnessing almost all multipartite entangled pure states. Moreover, we provide one example for witnessing generalized tripartite entangled pure states in Subsec.J.2. We present a general method for testing the $k$-separablity of multipartite pure states in Subsec.J.3.

\subsection{Proof of Corollary 2}

Consider an $n$-partite state $|\Phi\rangle_{A_1\cdots{}A_n}$ on Hilbert space $\otimes_{j=1}^n\mathbb{H}_{j}$.

\begin{figure}
\begin{center}
\resizebox{220pt}{120pt}{\includegraphics{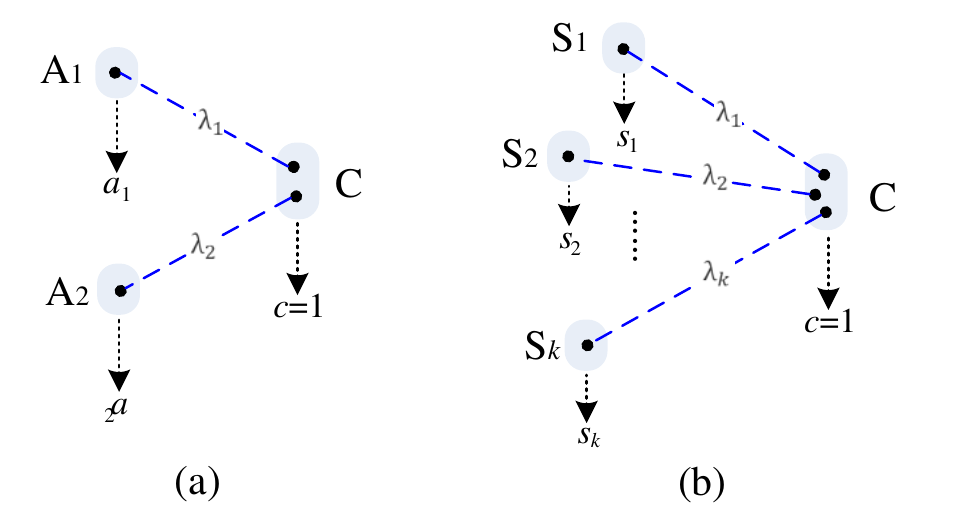}}
\end{center}
\caption{\small (Color online) (a) An equivalent tripartite chain network. Each pair of $A_1$ and $C$ (or $C$ and $A_2$ shares an auxiliary source $\lambda_1$ (or $\lambda_2$). (b) An equivalent $k$-partite chain network. Each pair of $S_i$ and $C$ share one auxiliary source $\lambda_i$, $i=1, \cdots, k$. The output $c$ of $C$ is fixed as $1$ with unit probability.}
\label{figs10}
\end{figure}

Assume $|\Phi\rangle_{A_1\cdots{}A_n}$ is separable. There is an integer $i$ such that $A_i$ and $A_{i+1}$ are separable, that is,
\begin{eqnarray}
\rho_{A_iA_{i+1}}=\rho_{A_i}\rho_{A_{i+1}}
\label{JJ0}
\end{eqnarray}
where $\rho_{A_iA_{i+1}}={\rm tr}_{A_j,j\not=i,i+1}(|\Phi\rangle\langle \Phi|)$, $\rho_{A_i}={\rm tr}_{A_j,j\not=i}(|\Phi\rangle\langle \Phi|)$ and $\rho_{A_{i+1}}={\rm tr}_{A_j,j\not={i+1}}(|\Phi\rangle\langle \Phi|)$ denote the reduced density matrices. Under the local projection under the computation basis $\{|j\rangle, \forall j\}$, the joint distribution $P(a_i,a_{i+1})$ satisfies
\begin{eqnarray}
P(a_i,a_{i+1})=p(a_i)p(a_{i+1})
\label{JJ1}
\end{eqnarray}
With this fact, it only needs to consider the reduced distribution $P(a_i,a_{i+1})$. Take $A_1$ and $A_{2}$ as an example. If they are separable, $A_1$, $A_2$ and auxiliary particle $C$ may be regarded as an equivalent tripartite chain network ${\cal N}_q$, as shown in Fig.\ref{figs10}(a), where each pair of $A_1$ and $C$ (or $C$ and $A_2$ shares an auxiliary source $\lambda_1$ (or $\lambda_2$). The output $c$ of $C$ is fixed as $1$ with unit probability. From Eq.(\ref{eqn-1}), the joint distribution satisfies
\begin{eqnarray}
P(a_1,c=1,a_2)=P(a_1,a_2)
\label{JJ2}
\end{eqnarray}
where $a_i$ is independent of the shared auxiliary source $\lambda_i$, $i=1,2$. Note $A_1$ and $A_2$ are separable. From Theorem 2, the joint distribution $P(a_1,c=1,a_2)$ satisfies the configuration inequality
\begin{eqnarray}
P(a_1,c=1,a_2)&\leq & p(a_1)^{\frac{m-1}{m}}p(c=1)^{\frac{1}{m}}p(a_2)^{\frac{m-1}{m}}
\nonumber
\\
&=& p(a_1)^{\frac{m-1}{m}}p(a_2)^{\frac{m-1}{m}}
\label{JJ3}
\end{eqnarray}
for any $m\geq 2$.

\textbf{Fact}. The inequality (\ref{JJ3}) holds if and only if the following inequality holds
\begin{eqnarray}
P(a_1,a_2)\leq p(a_1)p(a_2)
\label{JJ4}
\end{eqnarray}

This can be proved by using the equality of $\lim_{m\to \infty}p(a_1)^{\frac{m-1}{m}}p(c=1)^{\frac{1}{m}}p(a_2)^{\frac{m-1}{m}}=p(a_1)p(a_2)$. Eq.(\ref{JJ1}) consists of one facet of correlations for bipartite entangled pure states. So, $a_i$ and $a_{i+1}$ are dependent ($A_i$ and $A_{i+1}$ are entangled in $|\Phi\rangle_{A_1\cdots{}A_n}$) if and only if the reduced probability distribution $P(a_1,a_2)$ is not independent for some $a_i$ and $a_2$, that is, $P(a_1,a_2)$ violates the inequality (\ref{JJ4}) for some $a_i$ and $a_2$.

Now, we provide an efficient method to witness any multipartite entanglement as follows.

\begin{minipage}{8.2cm}
\begin{algorithm}[H]
	\centering
	\caption{Witnessing multipartite entanglement}
\begin{itemize}
\item[Input:] An $n$-partite entangled state $|\Phi\rangle_{A_1\cdots{}A_n}$.
\item[Output:] $n$-partite entanglement.
\item{} For $i$ from $1$ to $n-1$.
\item{} Obtain the probability distribution $P(a_i,a_{i+1})$ under the local measurement basis of $\{|s\rangle_{A_i}\}$ and $\{|t\rangle_{A_{i+1}}\}$.
\item{} Test the configuration inequality (\ref{JJ4}).
\item{} Return: $|\Phi\rangle$ is $n$-partite entanglement if all the distributions  $P(a_i,a_{i+1})$'s violate the inequality (\ref{JJ2}) for some $a_i$ and $a_{i+1}$.
\end{itemize}

\label{alg4}
\end{algorithm}
\end{minipage}

The proof is completed by verifying the inequality (\ref{JJ4}) for all the reduced probability distributions $P(a_i,a_{i+1})$, $i=1, \cdots, n-1$, that is, testing $n-1$ inequalities for all the distributions $P(a_i,a_{i+1})$'s.  Consider a joint probability distribution $P(a_i,a_{i+1})$. If $P(a_i,a_{i+1})$ violates the inequality (\ref{JJ4}) for some specific values of $a_i$ and $a_{i+1}$, the particles $A_i$ and $A_{i+1}$ are entangled. Otherwise, all the probability distributions of $P(a_i,a_{i+1})$ do not violate the inequality (\ref{JJ4}). Note that
\begin{eqnarray}
\sum_{a_i,a_{i+1}}P(a_i,a_{i+1})=\sum_{a_i,a_{i+1}}p(a_1)p(a_2)=1
\label{JJ5}
\end{eqnarray}
It follows Eq.(\ref{JJ1}) holds for all the values of $a_i$ and $a_{i+1}$. These equalities may hold for some entangled states (fo example, $|\Phi\rangle=\frac{1}{2}|00\rangle+|01\rangle+|10\rangle-|11\rangle$). Fortunately, these equalities has only defined a zero-measure of subset in all the values of $a_i$ and $a_{i+1}$.

Note each distribution $P(a_i,a_{i+1})$ can be obtained with constant complexity under the local projection measurements. Moreover, for each probability distribution $P(a_i,a_{i+1})$, the test of the configuration inequality (\ref{JJ4}) has constant complexity. It follows the total time of Algorithm 4 is linear complexity because only $n-1$ number of two-partite inequalities will be tested. This completes the proof.

\subsection{Witnessing a general three-qubit state}

Consider a general three-qubit state $|\Phi\rangle_{ABC}$ with the following normal form \cite{Acin} of
\begin{eqnarray}
|\Phi\rangle_{ABC}&=&\alpha_{0}|000\rangle+\alpha_1e^{i\phi}|100\rangle
+\alpha_2|101\rangle
\nonumber\\
&&+\alpha_3|110\rangle+\alpha_4|111\rangle
\label{I1}
\end{eqnarray}
where $\alpha_i$ satisfy $\sum_i\alpha_i^2=1$. Its entanglement can be verified by using the bipartitions $\{A\}$ and $\{B,C\}$, $\{B\}$ and $\{A,C\}$, and $\{C\}$ and $\{A,B\}$ by using the inequality (\ref{fin4'}).

After the projection measurement under the basis $\{|0\rangle, |1\rangle\}$, the joint distribution is given by
\begin{eqnarray}
P(a,b,c)&=&\alpha_{0}^2[000]+\alpha_1^2[100]+\alpha_2^2[101]
\nonumber\\
&&+\alpha_3^2[110]+\alpha_4^2[111]
\label{I2}
\end{eqnarray}
It follows that
\begin{eqnarray}
&&p_a(0)=\alpha_{0}^2,p_a(1)=1-\alpha_{0}^2,
\nonumber\\
&&p_b(0)=\alpha_{3}^2+\alpha_4^2, p_b(1)=1-\alpha_{3}^2-\alpha_4^2,
\nonumber\\
&&p_c(0)=\alpha_{2}^2+\alpha_4^2, p_c(1)=1-\alpha_{2}^2-\alpha_4^2,
\nonumber\\
&&P_{ab}(0,0)=\alpha_{0}^2, P_{ab}(1,0)=\alpha_{1}^2+\alpha_2^2,
\nonumber\\
&&P_{ab}(11)=\alpha_{3}^2+\alpha_{4}^2, P_{bc}(0,0)=\alpha_{0}^2+\alpha_1^2,
\nonumber\\
&&P_{bc}(0,1)=\alpha_{2}^2, P_{bc}(1,0)=\alpha_3^2,
\nonumber\\
&& P_{bc}(1,1)=\alpha_{4}^2, P_{ac}(0,0)=\alpha_{0}^2,
\nonumber\\
&&
P_{ac}(1,0)=\alpha_1^2+\alpha_3^2, P_{ac}(11)=\alpha_2^2+\alpha_{4}^2
\label{I3}
\end{eqnarray}

Firstly, if $|\Phi\rangle_{ABC}$ is separable in terms of the bipartition $\{A\}$ and $\{B,C\}$, the reduced distribution $\{P(a,b)\}$ satisfies the inequality (\ref{JJ4}). From Eqs.(\ref{I2}) and (\ref{I3}), it follows that
\begin{eqnarray}
v_{00}&:=&P_{ab}(0,0)-p_a(0)p_b(0)
\nonumber\\
&=&\alpha_0^2-\alpha_0^2(\alpha_{3}^2+\alpha_4^2)
\nonumber\\
&>&0
\label{I5}
\end{eqnarray}
if $\alpha_0^2$ and $\alpha_{3}^2+\alpha_4^2\not=0$, which holds for $\alpha_{0},\alpha_3, \alpha_4 \in (0,1)$ and $\alpha_{1},\alpha_2\in [0,1)$. The distribution $P_{ab}(0,0)$ in Eq.(\ref{I2}) violates the inequality (\ref{JJ4}) for $\alpha_{0},\alpha_3, \alpha_4 \in (0,1)$ and $\alpha_{1},\alpha_2\in [0,1)$.

Secondly, if $|\Phi\rangle_{ABC}$ is separable in terms of the bipartition $\{B\}$ and $\{A,C\}$, the reduced distribution $\{P(a,c)\}$ satisfies the inequality (\ref{JJ4}). From Eqs.(\ref{I2}) and (\ref{I3}), it follows that
\begin{eqnarray}
v_{00}&:=&P_{ac}(0,0)-p_a(0)p_c(0)
\nonumber\\
&=&\alpha_{0}^2-\alpha_0^2(\alpha_{2}^2+\alpha_4^2)
\nonumber\\
&>&0
\label{I10a}
\end{eqnarray}
if $\alpha_0^2$ and $\alpha_{2}^2+\alpha_4^2\not=0$, which holds for $\alpha_{0},\alpha_2, \alpha_4 \in (0,1)$ and $\alpha_{1},\alpha_3\in [0,1)$. The distribution $P_{ac}(0,0)$ in Eq.(\ref{I2}) violates the inequality (\ref{JJ4}) for $\alpha_{0},\alpha_2, \alpha_4 \in (0,1)$ and $\alpha_{1},\alpha_3\in [0,1)$.

Thirdly, if $|\Phi\rangle_{ABC}$ is separable in terms of the bipartition $\{A\}$ and $\{B,C\}$, the reduced distribution $\{P(b,c)\}$ satisfies the inequality (\ref{JJ4}). From Eqs.(\ref{I2}) and (\ref{I3}), it follows that
\begin{eqnarray}
v_{00}&:=&P_{bc}(0,0)-p_b(0)p_c(0)
\nonumber\\
&=&\alpha_{0}^2+\alpha_1^2-(\alpha_{3}^2+\alpha_4^2)(\alpha_{2}^2+\alpha_4^2)
\label{I10}
\\
v_{01}&:=&P_{bc}(0,1)-p_b(0)p_c(1)
\nonumber\\
&=&\alpha_{2}^2-(\alpha_{3}^2+\alpha_4^2)(1-\alpha_{2}^2-\alpha_4^2)
\label{I11}
\\
v_{10}&:=&P_{bc}(1,0)-p_b(1)p_c(0)
\nonumber\\
&=&\alpha_{3}^2-(1-\alpha_{3}^2-\alpha_4^2)(\alpha_{2}^2+\alpha_4^2)
\label{I12}
\\
v_{11}&:=&P_{bc}(1,1)-p_b(1)p_c(1)
\nonumber\\
&=&\alpha_{4}^2-(1-\alpha_{3}^2-\alpha_4^2)(1-\alpha_{2}^2-\alpha_4^2)
\label{I13}
\end{eqnarray}
Note $\sum_{b,c}P(b,c)=\sum_{b,c}p(b)p(c)$. It follows $v_{ij}>0$ for some $i,j\in \{0,1\}$, except for $v_{00}=v_{01}=v_{10}=v_{11}=0$. The distribution $\{P(b,c)\}$ in Eq.(\ref{I2}) violates the inequality (\ref{JJ4}) for almost $\alpha_i$'s, except for a subset defined by $v_{00}=v_{01}=v_{10}=v_{11}=0$ with zero-measure on the sphere of $\sum_i\alpha_i^2=1$.

To sum up, the tripartite entanglement of $|\Phi\rangle_{ABC}$ can be almost verified except for a subset defined by
\begin{eqnarray}
\left\{
\begin{array}{lll}
\alpha_{0}^2+\alpha_1^2-(\alpha_{3}^2+\alpha_4^2)(\alpha_{2}^2+\alpha_4^2)=0
\\
\alpha_{2}^2-(\alpha_{3}^2+\alpha_4^2)(1-\alpha_{2}^2-\alpha_4^2)=0
\\
\alpha_{3}^2-(1-\alpha_{3}^2-\alpha_4^2)(\alpha_{2}^2+\alpha_4^2)=0
\\
\alpha_{4}^2-(1-\alpha_{3}^2-\alpha_4^2)(1-\alpha_{2}^2-\alpha_4^2)=0
\\
\sum_{i}\alpha_i^2=1
\end{array}
\right.
\end{eqnarray}
which have the soluations of
\begin{eqnarray}
\alpha_{0}^2+\alpha_1^2=\alpha_{2}^2=\alpha_{3}^2=\alpha_{4}^2=\frac{1}{4}
\end{eqnarray}

\subsection{Testing the $k$-separability of multipartite pure states}

Algorithm 4 is easily extended to verify the $k$-separability of $|\Phi\rangle_{A_1\cdots A_n}$ in terms of $k$ partition $S_1, \cdots, S_k$ of $\{A_1, \cdots, A_n\}$. Here, $|\Phi\rangle$ is regarded as an equivalent $k+1$-partite star network ${\cal N}_q$ consisting of particles in $S_1, \cdots, S_k$ and auxiliary particle $C$,  as shown in Fig.\ref{figs10}(b), where each pair of $S_i$ and $C$ share one auxiliary source $\lambda_i$, $i=1, \cdots, k$. The output $c$ of $C$ is fixed as $1$ with unit probability. From Eq.(1) the joint distribution satisfies $P(S_1,\cdots, S_k, C=1)=P(S_1,\cdots, S_k)$. From Theorem 2, it follows that
\begin{eqnarray}
P(S_1,\cdots, S_k)\leq \prod_{i=1}^kP(s_i)^{\frac{m-1}{m}}
\label{star}
\end{eqnarray}
This inequality is useful for testing the $k$-separability of $|\Phi\rangle_{A_1\cdots A_n}$.

\section{Proof of Theorem 3}

In this section, we consider the correlations achievable in a generalized network with only the no-signaling (NS) principle. In this model, the resources are not specified as classical variables or quantum states, but general NS boxes. Each party having an arbitrary number of NS boxes shared with others, can locally ''wire" these boxes in the most general way to determine an output. The inputs of certain boxes can be chosen by the party, while others can be determined by wirings, the output of one box being used as in the input for another one. All theories in this framework share the following features with classical and quantum theory \cite{PR}.
\begin{itemize}
\item{}\textbf{Axiom 1}(\textit{Convexity}) The system and local measurements are represented by some space, that is, $\mathbb{H}=\{\rho\}$ consists of all the system $\rho$, ${\cal M}_a=\{M_a\}$ and ${\cal M}_b=\{M_b\}$ consist of all the local measurement devices $M_a$ and $M_b$ respectively. All of these spaces are convex, that is, $\rho\in {\cal H}$ with $\rho=p_1\rho_1+p_2\rho_2$ and $M\in {\cal M}$ with $M=p_1M_1+p_2M_2$, where $\rho_1,\rho_2\in \mathbb{H}$, $M_1,M_2\in {\cal M}$ and $\{p_1,p_2\}$ is a probability distribution.
\item{}\textbf{Axiom 2}(\textit{Distinguishability}) Local measurements are distinguishable for each subsystem, that is, $M_a$ and $M_b$ can be labeled as different indexes $a,b\in \mathbb{Z}$.
\item{} \textbf{Axiom 3}(\textit{Commutativity}) Local operations on distinct subsystems are commutative. In the case of a bipartite system $A$ and $B$, for example, if an operation is performed on the system $A$ alone, and an operation on the system $B$ alone, the order of the operations being performed does not affect the result, that is, $G(U_a,U_b,\rho)=G(U_b,G(U_a,\rho))=G(U_a,G(U_b,\rho))$, where $G(U_a,U_b,\rho)$ denotes local operations $U_a$ and $U_b$ simultaneously performed on system $\rho$; $G(U_b,G(U_a,\rho))$ denotes that local operation $U_a$ is first performed and then $U_b$ is performed; $G(U_a,G(U_b,\rho))$ denotes that local operation $U_b$ is first performed and then $U_a$ is performed, and $G$ denotes the general operation mapping.
\item{}\textbf{Axiom 4} (\textit{Statistics}) The measurement statistical distribution satisfies the large number rule, that is, one can get a positive number of $P_{ns}(a,b)\geq 0$ from a large number of the same measurement on the identity  and independent system. In this case, the composite system is determined by correlations between local measurements, that is, $P_{ns}(a,b)=F_{ns}(\rho,M_a,M_b)$ with some mapping $F_{ns}$. All the measurement outcomes satisfy a probability distribution, $\sum_{a,b}P_{ns}(a,b)=1$, and $P_{ns}(a,b)\geq 0$.
\item{}\textbf{Axiom 5}(\textit{No-signalling principle}) All the local measurements satisfy the no-signalling principle that requires local operations on local systems cannot instantaneously relay information to other systems, that is, $p_{ns}(a)=\sum_{b}P_{ns}(a,b)$ and $p_{ns}(b)=\sum_{a}P_{ns}(a,b)$.
\item{}\textbf{Axiom 6} (\textit{Linearity}) $F_{ns}$ is linear for the system $\rho$ and  local measurements $M_a,M_b$, that is,
\begin{eqnarray}
F_{ns}(p_1\rho_1+p_2\rho_2,M_a,M_b) &=& p_1F_{ns}(\rho_1,M_a,M_b)
 \nonumber\\
& &+p_2F_{ns}(\rho_2,M_a,M_b)
 \nonumber\\
F_{ns}(\rho,p_1M_a+p_2M_a',M_b) &=& p_1F_{ns}(\rho,M_a,M_b)
 \nonumber\\
& &+p_2F_{ns}(\rho,M_a',M_b)
 \nonumber\\
F_{ns}(\rho,M_a,p_1M_b+p_2M_b') &=& p_1F_{ns}(\rho,M_a,M_b)
\nonumber\\
& & +p_2F_{ns}(\rho,M_a,M_b')
\label{H1}
\end{eqnarray}
where $p_1\rho_1+p_2\rho_2$ denotes the superposition of two systems $\rho_1$ and $\rho_2$, and $p_1M_a+p_2M_a'$ and $p_1M_b+p_2M_b'$ denote the superposition of two local measurements $M_b$ and $M_b'$ with the probability distribution $\{p_1,p_2\}$.
\end{itemize}

Here, Axiom 1 shows the convexity of the spaces for the representations of the involved system and local measurements. The main reason is from the classical probability. One can always represent a probabilistic mixture of different states or measurements. Axiom 2 shows the distinguishability of local measurements, which may be derived from the discrete features of times for the measurements. Otherwise, one can discretize it by observing the involved system. Axiom 3 is used for the commutativity of local measurements on different systems. This implies the joint statistics derived from two local measurements are independent of the measurement order being performed. Otherwise, the measurement outcomes depend on the measurement order. Axiom 4 is used for the statistical distribution of measurement outcomes, which can be described by the large number law. It means one can get the measurement outcome statistics (distributions) by using larger number of experiments with independent and identically distributed systems and measurements. Axiom 5 shows the no-signalling principle, which means local statistics is completed determined by local measurements. Axiom 5 shows the linearity of statistical representations of measurement outcomes. Generally, these axioms are all related to the measurement statistics which is involved in the main result in what follows. These axioms hold for the classical model, hidden variable model, quantum mechanics and PR-box \cite{PR,BLM,Barrett}. However, pseudo-probable model \cite{Bartlett1,Feynman} which may result in negative distributions does not satisfy Axiom 4. With these axioms, the generalized Finner inequality holds for no-signaling networks as follows

\textbf{Theorem 3' (No-signalling Finner inequality)}. \textit{Let $s=\{s_1, \cdots, s_n\}$ be a fractional independent set of ${\cal N}_{ns}$. $f_j$ are real positive local post-processing (functions) of the classical output of the party $\textsf{A}_j$. Then, any correlation $P$ achievable in the no-signalling network ${\cal N}_{ns}$ satisfies
\begin{eqnarray}
\mathbb{E}_{ns}[\prod_{j=1}^nf_j]\leq \prod_j\|f_j\|_{\frac{1}{s_j}}
\label{H2}
\end{eqnarray}
In particular, if $f_j$ is the indicator function of the output of $\mathsf{A}_j$ being $a_j$, we have
\begin{eqnarray}
P_{ns}({\bf a}) \leq \prod_{j=1}^nP_{ns}(a_j)^{s_j}
\label{H3}
\end{eqnarray}}

\textbf{Proof of Theorem 3'}. Similar to proof of Theorem 2, it only need to consider weights $\textbf{s}=(s_1, \cdots, s_n)$ with $s_j\in (0,1)$ for all $j$. Assume $f_j$ is an arbitrary positive function applied on the output of the $j$-th party $\textsf{A}_j$. From Axioms 1 and 2, each local measurement may be linearly represented by some basic measurements $M_{a_j}$, that is, $X_j=\sum_{a_j}p_{a_j}M_{a_j}$ with a probability distribution $\{p_{a_j}\}$. Note each system $\rho\in {\cal H}$ can be linearly represented by some basic states $\rho_j\in {\cal H}$, that is, $\rho=\sum_jq_j\rho_{j}$ with a probability distribution $\{q_j\}$. In this case, we define
\begin{eqnarray}
 X_j=\sum_{a_j}f_j(a_j)M_{a_j}
\label{H5}
\end{eqnarray}
where $\{M_{a_j}\}$ are local measurement devices, $f_j$ denotes positive the post-processing function of the $j$-th party. From the linearity of the measurement assumption in Axiom 5 we get
\begin{eqnarray}
& & F_{ns}(\rho,X_1, \cdots, X_n)
 \nonumber
 \\
 &=&\sum_{{\bf a}}\prod_{j=1}^nf_j(a_j)F_{ns}(\rho, M_{a_1}, \cdots, M_{a_n})
 \nonumber
 \\
 &=&\sum_{{\bf a}}\prod_{j=1}^nf_j(a_j)P_{ns}({\bf a})
\label{H6}
 \\
 &\leq &\mathbb{E}_c[\prod_{j=1}^nf_j]
\label{H61}
\end{eqnarray}
where $\{P_{ns}({\bf a})\}$ satisfies $P_{ns}({\bf a})\geq0$ and $\sum_{{\bf a}}P_{ns}({\bf a})\leq 1$ from Axioms 3 and 4. It means $\{P_{ns}({\bf a})\}$ can be extended to a classical probability distribution $\{P_c({\bf a})\}$, by defining $P_c({\bf a})=P_{ns}({\bf a})$ for any $a_1, \cdots, a_n$ and $P_c(a_1=\clubsuit,\cdots, a_n=\clubsuit)=1-\sum_{{\bf a}}P_{ns}({\bf a})$ with $\clubsuit\not\in \cup_i\{a_i\}$. The inequality (\ref{H61}) follows from the inequalities of $f_j(a_j)\geq0$. Here, $\mathbb{E}_c[\prod_{j=1}^nf_j]$ denotes the classical expect of the positive function $\prod_{j=1}^nf_j$ associated with the classical distribution $\{P_c({\bf a})\}$. Note each $m$-partite system $\rho$ can be regarded as an $m$-hype edge connected by $m$ vertexes. With the equivalence, we get from the generalized Finner inequality (\ref{E1}) that
\begin{eqnarray}
F_{ns}(\rho,X_1, \cdots, X_n) \leq \prod_{j=1}^n\|f_j\|_{\frac{1}{s_j}}
\label{H7}
\end{eqnarray}

In what follows, it is easy to prove
\begin{eqnarray}
\mathbb{E}_{ns}[\prod_{j=1}^nf_j]=F_{ns}(\rho,X_1, \cdots, X_n)
\label{H8}
\end{eqnarray}
In fact, from the expect $\mathbb{E}$ of the definition, we have
\begin{eqnarray}
\mathbb{E}_{ns}[\prod_{j=1}^nf_j]=\sum_{{\bf a}}P_{ns}({\bf a})\prod_{j=1}^nf_j(a_j)
\label{H9}
\end{eqnarray}
where $\{a_j\}$ denotes the measurement outcomes of the $j$-th party, and $P_{ns}({\bf a})$ denotes the probability that the $j$-th party gets the measurement outcome $a_j$, $j=1, \cdots, n$. From Eqs.(\ref{H6}) and (\ref{H9}), it implies the equality (\ref{H8}). This completes the proof.

\section{Facet inequalities}

From Theorem 2, we can obtain some interesting facet inequalities.

\textbf{Example S10 (Star Network)}. Consider an $n$-partite star network, as shown in Fig.\ref{figs2}. Each pair of two parties $\textsf{A}_j$ and $\textsf{A}_n$ shares one bipartite entanglement, $j=1, \cdots, n-1$. By using the fractional independent set given by in Eq.(\ref{D3}), it follows that
\begin{eqnarray}
P(\textbf{a})\leq p(a_n)
\label{L1}
\end{eqnarray}
when $m\to\infty$. Another method is using the fractional independent set $\textbf{s}$ given in Eq.(\ref{D4}) to get the facet inequality as
\begin{eqnarray}
P(\textbf{a})\leq \prod_{i=1}^{n-1}p(a_i)
\label{L2}
\end{eqnarray}
Similar result holds for generalized $k$-independent networks \cite{Luo2018}.

\textbf{Example S11 (Chain network)}. Consider an $n$-partite chain network. Here, each pair of two parties $\textsf{A}_j$ and $\textsf{A}_{j+1}$ shares one bipartite source, $j=1, \cdots, n-1$. By using the fractional independent set given in Eq.(\ref{D5}) or (\ref{D6}) we get configuration inequalities as
 \begin{eqnarray}
&&P(\textbf{a})\leq \prod_{\rm{even}\, i}p(a_i),
\\
&&P(\textbf{a})\leq \prod_{\rm{odd}\, i}p(a_i).
\label{L3}
\end{eqnarray}
These inequalities also hold for $n$-partite cyclic networks, shown as Fig.\ref{figs4}.

For other networks, the facet inequalities may be followed from Theorem 2 for large parameter $m$.

\end{document}